\def\huawei{0}

\if\huawei1
\documentclass[acmsmall,authordraft,nonacm]{acmart}
\else
\documentclass[acmsmall]{acmart}
\fi

\usepackage{amsfonts} 

\usepackage{array}
\usepackage{lipsum}

\usepackage{xspace}
\usepackage{algorithm}
\usepackage[noend]{algpseudocode}
\usepackage{amsmath}
\usepackage{caption}
\usepackage{subcaption}
\usepackage{relsize}
\usepackage{xcolor}

\newcolumntype{L}[1]{>{\raggedright\let\newline\\\arraybackslash\hspace{0pt}}m{#1}}
\newcolumntype{C}[1]{>{\centering\let\newline\\\arraybackslash\hspace{0pt}}m{#1}}
\newcolumntype{R}[1]{>{\raggedleft\let\newline\\\arraybackslash\hspace{0pt}}m{#1}}
\usepackage{multirow}

\newcommand{\argmin}{\mathop{\mathrm{arg\,min}}}
\newcommand{\argmax}{\mathop{\mathrm{arg\,max}}}

\newcommand{\MAJOR}[2]{#2}

\newcommand{\MAJORSTARTS}[0]{}
\newcommand{\MAJORENDS}[0]{}

\newcommand{\NSGAP}{$\textit{NSGA-II}'$\xspace}

\newcommand{\APPR}{SEDE\xspace} 
\newcommand{\GAAlg}{PaiR\xspace}
\newcommand{\nsga}{NSGA-II\xspace}
\newcommand{\dnsga}{DeepNSGA-II\xspace}

\newcommand{\HUDD}{HUDD\xspace}

\definecolor{mygreen}{rgb}{0.0, 0.2, 0.13}

\usepackage{marginnote}
\usepackage[backgroundcolor=white,bordercolor=blue,linecolor=blue]{todonotes}

\newcommand{\IEE}{IEE\xspace}

\newcommand{\EquationsSize}{\small}

\usepackage{pifont}
\begin{document}

\title{Simulator-based explanation and debugging of hazard-triggering events in DNN-based safety-critical systems}



\author{Hazem Fahmy}
\affiliation{%
  \institution{SnT Centre, University of Luxembourg}
  \streetaddress{JFK 29}
  \city{Luxembourg}
  \country{Luxembourg}}
\email{hazem.fahmy@uni.lu}

\author{Fabrizio Pastore}
\affiliation{%
  \institution{SnT Centre, University of Luxembourg}
  \streetaddress{JFK 29}
  \city{Luxembourg}
  \country{Luxembourg}}
\email{fabrizio.pastore@uni.lu}

\author{Lionel Briand}
\affiliation{%
  \institution{SnT Centre, University of Luxembourg}
  \streetaddress{JFK 29}
  \city{Luxembourg}
  \country{Luxembourg}}
  \affiliation{%
  \institution{School of EECS, University of Ottawa}
  \city{Ottawa}
  \country{Canada}}
\email{lionel.briand@uni.lu}

\author{Thomas Stifter}
\affiliation{%
  \institution{IEE S.A.}
  \streetaddress{}
  \city{Luxembourg}
  \country{Luxembourg}}
\email{thomas.stifter@iee.lu}



\begin{abstract}
When Deep Neural Networks (DNNs) are used in safety-critical systems, engineers should determine the safety risks associated with failures (i.e., erroneous outputs) observed during testing. For DNNs processing images, engineers visually inspect all failure-inducing images to determine common characteristics among them. Such characteristics correspond to hazard-triggering events (e.g., low illumination) that are essential inputs for safety analysis. 
Though informative, such activity is expensive and error-prone.

To support such safety analysis practices, we propose \APPR, a technique that generates readable descriptions for commonalities in failure-inducing, real-world images and improves the DNN through effective retraining. 
\APPR leverages the availability of simulators, which are commonly used for cyber-physical systems. 
It relies on genetic algorithms to drive simulators towards the generation of images that are similar to failure-inducing, real-world images in the test set; it then employs
rule learning algorithms to derive expressions that capture  commonalities in terms of 
simulator parameter  values. The derived expressions are then used to generate additional images to retrain and improve the DNN.

With DNNs performing in-car sensing tasks, \APPR successfully characterized hazard-triggering events leading to a DNN accuracy drop. Also, \APPR enabled retraining leading to significant improvements in DNN accuracy, up to $18$ percentage points.
\end{abstract}





\setcopyright{acmcopyright}
\acmJournal{TOSEM}
\acmYear{2022} \acmVolume{1} \acmNumber{1} \acmArticle{1} \acmMonth{10} \acmPrice{}\acmDOI{10.1145/3569935}

\maketitle


\section{Introduction}

Deep Neural Networks (DNNs) are increasingly common building blocks in many modern software systems, including safety-critical cyber-physical systems.
This is true for automotive systems, where DNNs are used for a range of activities, from automating driving tasks, such as emergency braking or lane changing~\cite{NVIDIADNN,TeslaDNN}, to supporting passenger safety through drowsiness detection and gaze detection systems~\cite{Naqvi2018}.

The DNNs used in computer-vision components of safety-critical autonomous systems are trained and tested using both images generated with simulators and real-world images. Because of the costs associated with manual labelling and data collection, the availability of real-world images is often limited, \MAJOR{R1.1}{which may negatively affect the accuracy of DNNs. As a result, DNN developers are increasingly relying on  simulator images to train DNN models, while using real-world images to fine-tune the DNN and then test it}
~\cite{Bird2020,Kim2017,Inoue2018,DiasDaCruz2022Syn2real}. 

\MAJOR{R3.28}{In the presence of DNN failures (i.e., DNN outputs that do not match the ground truth \footnote{In this paper, we use the term \emph{DNN output} to refer to either the predicted class, for classifier DNNs, or the predicted numerical value, for DNNs addressing regression problems. For classifier DNNs, we observe a DNN failure when the predicted class does not match the expected one. For regression DNNs, a DNN failure is observed when the difference between the predicted and the actual value is above a threshold specified by a domain expert.}}), engineers visually inspect the failure-inducing images to perform root cause analysis. In a safety-critical context, the objective of such root cause analysis is to identify 
the events that have triggered DNN failures, which are referred to as the \emph{hazard-triggering events}. Indeed, such identification is part of safety standards (e.g., ISO/PAS 21448 Chapter 7~\cite{SOTIF}) and enables engineers to evaluate the risk associated to potentially hazardous behaviors of the DNN-based system. 
For example, when DNN inputs are images, 
by iteratively inspecting multiple images, engineers should be able to group images presenting common characteristics and therefore identify the hazard-triggering events among such commonalities (e.g., the drivers' head is turned above a certain angle and there is a shadow on their face)\MAJOR{R3.26}{\footnote{Although hazards are triggered by specific inputs (e.g., head turned $43$ degrees leads to misclassification), when identifying hazard-triggering events, engineers are interested in generalizing from the characteristics of multiple failure-inducing inputs (e.g., head turned more than 40 degrees).}}.

In a safety context, the identification of hazard-triggering events enables engineers to estimate the \emph{probability of exposure} to a specific hazard (e.g., a system failure caused by an erroneous output from the DNN), which is necessary for risk assessment. For our example above, engineers
may determine how likely it is for a shadow to partially cover the driver’s face while her head is turned.  

Hazard-triggering events represent the root causes of DNN failures. 
Unfortunately, their manual identification is expensive and error-prone because it requires the inspection and comparison of many DNN inputs (e.g., images). 
To simplify the identification of DNN failures, engineers may rely on visualization techniques to generate heatmaps, i.e., images that use colors to capture the relevance of pixels in their contribution to a DNN output~\cite{Selvaraju17,Montavon2019}. 
Although a human operator may determine the root cause of a DNN failure by noticing that multiple heatmaps highlight the same objects (e.g., long hair~\cite{Selvaraju17}), the analysis of a large set of failure-inducing images is error-prone (e.g., engineers may not notice some of the failure causes). To overcome this problem,
our previous work,
Heatmap-based Unsupervised Debugging of DNNs (HUDD)~\cite{HUDD:TRel}, automatically groups images showing a same root cause by analyzing heatmaps derived from DNN outputs and neuron activations --- such groups of images are named root cause clusters (RCCs). The rationale behind HUDD is that images sharing the same root causes (e.g., a face turned left) should present similar neuron activations and, consequently, similar heatmaps. However, 
with HUDD, root cause analysis still remains error-prone because it relies on the capability of the engineer to interpret the generated outputs (e.g., she may not notice that DNN failures depend not only on a face being turned left but also on a specific illumination angle).

In this paper, we address the problem of automatically generating explicit descriptions for hazard-triggering events from real-world images. Such descriptions are provided in terms of logical expressions constraining the configuration parameters of the simulator used to train the DNN (e.g., rotation angle of the driver’s head and illumination angle). For example, we may report that the hazard-triggering event that prevents the gaze angle from being correctly estimated is the driver's head turned by more than $60$ degrees with an illumination angle above $45$ degrees (i.e., both eyes are barely visible and covered by a shadow).
We name our approach Simulator-based Explanations for DNN failurEs (\APPR).

Given a set of failure-inducing real-world images (e.g., the failure-inducing images in the test set), \APPR relies on HUDD to generate RCCs. We assume that all the images belonging to a RCC present the same hazard-triggering events, as suggested by HUDD's empirical results.
For each RCC identified by HUDD, 
\APPR relies on the simulator used for DNN training to generate more images that belong to the RCC.
Since a RCC characterizes a small portion of the input space, \APPR relies on evolutionary algorithms to efficiently generate RCC images\MAJOR{R3.29}{; indeed, evolutionary algorithms explore the input space guided by fitness functions measuring how close an input is to satisfying a given objective.} To better identify the commonalities among RCC images and avoid generating images that present characteristics that are accidentally shared \MAJOR{R3.30}{\footnote{An evolutionary algorithm may get stuck in a local optimum and, for example, generate images of persons  with blue eyes whose head is turned $45$ degrees, though the eye color does not affect the DNN outputs.}}, we propose \GAAlg (Pairwise Replacement), a genetic algorithm that not only generates images belonging to the RCC but also maximizes their diversity. 

Once \GAAlg has generated images belonging to the RCC, 
to ensure that these images include hazard-triggering events, \APPR produces additional images that are mispredicted, in addition to belonging to the RCC. This is achieved by relying on a modified version of the multi-objective algorithm \nsga (hereafter, \NSGAP). 
\MAJOR{R3.3}{Different from recent extensions of \nsga~\cite{DeepMetis,Riccio2020} that aim to maximize diversity (hereafter, \dnsga), \GAAlg does not rely on an archive and does not require the definition of thresholds for selecting the final set of images.}
Finally, to precisely characterize hazard-triggering events (i.e., to distinguish between commonalities that cause DNN failures and accidental similarities), \APPR relies on \MAJOR{R3.31}{one additional execution of \NSGAP} to generate additional images that are similar to failure-inducing images but do not cause a DNN failure.


The availability of failing and passing images enables \APPR to derive, leveraging the PART decision rule learning algorithm~\cite{PART}, a set of expressions for the simulator parameters that capture commonalities among failure-inducing images. 
These expressions characterize hazard-triggering events and also delimit part of the input space that shall be considered unsafe. The expressions generated by \APPR can be used to either estimate the probability of exposure to the hazard-triggering events (based on domain knowledge) or improve the DNN.  Improvements in DNN accuracy can be achieved by retraining the model using additional images generated using a simulator, configured with parameters that match the expressions generated by \APPR. 

We have performed an empirical evaluation of our approach with three relevant case study subjects provided by our industry partner in the automotive domain, IEE Sensing~\cite{IEE}. Our subjects concern head pose classification and facial landmarks detection, which are important features for in-car sensing solutions such as drowsiness detection. Our empirical results show that (1) \GAAlg outperforms \nsga \MAJOR{R3.3}{and \dnsga} in identifying a larger set of diverse images belonging to a higher number of RCCs, (2) \APPR effectively identifies a set of diverse images with common characteristics, (3) \APPR derives expressions that identify unsafe parts of the input space where the DNN accuracy drops up to $80\%$, and, (4) the expressions generated by \APPR can be used to improve the DNN accuracy (up to $18$ percentage points in accuracy).

The rest of the paper is structured as follows. In Section \ref{sec:background}, we present the required background (heatmaps generation, HUDD, many-objective search algorithms, and rule learning algorithms). In Section \ref{sec:approach}, we describe \APPR. In Section \ref{sec:empirical}, we present our empirical evaluation, which includes a comparison with \nsga, \MAJOR{R3.3}{\dnsga}, HUDD, and a random baseline. In Section \ref{sec:related}, we discuss related work. Finally, we conclude this paper in Section \ref{sec:conclusion}.

\section{Background}
\label{sec:background}

\subsection{DNN Explanation and Heatmaps}
 \label{sec:background:explanation}
  
Most of the approaches that aim to explain DNN outputs~\cite{GARCIA2018} concern the generation of heatmaps that capture the importance of pixels in image predictions. 
\MAJOR{R3.32}{They include both approaches that are black-box (i.e., determine important pixels by observing how the DNN output changes after semi-randomly modifying the input pixel values, without relying on other information~\cite{Petsiuk2018rise,Dabkowski17}) and  white-box (i.e., rely on information captured from internal layers~\cite{Montavon2019,Selvaraju17,Zeiler14,DB15a,Zhou16}).}


Among the existing heatmap generation approaches, HUDD relies on the Layer-Wise Relevance Propagation (LRP)~\cite{Montavon2019}, a white-box approach. First, different from black-box approaches, LRP assesses the relevance of neurons belonging to internal DNN layers\MAJOR{R3.33}{, which enables HUDD to cluster images based on neuron relevance in internal DNN layers. This can be very effective as internal DNN layers act as an abstraction over the inputs and can filter out irrelevant information such as image background}. 
\MAJOR{R3.34, R3.38}{Second, different from white-box approaches that generate class activation maps (CAM) by backpropagating the softmax output to feature maps\footnote{In a Convolutional Neural Network (CNN), a feature map (or activation map) is the matrix resulting from the application of a kernel to the input tensor~\cite{CNNsurvey}.} and upsampling~\cite{Zhou16}, it accounts for all the neurons and layers in the DNN.}
Third, different from 
deconvolutional networks~\cite{Zeiler14} and guided backpropagation~\cite{DB15a}, it generates precise, non-sparse heatmaps\MAJOR{R3.36}{, which should help achieve better clustering results}. Fourth, different from Grad-CAM~\cite{Selvaraju17}, it also works with convolutional DNN layers and regression DNNs.


\begin{figure}[b]
\includegraphics[width=8.4cm]{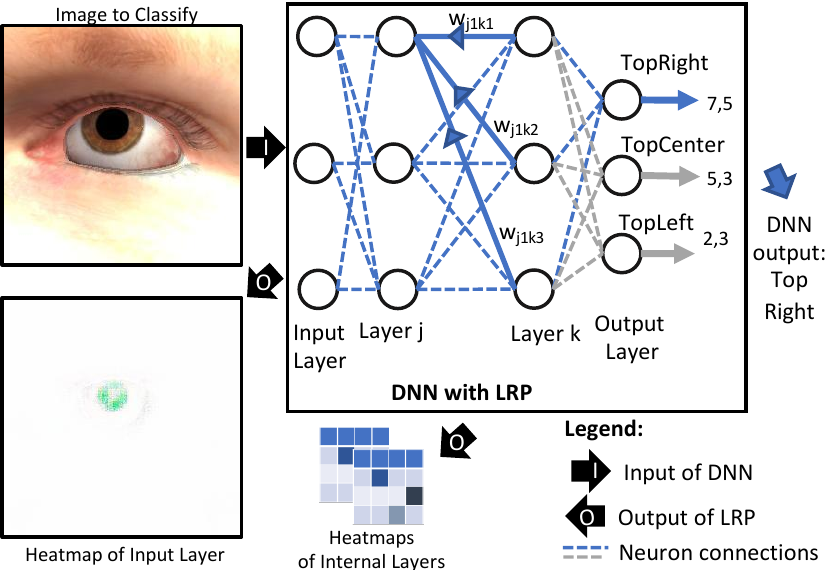}
\caption{Layer-Wise Relevance Propagation.}
\label{fig:LRP}
\end{figure}

LRP redistributes the relevance scores of neurons in a higher layer to scores of the lower layer. Assuming $j$ and $k$ to be two consecutive layers of the DNN, LRP propagates the relevance scores computed for a given layer $k$ to a neuron of the lower layer $j$. It has been theoretically justified as a form of Taylor decomposition~\cite{MONTAVON2017DTD}.
Fig.~\ref{fig:LRP} illustrates the execution of LRP on a fully connected network used to classify inputs. 
\MAJOR{R3.37}{LRP formulas for well-known layer types are available~\cite{Montavon2019}.}

A heatmap is a matrix with entries in $\mathbb{R}$, i.e., it is a triple $(N,M,f)$ where $N,M \in \mathbb{N}$ and $f$ is a map $[N] \times [M] \rightarrow \mathbb{R}$. 
Hereafter, we use the syntax $H_{i,j}^L$ to refer 
 to an entry in row $i$ ($i < N$) and column j ($j < M$) of a heatmap $H$ computed on layer $L$. 
 The size of the heatmap matrix (number of entries) is $N \cdot M$, with $N$ and $M$ depending on the dimensions of the DNN layer L. For convolution layers, $N$ captures the number of neurons in the feature map, while $M$ captures the number of feature maps. 

\subsection{HUDD}
\label{sec:background:HUDD}

HUDD~\cite{HUDD:TRel} identifies the different situations in which an image-processing DNN is likely to trigger an erroneous output by generating clusters (i.e., root cause clusters, RCCs) containing misclassified input images. The images belonging to the same RCC share a common set of characteristics that are plausible causes for failures. 
HUDD is based on the intuition that since heatmaps capture the relevance of each neuron on DNN outputs, the processing of inputs sharing the same root cause (or hazard-triggering event) should lead to similar heatmaps. 
For this reason, to identify the root causes of DNN failures, HUDD relies on clustering based on heatmaps.
 
HUDD works as follows. First, for each failure-inducing image in the test set, it relies on LRP to generate heatmaps of internal DNN layers. Each  \MAJOR{R3.39}{entry of a heatmap for a specific DNN layer captures the relevance score of a neuron in that layer}.
Then, \HUDD relies on a hierarchical agglomerative clustering algorithm~\cite{King:2014} to group images. To measure the distance between two images, which is necessary for clustering, it computes the Euclidean distance between their heatmaps. To determine the number of clusters to generate, it relies on the knee-point method to identify the configuration in which the weighted intra-cluster distance (i.e., the distance between pair of images within a cluster, weighted by the relative size of the cluster) stops decreasing significantly. 

Since certain input space regions might be less dense than others (i.e., include fewer data items), the weighted intra-cluster distance computed by HUDD balances cohesion and cluster size (i.e., it is acceptable to have a larger intra-cluster distance when the cluster has less elements). Also, since LRP can generate heatmaps for each DNN layer, HUDD generates a set of clusters for each DNN layer and selects the one that minimizes the weighted intra-cluster distance (i.e., the one with the most cohesive clusters). 

HUDD has proven useful in identifying root causes (or hazard-triggering events) due to a diverse set of problems, including an incomplete training set, an incomplete definition of the predicted classes, and limitations in the simulator controls.


\subsection{Multi-objective and Many-objective Optimization}

Evolutionary methodologies are state-of-the-art solutions to find a set of nondominated solutions in multi-objective optimization problems. In software engineering, they are widely adopted to identify test inputs that maximize some coverage criterion~\cite{MOSA}. 
Also, they have been successfully used to test autonomous driving systems~\cite{fitash,raja,Abdessalem2018,Waeselynck:2022}.

\nsga~\cite{NSGA2,Abdessalem2018,raja,Waeselynck:2022} is a state-of-the-art solution for optimization problems with up to three objectives. It is shown in Fig.~\ref{algo:nsga}. \nsga works by first generating a random population (Line~\ref{algo:nsga:pop}), which is evolved in a number of iterations (Lines~\ref{algo:nsga:mainCycle:start} to \ref{algo:nsga:mainCycle:end}). A new offspring ($Q_t$) is generated by applying mutation and crossover operators (Line~\ref{algo:nsga:off}). The algorithm preserves elitism since the new population ($P_{t+1}$) is generated by selecting the best individuals in the union of the current population ($P_t$) and the offspring (Line~\ref{algo:nsga:union}). The best individuals are selected by first ranking individuals based on their belonging to the l-th nondominated fronts (Line~\ref{algo:nsga:sort}), which is efficiently performed by the fast-nondominated-sorting procedure~\cite{NSGA2}. Then, the algorithm adds individuals to the population according to their rank (Line~\ref{algo:nsga:add}), till it finds a ranked subset that cannot be fully added (Line~\ref{algo:nsga:add:cond}). Finally, to preserve diversity in the population, \nsga selects the items belonging to the last ranked subset according to a crowding distance function that prioritizes individuals being more distant from others
along with individuals at the boundaries of the objective space
(Lines~\ref{algo:nsga:crowding:start} to~\ref{algo:nsga:crowding:end}).
Fig.~\ref{algo:crowding} provides the pseudocode for the function that computes the crowding-distance for a set of individuals. Lines~\ref{algo:crowding:dist1} and~\ref{algo:crowding:worst} prioritize the individuals with the best and worst fitness, respectively, by assigning them infinite distance.



\begin{figure}

\newcommand{\NSGAT}[1]{\textcolor{red}{#1}}

\begin{algorithmic}[1]

\scriptsize
\Require O = {o1, o2, ... } the objectives
\Require M, population size
\Require T, search budget (i.e., number of iterations to perform)

\Ensure a population whose Pareto front includes the best found solutions

\State $t  \gets 0$ //current generation
\State $P_t  \gets \mathit{CREATE\ RANDOM\ POPULATION\ WITH\ SIZE\ }M$ \label{algo:nsga:pop}
\While {$t < T$} \label{algo:nsga:mainCycle:start}
\State $Q_t  \gets \mathit{GENERATE\ OFFSPRING\ FROM\ }P_t$ \label{algo:nsga:off}
\State $R_{t} \gets P_{t} \cup Q_{t}$  \label{algo:nsga:union}
\State $(F_1,F_2, ..., F_l ) \gets \mathit{APPLY\ }\texttt{fast-nondominated-sorting}\mathit{\ TO\ }R_{t}$ \label{algo:nsga:sort}
\State $r  \gets 0$ {\color{gray}//counter for the Pareto front rank}
\While {$|P_{t+1}| + |F_r| \le M$} \label{algo:nsga:add:cond} {\color{gray}//iterate over all the Pareto fronts till they fit into the population}
\State $P_{t+1} \gets P_{t+1} \cup F_r$ \label{algo:nsga:add}
\State $r \gets r + 1$
\EndWhile
\State $\mathit{ASSIGN\ CROWDING\ DISTANCE\ TO\ INDIVIDUALS\ IN\ }F_{r}$ {\color{gray}//see Fig.~\ref{algo:crowding}; $F_{r}$ is the front that does not fully fit into $P_{t+1}$} \label{algo:nsga:crowding:start}
\State $\mathit{SORT\ } F_{r} \mathit{\ BASED\ ON\ CROWDING\ DISTANCE}$
\While {$|P_{t+1}| \le M$} {\color{gray}//till $P_{t+1}$ is not full, add the individuals in the sorted $F_{r}$}
\State $\mathit{ind} \gets \mathit{EXTRACT\ FIRST\ INDIVIDUAL\ IN\ }F_{r}$  \label{algo:nsga:crowding}
\State $P_{t+1} \gets P_{t+1} \cup \mathit{ind}$ \label{algo:nsga:crowding:end}
\EndWhile
\State $t  \gets t + 1$ 
\EndWhile \label{algo:nsga:mainCycle:end}

\State {\textbf{Return} $P_t$ }

\end{algorithmic}
\caption{NSGA-II }
\label{algo:nsga}
\end{figure}



\begin{figure}

\newcommand{\NSGAT}[1]{\textcolor{red}{#1}}

\begin{algorithmic}[1]

\scriptsize

\Require I, vector whose components represent an individual

\State $l = |I|$
\For {$i = 0; i < |I|$ }
\State $I[i] = 0$
\EndFor

\For {\textbf{each} objective $o$ }
\State $I \gets $ sort $I$ according to objective $o$
\State $I[1]_{distance} \gets \infty$ \label{algo:crowding:dist1}
\State \textcolor{red}{$I[l]_{distance} \gets \infty$}
\label{algo:crowding:worst}
\For {$i=2$ to $(l-1)$ }
\State $I[i]_{distance} \gets I[i]_{distance} + \frac{( I[i+1]_{m} - I[i-1]_{m})}{(\mathit{max\ fitness\ for\ o}) - (\mathit{min\ fitness\ for\ o})}$
\EndFor
\EndFor

\end{algorithmic}
\caption{Crowding-distance-assignment function for NSGA-II. Red part indicates the instruction commented out in \NSGAP to ensure that, for each objective, the best individual is preserved.}
\label{algo:crowding}
\end{figure}

The term many-objective is used for algorithms that tackle problems with more than three objectives~\cite{MOEA:Survey}. 
A popular many-objective algorithm used in software testing is MOSA~\cite{MOSA}. Given test cases (i.e., sequences of inputs for program APIs) as individuals, MOSA models branch coverage as a search objective but does not account for the length of individuals. Shorter test cases are, however, easier to read than longer ones and should be prioritized. Therefore, MOSA extends \nsga by relying on an archive that is used to keep track of the shortest test cases accidentally identified during the search. In addition, since MOSA aims to fulfil all the objectives and not only a subset of them, instead of relying only on the nondominance relation to select elitist individuals (i.e., Line~\ref{algo:nsga:sort} in Fig.~\ref{algo:nsga}), it makes use of the \emph{preference criterion}, 
a strategy that ensures preserving individuals that minimize the objective score for uncovered objectives (i.e., objectives without an individual already in the archive).
\MAJOR{R3.44}{
To generate inputs for DNN-based systems with MOSA~\cite{Fitash:ICSE:2022}, it is necessary to specify a fitness threshold to determine objective coverage (e.g., a safety violation occurs if the ego vehicle gets too close to the vehicle in front). At every iteration, the archive is updated to keep the best individual found during the search; however, the probability of replacing an individual already in the archive is low. Indeed, thanks to the preference criterion, the population includes only the best individuals for the objectives not covered yet. When it is not possible to specify a threshold to determine when an objective is covered (e.g., our context), the main benefit of MOSA is to ensure that the best individual for each objective is preserved; indeed, the archive becomes useless because the best individuals are preserved directly in the population. 
However, a simple modification of \nsga can achieve the same purpose (see Section~\ref{sec:approach:ga:mnsgaii}).}

\subsection{Diversity optimization}
\label{sec:diversity}

In our work, we aim to rely on evolutionary algorithms to generate a set of diverse images belonging to a RCC.
As described in the related literature~\cite{fitash,Fitash:ISSTA:2021,Haq:EMSE:2021,Riccio2020}, an image can be modelled as an individual (or chromosome) of the population, represented as a vector whose components capture the values assigned to simulator parameters used to generate the image. 

Although, in principle, genetic algorithms can be used to evolve a population of individuals 
and achieve our objectives 
(i.e., generate individuals belonging to the RCC and maximize the diversity between them), 
the use of 
\MAJOR{R3.1}{well-known} algorithms for this purpose (e.g., \nsga) is not feasible because it is not possible to define a fitness function for diversity, as explained next. 


To maximize diversity, since it is the  property of a set of images, it is not possible to define a fitness function that assesses how much an offspring individual contributes to diversity prior to knowing what the final set will be.
Also, note that selection based on crowding distance, which is part of \nsga,  does not help because such distance is defined over the objective space, as opposed to the simulator parameter space.
Consequently, the \nsga algorithm cannot simply be adopted to address our problem.

\MAJOR{R3.43}{We can represent a solution for one of our objectives (i.e., a set of $n$ diverse images belonging to a RCC) as an individual where each image is modelled with a subsequence of the chromosomes (e.g., the first subsequence of $m$ chromosomes are for the first images, the second for the second image, and so on). 
This would enable \emph{whole test suite generation} (i.e., the generation, within a same search run, of all the solutions for our objectives)~\cite{WholeTestSuite}. However, different from traditional software testing, individuals achieving different objectives are unlikely to present dependencies. For example, in traditional software testing, nested conditions lead to dependencies between search objectives, whereas two RCCs should include images that are diverse (e.g., one picture of a person looking left and another picture with a person looking right); therefore, in our context, generating an image that fulfills one objective should not help with achieving another objective.
Further, modelling multiple images as a single individual would lead to many simulator runs to generate a single individual, which leads to an inefficient exploration of the search space and, therefore, entails a large budget to generate the desired solutions. For these reasons we chose to model a single image as an individual.}

An alternative solution consists of relying on an archive where, at each iteration of the search algorithm, we add the individuals in the \MAJOR{R3.42}{first non-dominated front} that contribute to increasing diversity. \MAJOR{R3.2}{Related work suggests adding to the archive the individuals with a \emph{sparseness} value (i.e., the distance from the nearest neighbour) that is above a set  threshold (hereafter, \emph{sparseness threshold})~\cite{Riccio2020,DeepMetis}.
DeepJanus~\cite{Riccio2020} and DeepMetis~\cite{DeepMetis} are two state-of-the-art solutions that rely on a \emph{sparseness threshold}. They extend the popular \nsga algorithm with an archive 
and with repopulation (to escape from stagnation by replacing the most dominated individuals with random ones); we refer to such algorithm as \emph{\dnsga}.}
Unfortunately, in our context, the sparseness threshold may directly affect the quality of the results; indeed, if the threshold is too low, most of the selected individuals will be similar to each other, thus making the learning algorithm derive rules that do not characterize the whole RCC. 
Similarly, if the threshold is too high, we end up selecting only a small number of individuals, which prevents the generation of accurate rules.
Unfortunately, identifying an appropriate threshold requires multiple executions of the algorithm, which, in our context, shall be repeated for every RCC; indeed, since it may be easier for the simulator to generate certain images than others, some portions of the input space may be denser than others and, therefore, different thresholds might be needed for different RCCs. In practice, this would lead to an expensive process, which is practically infeasible. \MAJOR{R3.2}{One simple solution is to rely on a sparseness threshold of zero (i.e., we add to the archive any image that differs from the ones already in it)~\cite{DeepMetis}. However, such choice may render the algorithm less efficient as it leads to larger archives and the larger the archive the more costly the distance is to compute. Further, in our context (Section~\ref{sec:approach}), the best individuals identified through the search (in the archive) are used as input for additional search iterations; therefore, the \emph{\dnsga} algorithm should be combined with a strategy to select the best individuals in the archive (Section~\ref{sec:empirical:gaalg} provides details about the strategy adopted in our empirical evaluation).} 

\MAJORSTARTS
 In addition to the limitations indicated above, please note that DeepJanus and DeepMetis
address problems that are different than ours. 
They both rely on two fitness functions, namely $F_1$ (to be maximized) and $F_2$ (to be minimized).
In both DeepJanus and DeepMetis, at every search iteration, every individual of the population is added to the archive if $F_1$ is above the sparseness threshold and if $F_2$ is below a user-defined threshold capturing if the individual addresses the other objective of the search further described below.
DeepJanus characterizes the frontier of DNN misbehaviours by identifying pairs of inputs that
are close to each other, with one input leading to a correct DNN output and the other to a DNN
failure~\cite{Riccio2020}.  Function $F_1$ focuses on sparseness and closeness of input pairs and is computed as follows:
\begin{equation*}
F_{1} =   \{\mathit{sparseness}(i,A) - k * \mathit{distance}(\mathit{i.m1},\mathit{i.m2})\}
\end{equation*}
with $x.m1$ and $x.m2$ being a pair of inputs, $\mathit{sparseness}$ the function that computes the distance from the nearest individual in the archive $A$, and $\mathit{distance}$ the distance between two individuals. Function $F_2$ focuses on the closeness to the frontier and is computed as 

\begin{equation*}
\EquationsSize
F_2 (i)=  
    \begin{cases}
      \mathit{eval}(i.m1) \cdot \mathit{eval}(i.m2) & \text{if > 0}\\   
      -1  & \text{otherwise}\\
    \end{cases}       
\end{equation*}

with $\mathit{eval}$ \emph{being the difference between (a) the confidence level associated with the expected label and (b) the maximum confidence level associated to any other class}. Based on the above, DeepJanus cannot be directly applied in our context because we do not aim to generate pairs of individuals.
DeepMetis aims to identify inputs leading to DNN failures in mutated DNNs (i.e., DNN models trained or modified to be less accurate than the DNN under test)~\cite{DeepMetis}. Different from DeepJanus, in DeepMetis, an individual is a single image.
 Function $F_1$ focuses only on sparseness 
\begin{equation*}
F_{1} =   \{\mathit{sparseness}(i,A)\}
\end{equation*}
Function $F_2$ measures how well an individual triggers inaccurate outputs in mutated DNNs and is computed as 
\begin{equation*}
F_{2} =   \sum\limits_{\mathit{mutant}\in\mathit{Mutant}}^{}{\mathit{eval}_\mathit{mutant}(i)}
\end{equation*}
with $\mathit{eval}_\mathit{mutant}$ returning the
\emph{difference between the confidence associated with the expected class and the maximum confidence associated with any other class when the prediction is correct (-1 if the prediction is wrong)}.
Although $F_{1}$ might also be used in our context to drive the generation of sparse solutions, $F_{2}$ needs to be modified to capture our goals (i.e., the input belongs to a \emph{RCC} and leads to a failure).

 To address our problem, we thus defined an algorithm (i.e., the \GAAlg algorithm, described in Section~\ref{sec:approach:ga}) that takes advantage of every image generated by the simulator in an iteration and does not require a sparseness threshold. 
 Further, we have separated the identification of images belonging to a RCC, which is achieved by \GAAlg, from the identification of failing images, which is achieved with additional search iterations performed through an extension of the \nsga algorithm. 
 In Section~\ref{sec:empirical:gaalg}, we compare \GAAlg with a \dnsga solution with a sparseness threshold of zero and $F_2$ matching the same fitness function  used by \GAAlg to identify individuals belonging to a \emph{RCC}.
\MAJORENDS
\MAJOR{R3.43}{\GAAlg does not support whole test suite generation but it is executed once for each RCC.}

\subsection{Rule learning algorithms}
\label{sec:rule:learning}

In our work, we aim to derive properties that characterize unsafe images.
Given a dataset capturing properties of safe and unsafe images (e.g., the parameter values used to generate an image with a simulator), our objective can be achieved by relying on machine learning algorithms that derive \emph{decision rules}~\cite{Molnar2011}.

A decision rule is an IF-THEN statement consisting of a condition and a prediction. For example, in our context, a decision rule may indicate that if the horizontal angle used to generate an image is above 50 degrees (i.e., the head of the person is so turned that her left eye is barely visible), then the image is failure-inducing. 
Decision rules are machine learning models that can be easily interpreted by humans~\cite{Molnar2011}, which motivates our choice.

The support of a rule indicates the percentage of instances to which the condition of a rule applies, while its accuracy measures the proportion of classes correctly predicted for the instances to which the condition of the rule applies.

To combine multiple rules, rule learning algorithms derive either decision lists or decision sets. A decision list is ordered, and we should rely on the prediction of the first rule whose condition evaluates to true.
In a decision set, predictions are based on some strategy (e.g., majority voting). Since we aim to rely on the generated rules to derive logical expressions, for our work, we selected algorithms deriving decision lists which can be directly translated into a single expression (see Section~\ref{sec:approach:part}).
In particular we rely on the PART algorithm~\cite{PART}, which combines the approach of RIPPER~\cite{RIPPER}, a rule learning algorithm, with decision trees, which have been successfully applied in related work to characterize DNN inputs~\cite{Fitash:ISSTA:2021}.

PART, similarly to RIPPER, relies on a \emph{separate-and-conquer} approach. 
It relies on a partial C4.5 decision tree to derive each rule. 
It starts by deriving a rule that provides accurate predictions for some of the data points. Such result is achieved by building a pruned C4.5 decision tree from all the data points, and then selecting the leaf with the largest support to transform it into a rule. The decision tree is then discarded and all the data points that are covered by the rule are excluded from the training set. 
This rule learning procedure is repeated until the whole data set is processed. 

Since, in our context, for each image under analysis we know (1) the DNN outcome (i.e., a DNN failure or a correct output) and (2) the set of simulator parameter values used to generate it, we can configure PART to consider the DNN outcome as the class to predict and the simulator parameter values as the features to derive rules from.

\begin{figure}
\begin{algorithmic}[1]
\scriptsize
\State $HeadPose_Y > 50.34 \ :\ \mathit{class}=\mathit{DNN-error}$
\label{example:part:1}

\State $HeadPose_Y < 13.34\ :\ \mathit{class}=\mathit{DNN-correct}$
\label{example:part:2}

\State $HeadPose_Z > 60 \ \& \  HeadPose_Y > 30 \ :\ \mathit{class}=\mathit{DNN-error}$
\label{example:part:3}

\State $HeadPose_Z \leq 60\ :\ \mathit{class}= \mathit{DNN-correct}$
\label{example:part:4}

\State (default)\ : \ $\mathit{class}= \mathit{DNN-error}$
\label{example:part:5}

\end{algorithmic}
\caption{Example PART output}
\label{example:part}
\end{figure}

\begin{figure}[t]
\includegraphics[width=4cm]{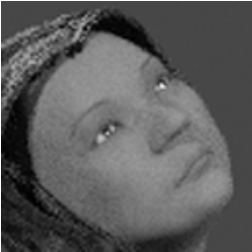}
\caption{Example image generated with IEE-Humans, one of the simulators used in our empirical evaluation. It is labeled as looking top-right (i.e., towards the top-right corner of the picture).}
\label{fig:HumanExample}
\end{figure}

Fig.~\ref{example:part} provides an example output generated by PART when processing images from our case study subjects. Each image depicts a person's head and is represented by a number of simulator parameters configured to generate the image (an example is shown in Fig.~\ref{fig:HumanExample}). 
The considered parameters include, among others, the orientation of the head (pitch, yaw, and roll captured by the parameters $HeadPose_X$, $HeadPose_Y$, and $HeadPose_Z$).


In Fig.~\ref{example:part}, Line~\ref{example:part:1} indicates that we observe a DNN failure if the value of the parameter $HeadPose_Y$ is above $50.34$ (i.e., the person is looking to the right of the viewer, with the head turned so much that her left eye is barely visible).
Line~\ref{example:part:2} indicates that, when the rule in Line~\ref{example:part:1} does not apply and the person is looking center or to her right (i.e., if $HeadPose_Y$ is below $13.34$), then the DNN produces a correct output.
Line~\ref{example:part:3} indicates that, when the two rules above do not hold, we observe a  
DNN failure if the value of parameter $HeadPose_Z$ is above $60$ and the value of parameter $HeadPose_Y$ is above $30$ (i.e., when the head is tilted and the person is looking to her left then the DNN makes a mistake even if her left eye is visible).
Line~\ref{example:part:4} indicates that, when the three rules above do not hold, the  
DNN output is correct if the value of parameter $HeadPose_Z$ is below $60$ (i.e., the head is not tilted).
Finally, Line~\ref{example:part:5} is the default rule, which indicates that, when all the rules above do not hold, the DNN output is incorrect.


\section{Simulator-based Explanations for DNN failures}
\label{sec:approach}

\MAJOR{R3.A}{In this paper, we propose \emph{Simulator-based Explanations for DNN failurEs (\APPR)},} an approach to characterize the root causes (events) leading to DNN failures; we call such events hazard-triggering events. It targets contexts in which DNNs are partly trained using simulators, which is common practice in safety-critical contexts with complex inputs~\cite{Bird2020,Kim2017,Inoue2018}; for example, DNNs implementing vision-based driving tasks or interpreting human postures. This is the case for our industry partner, which relies on a simulator capable of generating images of human bodies, seated in a car environment, to train DNNs that interpret human postures (e.g., determine gaze or drowsiness).

\begin{figure}[b]
\includegraphics[width=0.8\textwidth]{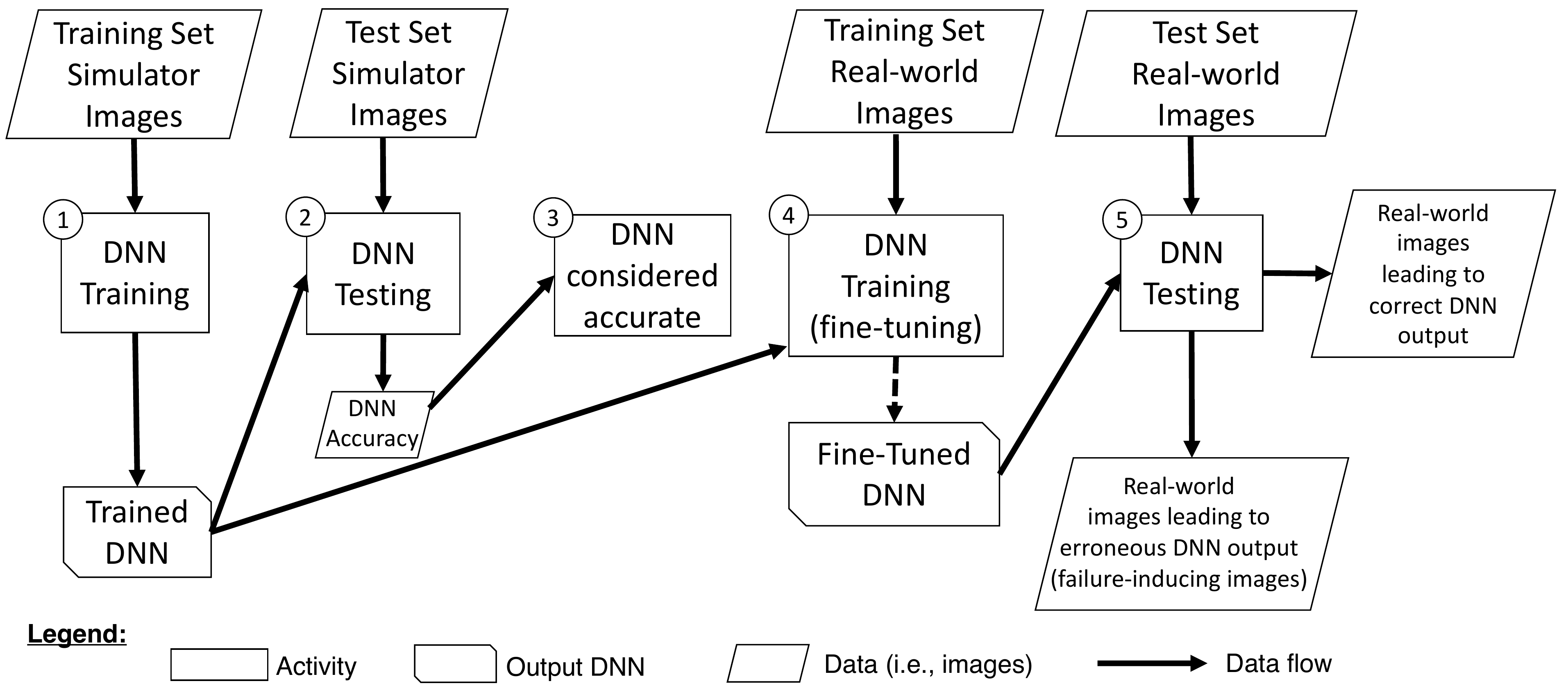}
\caption{DNN training and testing}
\label{fig:fineTuning}
\end{figure}

The reference DNN training process is depicted in Fig.~\ref{fig:fineTuning}. It leverages simulators
to reduce the cost related to collecting and labelling a large number of images. In general, engineers tend to initially train the DNN using images automatically generated with a high-fidelity simulator (e.g., a human body simulator). \MAJOR{R3.5}{When the DNN is deemed accurate,} to enable the correct processing of real-world images, engineers fine-tune the trained DNN using real-world images (e.g., images of real people seated in a car). Fine-tuning consists of relying on the DNN training algorithm to update the weights of the DNN trained with simulator images; retraining may concern the whole set or a subset of the DNN weights (e.g., the fully connected layers in a CNN). \MAJOR{R3.5}{The fine-tuned DNN is then tested with real-world images; in current practice, engineers visually inspect the images leading to DNN failures and determine what are the hazard-triggering events.
\APPR, instead, automatically characterizes the hazard-triggering events observed in a} test set with real-world images; such objective is achieved by generating an expression that constrains the parameters of the simulator used to generate the training set.

\begin{figure}[b]
\includegraphics[width=0.7\textwidth]{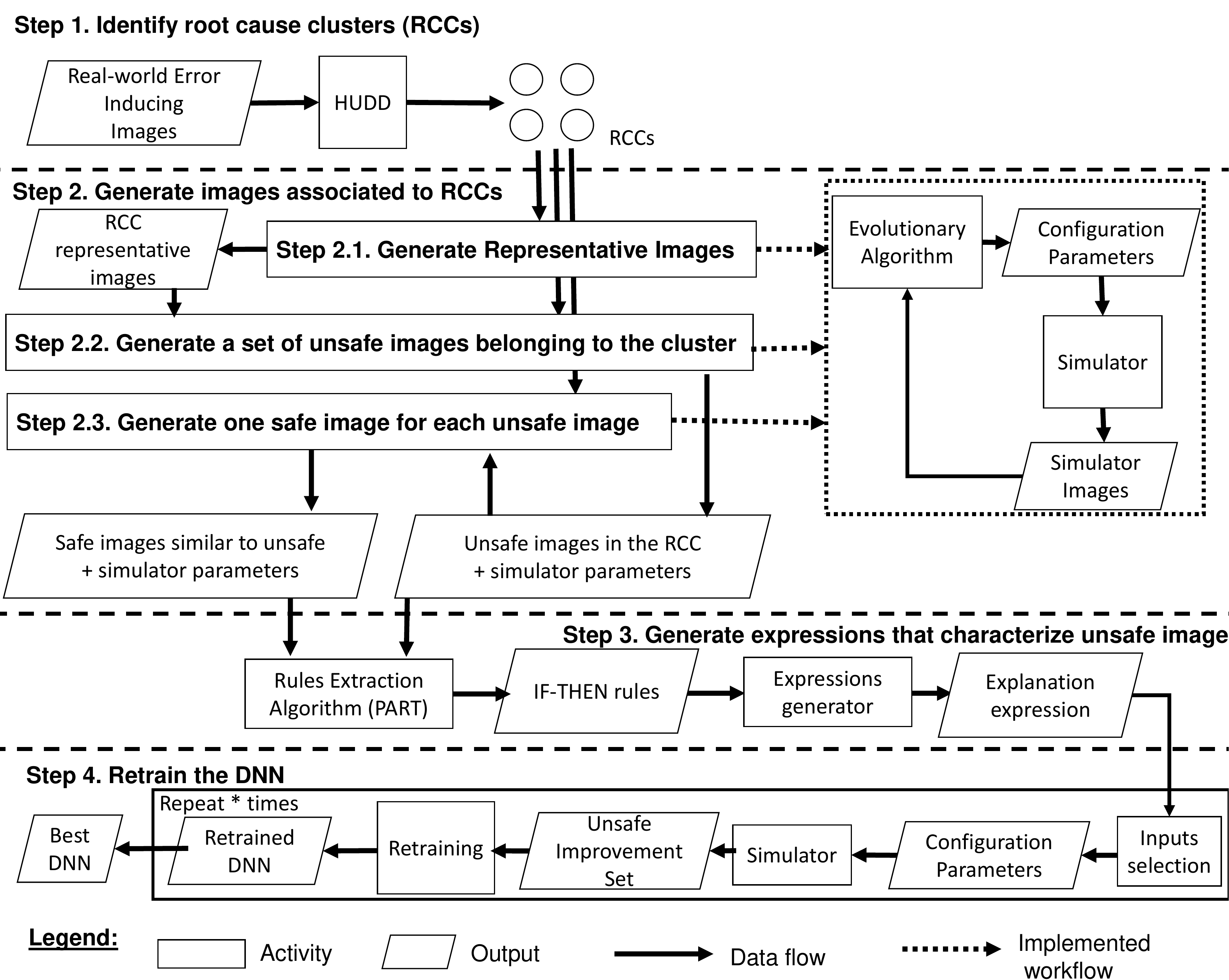}
\caption{The \APPR workflow}
\label{fig:approach}
\end{figure}

\APPR works in four steps depicted in Fig.~\ref{fig:approach}.
In the \emph{first step}, \APPR relies on a state-of-the-art solution (HUDD, see Section~\ref{sec:background:HUDD}), to automatically identify clusters of images (root cause clusters, RCCs) leading to a DNN failure because of common hazard-triggering events.

In the \emph{second step}, \APPR relies on evolutionary search driven by the analysis of heatmaps and simulator parameters to generate images that are associated to RCCs, enabling accurate learning to characterize unsafe scenarios. To precisely derive constraints on simulator parameter values (configurations) for distinct hazard-triggering events, while avoiding overfitting, \MAJOR{R3.45}{for each RCC, we are interested in generating many diverse images leading to correct and erroneous DNN outputs.} 
Addressing multiple hazard-triggering events entails multiple search runs, as justified in Section~\ref{sec:approach:ga}. More precisely, for each RCC, our algorithm executes 
one search to generate a set of diverse images belonging to that RCC,
one search to generate a set of unsafe (i.e., failure-inducing) images that are close to each of the diverse images and belonging to the RCC, 
and another search to identify a set of safe images (i.e., not failure-inducing) that are close to the unsafe ones.
For classifier DNNs, a failure is triggered when the output class differs from the ground truth, where the latter is automatically derived from the parameters of the simulator. In the case of regression DNNs, a DNN failure is triggered when the difference between the DNN output and the reference value (ground truth) is above a set threshold, which is defined by engineers based on domain knowledge.


In the \emph{third step}, \APPR relies on the availability of a sufficiently large number of diverse safe and unsafe images, generated in the previous step, to identify those parameter ranges that characterize unsafe images.
More precisely, \APPR relies on a machine-learning algorithm (i.e., PART) to extract such ranges, under the form of logical expressions that accurately predict and therefore represent unsafe images.

In the \emph{fourth step}, \APPR relies on the derived expressions to generate an additional set of images that are used to retrain and improve the DNN.

\MAJOR{R3.C, R3.4}{Please note that \APPR can also be applied to test sets generated using simulators; however, compared to test sets with real-world images, simulator-based failure-inducing test sets are less challenging to characterize. Indeed, when test set images are generated using simulators, decision trees or any other rule learning algorithm can be used to identify the commonalities among the failure-inducing images~\cite{abdessalem2018testing,Haq:EMSE:2021}. For example, it is sufficient to generate RCCs with HUDD and then extract PART rules from them. However, if the number of available failure-inducing images is limited, learning may not lead to accurate results. In such cases, \APPR may be useful to generate additional failure-inducing images, thus enabling the generation of more accurate explanatory rules.} 

Below, we describe Steps 2 to 4 since Step 1 simply involves the execution of HUDD to generate RCCs.


\subsection{Step 2 - Automated Generation of Unsafe and Safe Images}
\label{sec:approach:ga}

In the second step of \APPR, for each RCC, we aim to automatically generate images that (goal 1) exhibit the same hazard-triggering events observed in the RCC and (goal 2) enable a learning algorithm to extract accurate rules predicting unsafe images. These rules shall ideally characterize parameter values that are observed only with unsafe images belonging to a specific RCC.

To achieve goal 1, it is sufficient to explore the input space by generating diverse images that are similar to the ones belonging to the RCC.

To achieve goal 2, instead, we need to generate both safe images and unsafe images. 
To be accurate, the rules derived by \APPR shall provide ranges for parameter values that are observed only or mostly with unsafe images and are as wide as possible. 
To this end, the unsafe images used to generate the rules must be as diverse as possible and for each unsafe image, we should identify a very similar safe image so that the algorithm learns to precisely distinguish unsafe images from safe ones, i.e., precisely learn the boundary between them.

To achieve the above, we rely on a divide-and-conquer approach that consists in executing three genetic algorithms for each RCC in a sequence.
We make use of genetic algorithms because they have been successfully used by related work to explore the input space of DNNs using simulators~\cite{Riccio2020}.


First (Step 2.1), we generate diverse images belonging to the RCC, safe or unsafe. The goal is to cover the cluster with representative images.  

Second (Step 2.2), we rely on the generated representative images to generate an additional set of unsafe and diverse images belonging to the cluster; this is achieved by generating, for each representative image, \MAJOR{R3.7}{one unsafe image that is close to it and 
belongs to the RCC}.

Third (Step 2.3), for each unsafe image derived by the second evolutionary search, we generate one safe image that is as close to it as possible. 

In summary, we aim to generate a sufficient number of diverse safe and unsafe images belonging to each RCC to enable effective learning. 

\subsubsection{Step 2.1 - Generate Representative Images}
\label{sec:approach:ga:sdga}\ \\

We have two objectives: (1) generate a set of images that belong to a cluster and (2) \MAJOR{R2.3}{minimize their similarity (i.e., maximize their diversity);} \MAJOR{R3.9}{however, since RCCs tend to contain images that are similar, the two objectives are unlikely to compete. Indeed, two dissimilar images are unlikely to belong to the same cluster. Further, for our purpose, it is useless to generate a set of diverse images if they do not belong to the RCC. Consequently, it is not desirable to rely on a multi-objective search algorithm to address them.} To drive the search process, we thus need a fitness function that enables an evolutionary search algorithm to first generate images that belong to a cluster and then decrease their similarity.

In Section~\ref{sec:diversity} we clarified why state-of-the-art approaches are inapplicable or ineffective when applied to achieve our objectives; in this section, we propose a dedicated fitness function and algorithm.

In the presence of a function that measures the distance of an individual $i$ from the center of the RCC $C$ (hereafter, $\mathit{RCC}_{\mathit{distance}}(C,i)$) and a function that measures how an individual contributes to 
\MAJOR{R2.3}{minimizing the similarity across cluster members (hereafter, $F_\mathit{similarity(i)}$)},
to obtain a mathematically adequate behavior, our fitness function $F_1^C(i)$ can be defined as follows:

\MAJORSTARTS
\begin{equation}
\label{eq:f:one:update}
\EquationsSize
F_1^C (i)=  
    \begin{cases}
      F_\mathit{similarity}(i) & \text{if }       \mathit{RCC}_{\mathit{distance}}(C,i) \le 1\\
      \mathit{RCC}_{\mathit{distance}}(C,i) & \text{otherwise}\\
    \end{cases}       
\end{equation}
\MAJORENDS

 Our goal is to minimize $F_1^C(i)$ and given that (1) $F_\mathit{similarity(i)} \le 1$ (see Equation~\ref{eq:f_ip} below)  and (2) $\mathit{RCC}_{\mathit{distance}} \le 1 $ is only true when the image belongs to the RCC, we obtain the targeted property: fitness is always lower in cases where an individual $i$ belongs to the RCC. Below, we describe the functions $F_\mathit{similarity}$ and $\mathit{RCC}_{\mathit{distance}}(C,i)$.

%
%
%


To determine if an individual belongs to a RCC, we can measure its distance from the RCC centroid and verify if it is shorter than the RCC radius (i.e., the max distance between the centroid and the farthest image).
For the distance metric, as in HUDD, we can use the Euclidean distance between two heatmaps (hereafter, $\mathit{HeatmapDistance}$); however,
to generate a heatmap we need a concrete image and, therefore, instead of relying on the cluster centroid, we identify its medoid.
The formula for the $\mathit{HeatmapDistance}$ function is:
\begin{equation}
\label{eq:HD}
  \mathit{HeatmapDistance}(A, B) = \sqrt{\sum\limits_{m=1}^{M}\sum\limits_{n=1}^{N}(A_{m,n}^L - B_{m,n}^L)^2}
\end{equation}
where A and B are two heatmap matrices generated with LRP, and ${m,n}$ indicate the row and column of a cell in a matrix ($M$ and $N$ indicate the total number of rows and columns, respectively). Since LRP generates one heatmap matrix for every neuron layer of the DNN, we rely on the heatmap generated for layer L as selected by \HUDD to generate the RCC.

Similar to related work~\cite{Fitash:ISSTA:2021,Fitash:ICSE:2022}, within our evolutionary algorithms, we represent an individual using a vector of simulator parameter values. For this reason, to compute the heatmap distance, our search algorithm, for every offspring individual, first generates one image using the simulator, then processes the generated image using the DNN under test, and finally generates its heatmap using the LRP algorithm.

The medoid of a RCC $C$ is the image that minimizes the average pairwise distance from the other images of the RCC; it is computed as follows:
\begin{equation}
\label{eq:M}
  \mathit{Medoid}(C) = \argmin_{y \in C} \Bigg \{ \frac{\sum\limits_{x \in C, x \ne y}^{} \mathit{HeatmapDistance}(x, y)}{|C| \cdot (|C|-1)} \Bigg \} 
\end{equation}

The radius of a RCC $C$ is the maximum distance between its medoid and any other image in $C$:
\begin{equation}
\label{eq:R}
\mathit{Radius}(C) = \argmax_{y \in C} \Bigg \{\mathit{HeatmapDistance}(y, \hspace{1pt} \mathit{Medoid}(C)) \Bigg \} 
\end{equation}

Since clusters may have different radiuses, we compute $\mathit{RCC}_{\mathit{distance}}(C,i)$ for a RCC $C$ as the heatmap distance between individual $i$ and $C$'s medoid $\mathit{Medoid}(C)$, normalized by $C$'s  radius $\mathit{Radius}(C)$:
\begin{equation}
\label{eq:D}
\mathit{RCC}_{\mathit{distance}}(C,i) = \frac{\mathit{HeatmapDistance}(i,\mathit{Medoid}(C))}{\mathit{Radius}(C)}
\end{equation}




To compute $F_{\mathit{similarity}}(i)$, we follow related work~\cite{Riccio2020} and measure how much an individual contributes to the diversity of the population by measuring its distance from the closest individual in the population.

We can measure the distance between two individuals $i$ and $j$ based on their chromosome vectors $v_i$ and $v_j$ (containing simulator parameter values), as follows:

\begin{align}
\label{eq:cos}
\mathit{ChromosomeDistance}(i,j) =& \cos(i,j)\\
\nonumber  &={v_i \cdot v_j \over \|v_i \|\|v_j \|} \\
\nonumber  &=\frac{\sum\limits_{p=1}^{|i,j|} ( v_i[p] \cdot  v_j[p] ) }{\sqrt{\sum\limits_{p=1}^{|v_i|} {v_i[p]}^2}\sqrt{\sum\limits_{p=1}^{|v_j|} {v_j[p]}^2}}
\end{align}

where $\cos(i,j)$ is the cosine similarity between $v_i$ and $v_j$, whose range is [0, 1]; $v_i[p]$ and $v_i[p]$ indicate the value of the p-th component in the vectors $v_i$ and $v_j$, respectively.



$F_1^C$ can be used to drive the search algorithm, described below.

\begin{figure}

\newcommand{\OT}[1]{\textcolor{red}{#1}}

\begin{algorithmic}[1]
\scriptsize
\Require $RCC_C$, the RCC under analysis
\Require s, population size
\Require r, number of random populations of images to generate initially
\Require b, search budget (i.e., number of iterations to perform)

\Ensure $P_1^C$, an optimal set of diverse images belonging to cluster $RCC_C$


\Repeat 
\label{algo:sdga:begin}
\State $R \gets R \cup \{ \mathit{RANDOM} \ \mathit{POPULATION}\ \mathit{with}\ \mathit{size}\ \mathit{s} \}$
\Until {$r$} random populations have been generated \label{algo:sdga:loop}
\State $t  \gets 0$ 
\State $P \gets$ selects the population in $R$ with the best fitness 
\label{algo:sdga:par-p}
\While {$t < b$} \label{algo:sdga:main}
\State EVALUATE the FITNESS of $P$ based on $RCC_i$
\State $O \gets$ GENERATE the OFFSPRING population from $P$  
\label{algo:sdga:off} 
\State EVALUATE the FITNESS of $O$ based on $P$ and $RCC_C$
\While {$\mathit{SIZE(}O) > 0$}
\label{algo:sdga:mainStart}
\label{algo:sdga:off-eval}
\State SORT $O$ based on decreasing FITNESS
\label{algo:sdga:off-replace1}
\State $i_o \gets$ extract and remove first individual in $O$
\If{there is at least one individual in $P$ that is outside the RCC}
\label{algo:sdga:ouside}
\State $i_p \gets$ IDENTIFY the individual in $P$ with the highest fitness
\label{algo:sdga:parentOutside}
\Else
\label{algo:sdga:inside}
\State $i_p \gets$ IDENTIFY the individual in $P$ that is closer to $i_o$ 
\label{algo:sdga:parent}
\EndIf
\If{fitness of ${i_o}$ $<$ fitness of ${i_p}$}
\label{algo:sdga:fitness:comparison}
\State {Within $P$, replace $i_p$ with $i_o$}
\State EVALUATE the FITNESS of $P$ based on $RCC_C$
\label{algo:sdga:recompute1}
\State EVALUATE the FITNESS of $O$ based on $P$ and $RCC_C$
\label{algo:sdga:recompute2}
\EndIf
\label{algo:sdga:off-replace2}
\EndWhile
\State $t \gets t + 1$
\EndWhile
\label{algo:sdga:mainEnd}
\State $P_1^j \gets$ all the individuals in $P$ that belong to $RCC_C$
\State {\textbf{return} $P_1^C$}
\label{algo:sdga:return}
\State \textbf{exit}
\label{algo:sdga:exit}
\end{algorithmic}
\caption{\GAAlg, the Genetic Algorithm integrated into \APPR}
\label{algo:sdga}
\end{figure}

\paragraph{The \APPR search algorithm: \GAAlg}


\emph{Pairwise Replacement} (\GAAlg) is our algorithm to generate representative RCC images; it is presented in Fig.~\ref{algo:sdga} and described below.

\GAAlg aims to evolve a whole population of individuals to generate a population of individuals that both belong to the RCC under analysis and are diverse. The size of the population is a parameter of the algorithm.
\GAAlg leverages all the images generated by the simulator at every search iteration by replacing one or more images in the parent population with offspring individuals having a better fitness.

\GAAlg works by looking for individuals that minimize $F_1^C$. At every iteration, \GAAlg creates a new population $P'$ obtained by replacing an individual $i_p$ with an individual $i_o$ having a better fitness (i.e., $F_1^C(i_o) < F_1^C(i_p)$), as described in the following.



The \GAAlg algorithm receives as input the RCC under analysis (hereafter, $RCC_C$), the configuration parameters $s$ \MAJOR{R3.48}{(population size)} and $r$ \MAJOR{R3.48}{(number of random populations to generate initially)}, and the search budget $b$ indicating the number of iterations to perform.

To maximize the chances of finding an optimal solution, as in related work~\cite{initPop}, the \GAAlg algorithm starts by generating several random populations (Line~\ref{algo:sdga:begin} to ~\ref{algo:sdga:loop}) and selecting the population including the individual with the lowest fitness value ($P$, Line~\ref{algo:sdga:par-p}). 


In its main loop (Lines~\ref{algo:sdga:main} to~\ref{algo:sdga:mainEnd}), \GAAlg computes the fitness values of individuals in the parent population $P$ 
and, from $P$, generates an offspring population $O$ by relying on the same strategy commonly adopted with \nsga in similar contexts
(i.e., tournament selection, simulated binary crossover, crossover probability $p_c$, and mutation probability $p_m$).



The offspring individuals are sorted to process first the ones with better fitness (Line~\ref{algo:sdga:off-replace1}).
\GAAlg considers each individual in $O$ ($i_o$) as a replacement of an individual in $P$ ($i_p$), if the former has a better fitness.
\GAAlg distinguishes between the cases in which P only contains individuals belonging to the RCC (Line~\ref{algo:sdga:inside}) and when at least one individual does not (Line~\ref{algo:sdga:ouside}). Indeed, if we do not have enough individuals belonging to the RCC, it is not important to maximize their diversity.

If $P$ contains at least one individual not belonging to the RCC, \GAAlg simply looks for the  individual in $P$ with the highest fitness (i.e., the one that is furthest from the cluster medoid).
Such individual will be replaced by $i_o$ if the latter has a lower fitness (Line~\ref{algo:sdga:fitness:comparison}), which is always the case if $i_o$ belongs to the RCC; otherwise, it depends on the value of $RCC_{distance}$:  $i_o$ replaces $i_p$ if $i_o$ is closer to the RCC medoid.

If all the individuals in $P$ belong to the RCC,
\GAAlg selects the individual in $P$ that is closer to $i_o$ (i.e., $i_p$, Line~\ref{algo:sdga:parent}). 
The offspring individual $i_o$ replaces  $i_p$ if $i_o$ has a lower fitness than $i_p$ (Line~\ref{algo:sdga:fitness:comparison}).
Since $i_p$ belongs to the RCC, according to Equation~\ref{eq:f:one:update}, 
$i_o$ has a lower fitness than $i_p$ 
when $i_o$ belongs to the RCC and has a lower $\mathit{F}_{\mathit{similarity}}$.

$F_{\mathit{similarity}}(i)$ is computed differently for individuals in  the parent and in the offspring population; the reason is that we must decide if an individual $i_o$ in the offspring population $O$, would lead to a population $P'$ with a larger diversity than $P$.
$P'$ is obtained by replacing an individual $i_p$  in the parent population $P$ with $i_o$;
therefore, we obtain a different $P'$ from a same $P$, depending on the selected individuals $i_p$  and $i_o$. 
To help identify the most diverse $P'$ and avoid computing the similarity for all possible $P'$ obtained from every individual in $O$, inspired by related work~\cite{Riccio2020}, 
we compare (a) the distance between 
$i_p$  and its closest neighbour in  $P$ \MAJOR{R3.49}{(i.e., the closest individual in $P$ excluding $i_p$ itself)} with (b) the distance between 
$i_o$ and the closest individual in  $P'$.
Note that $P'$ includes $i_o$ and $P \setminus \{i_p\}$. 
Therefore, 
for an individual $i_p$  in the parent population P, we compute one minus its distance from 
$\mathit{closest}_{i_p}$, \MAJOR{R3.49}{the closest neighbour to $i_p$}
\begin{equation}
F_{\mathit{similarity}(i_p)} = 1 - \mathit{ChromosomeDistance}(i_p,\mathit{closest}_{i_p})
\label{eq:f_ip}
\end{equation}
For an individual $i_o$ in the offspring population $O$, we compute one minus its distance from 
$\mathit{closest}_{i_o}$, the individual in $P \setminus \{i_p\}$ that is closest to $i_o$:
\begin{equation}
F_{\mathit{similarity}(i_o)} = 1 - \mathit{ChromosomeDistance}(i_o,\mathit{closest}_{i_o})
\label{eq:f_io}
\end{equation}

Based on the above, given two individuals $i_o$ and $i_p$ in the offspring and parent populations, respectively, the individual $i_o$ has a better fitness than $i_p$ if they both belong to the RCC and
\MAJOR{R2.12}{the similarity between $i_o$ and the closest individual in $P \setminus \{i_p\}$ is lower than the similarity between $i_p$ and its closest individual in $P$.}
Since the distance from the closest individual should increase (lower similarity) when replacing $i_p$ with $i_o$ in the population, we can assume that a population $P'$ will have higher diversity than P, which we demonstrate in Section~\ref{sec:empirical:gaalg}.

The process is repeated until $O$ is empty (Lines~\ref{algo:sdga:mainStart} to ~\ref{algo:sdga:off-replace2}); since $F_1^C$ depends on the distance between individuals being close to each other, given that both $O$ and $P$ vary over iterations, the fitness of the two populations is recomputed after every replacement (Lines~\ref{algo:sdga:recompute1} and~\ref{algo:sdga:recompute2}).

After $b$ iterations, the algorithm has generated a population $P_1^C$ of individuals that are diverse and very likely to belong to $RCC_C$.

\subsubsection{Step 2.2 - Generate a set of unsafe images belonging to the cluster}
\label{sec:approach:ga:mnsgaii}\ \\

For each $RCC_C$, we aim to generate a set of diverse, unsafe images, belonging to the cluster.
To preserve the diversity of the images generated by \GAAlg in Step 2.1 ($P_1^C$), we can identify, for each image in $P_1^C$, an image that is close to it and makes the DNN fail. We can thus model our problem as a many-objective optimization problem with $q$ objectives, where $q$ is the number of images in $P_1^C$.

We define $q$ fitness functions, one for each individual in $P_1^C$.
An individual is an image generated with the simulator, modelled as described in Section~\ref{sec:approach:ga:sdga}. 
All fitness functions share the same parameterized formula $F_{2}^{C,t}$, where $t \in \{1, ..., q\}$ :
\begin{equation}
\label{eq:f:two}
\EquationsSize
F_{2}^{C,t} (i) =  
    \begin{cases}
      \mathit{ChromosomeDistance}(i, P_1^C[t]) & if \text{ }  (\mathit{RCC}_{\mathit{distance}}(\mathit{RCC}_C,i) \le 1) 
       \hspace{1pt} \land ( \mathit{CLOSEST}(i,P_1^C) = P_1^C[t] )\\
      & \hspace{5pt} \land   (DNN_{\mathit{failure}}(i))\\
      1 \hspace{1pt} + \hspace{1pt} F_{\mathit{uncertainty}}(i) & if \text{ } (\mathit{RCC}_{\mathit{distance}}(\mathit{RCC}_C,i) \le 1) 
       \hspace{1pt} \land ( \mathit{CLOSEST}(i,P_1^C) = P_1^C[t] )\\
      & \hspace{5pt} \land   (DNN_{\mathit{correct}}(i))\\
      2 \hspace{1pt} + \hspace{1pt} \mathit{ChromosomeDistance}(i, \hspace{1pt} P_1^C[t]) & if \text{ }  (\mathit{RCC}_{\mathit{distance}}(\mathit{RCC}_C,i) \le 1) 
       \hspace{1pt} \land ( \mathit{CLOSEST}(i,P_1^C) \neq P_1^C[t] )\\
      3 \hspace{1pt} + \hspace{1pt} \mathit{RCC}_{\mathit{distance}}(\mathit{RCC}_C,i) & if \text{ } \mathit{RCC}_{\mathit{distance}}(\mathit{RCC}_C,i) > 1
    \end{cases}       
\end{equation}

$F_{2}^{C,t}$ is a piecewise-defined function implemented through four sub-functions that return a value in the range [0, 1] incremented by a constant (from $0$ for the first sub-function to $3$ for the fourth sub-function).

The fourth sub-function of the equation drives the search towards generating an individual belonging to $RCC_C$; indeed, when this is not the case, $F_{2}^{C,t}$ measures how far the individual is from the RCC medoid with  
$\mathit{RCC}_{\mathit{distance}}(\mathit{RCC}_C,i)$. Since not belonging to $RCC_C$ is the worst case for an individual, $F_{2}^{C,t}$ must be higher than in other cases. This is achieved by returning $3 + \mathit{RCC}_{\mathit{distance}}(\mathit{RCC}_C,i)$, given that the codomain of the first three sub-functions referred to in $F_{2}^{C,t}$ lay in the range [0, 3].

The third sub-function of the equation ensures that we generate one image for each individual in $P_1^C$; indeed, if the individual $i$ belongs to $RCC_C$ but the individual in $P_1^C$ that is closest to $i$ is not the t-th individual (i.e., $CLOSEST(i,P_1^C) \neq P_1^C[t]$), $F_{2}^{C,t}$ returns the cosine distance between $i$ and the t-th individual in $P_1^C$, incremented by 2.

The second sub-function helps generating an individual that makes the DNN fail. Indeed, if the individual $i$ belongs to $RCC_C$, is the closest to the t-th individual targeted by $F_{2}^{C,t}$, but does not make the DNN fail, then the fitness function returns a measure of DNN uncertainty increased by one. 

To compute the uncertainty component of the fitness, 
we return one minus the cross-entropy loss function, which is commonly used to measure uncertainty:
\begin{equation}
\label{eq:e-adj}
\EquationsSize
  F_{\mathit{uncertainty}}(i) = 
  1 - \mathit{entropy}(i)
\end{equation}

Cross-entropy measures how uncertain is the DNN about the provided output where lower cross-entropy values indicate higher certainty; therefore, by using one minus cross-entropy loss, we direct the search (minimization) towards the identification of images for which the DNN is less certain about outputs and, therefore, likely to produce an erroneous output.

The first sub-function of $F_{2}^{C,t}$, instead, for individuals that make the DNN fail, helps select the individual that is closest to 
the individual targeted by $F_{2}^{C,t}$; indeed, it returns the chromosome distance between the individual $i$ and the individual $P_1^C[t]$.


\emph{Search Algorithm.} To address our problem, we rely on a modified version of \nsga. \nsga is known to underperform in the presence of a large number of objectives ($>$3), mainly because of the exponential growth in the number of non-dominated solutions required for approximating the Pareto front~\cite{Ishibuchi:2009}. However,
we do not aim to find, as a solution to our problem, one single individual (i.e., image) that finds the best balance among different objectives but a set of images, each optimizing one independent objective (indeed, each image shall be close to a reference image in $P_1^C$); therefore, we are unlikely to find a set of nondominated solutions larger than $|P_1^C|$ (i.e., we will find one solution for each objective).
Also, by definition, $F_{2}^{C,t}$ is always different than zero, which renders ineffective algorithms that look for an individual
that \emph{covers} an objective \MAJOR{R3.49}{like MOSA. A solution to enable the adoption of MOSA would be to set a threshold to determine when the objective has been \emph{covered}; however, since RCCs differ with respect to their radius, we choose not to introduce a threshold that may lead to varying performance results across RCCs.}
Since we aim to optimize all the objectives, we rely on an extended version of \nsga with a modified crowding-distance function that ensures we select the minimal value for each objective, thus resembling the preference criterion of MOSA.
Finally, we do not require an archive because the population will always include the best individual found for each objective and we do not need to evaluate an individual according to objectives not explicitly modeled (e.g., inputs' length in MOSA).

We propose a modified version of \nsga (hereafter, \NSGAP) that includes a modified crowding-distance-assignment which
ensures preserving, for each objective, the individual with the lowest fitness value.
Our modified crowding-distance-assignment is shown in Fig.~\ref{algo:crowding} (Page~\pageref{algo:crowding}); different from the original crowding-distance-assignment used by \nsga, for each objective, we assign infinite crowding distance only to the individual that minimizes the fitness for the objective. \nsga, instead, assigns infinite distance also to the individual that maximizes the fitness for the objective (Line~\ref{algo:crowding:worst} in Fig.~\ref{algo:crowding}). \MAJOR{R3.49}{As in \nsga, if more than one individual has the same minimal fitness, \NSGAP assigns infinite distance only to one, randomly selected individual.} Our choice ensures that, 
when the Pareto front includes a number of individuals larger than the population size $s$, 
for each objective, we preserve the individual that minimizes the fitness value.
When the Pareto front has a number of individuals lower than the population size $s$, our crowding-distance assignment ensures the selection of individuals that minimize the fitness
\MAJOR{R3.49}{and are likely unsafe}, which is what we intend to preserve in the final population.
 However, this situation is unlikely. Indeed, it may occur only when one individual has the same fitness value for several (i.e., $z$, with $z \leq q$) objectives, in one of the following unlikely scenarios:
\begin{itemize}
    \item one image has the lowest $\mathit{RCC}_{\mathit{distance}}$ and no other individual falls in the cases covered by the first three sub-functions in $F_{2}^{C,t}$, for the same $z$ objectives;
    \item one image has the lowest $\mathit{ChromosomeDistance}(i, \hspace{1pt} P_1^C[t])$ for $z$ reference images and no other individual falls in the cases covered by the first two sub-functions in $F_{2}^{C,t}$, for the same $z$ objectives;
    \item one image has the lowest $F_\mathit{uncertainty}(i)$ and no other individual falls in the cases covered by the first sub-function in $F_{2}^{C,t}$, for the same $z$ objectives;
     \item one image is failure-inducing and has the lowest $\mathit{ChromosomeDistance}(i,P_1^C[t])$, for the same $z$ objectives.
\end{itemize}

We apply \NSGAP for $k$ iterations.
\MAJOR{R3.6}{Please note that when $P_1^C$ already includes unsafe images belonging to the RCC $C$, \NSGAP simply retains such images at every iteration; indeed, their fitness is already optimal according to $F_{2}^{C,t}$.}
After the k-th iteration of \NSGAP, 
\APPR generates a population of individuals that likely lead to a DNN failure, which we refer to as $P_2^C$; \MAJOR{R3.8}{at the end of the search, images not leading to DNN failures are removed from $P_2^C$}. 

\subsubsection{Step 2.3 - Generate one safe image for each unsafe image}
\ \\

We aim to generate a set of images that are similar to the unsafe images in $P_2^C$ but do not lead to a DNN failure.
As for the previous case,
we model our problem as a many-objective optimization problem with $q$ objectives, where $q$ is the number of images in $P_2^C$; for each image in $P_2^C$, we aim to generate an image that is close to it and makes the DNN pass. We rely on the same \NSGAP configuration used for Step 2.2 except for the fitness function, which in this step needs to be adapted to drive the generation of safe images; also, it is not necessary for these generated images to belong to $RCC_C$ (indeed, we may not have any safe image within $RCC_C$). In Step 2.3, \NSGAP evolves a population $P_3^C$ that initially matches $P_2^C$.

As for Step 2.2, we define $q$ fitness functions, one for each image in $P_2^C$; these functions share the same parameterized formula, for $t \in \{1, ..., q\}$:

\begin{equation}
\label{eq:f:three}
\EquationsSize
  F_3^{C,t} (i) =
    \begin{cases}
      \mathit{ChromosomeDistance}(i, \hspace{1pt} P_2^C[t]) & \text{if } (\mathit{CLOSEST}(i,P_2^C) = P_2^C[t])
      \hspace{1pt} \land (DNN_{\mathit{correct}}(i))\\
      1 \hspace{1pt} + \hspace{1pt} \mathit{entropy}(i) \hspace{1pt}  +  |1-\mathit{RCC}_{\mathit{distance}}(\mathit{RCC}_C,i)|  & \text{if } (\mathit{CLOSEST}(i,P_2^C) = P_2^C[t]) \hspace{1pt} \land (DNN_{\mathit{failure}}(i))\\ 
            2 \hspace{1pt} + \hspace{1pt} \mathit{ChromosomeDistance}(i, \hspace{1pt} P_2^C[t]) & \text{if } \mathit{CLOSEST}(i,P_2^j) \neq P_2^C[t]\\
    \end{cases}       
\end{equation}

The third sub-formula in  $F_3^{C,t}$ matches the third formula in  $F_2^{C,t}$ and aims to generate one image close to each 
each unsafe image in $P_2$.

The definition of the second sub-formula in  $F_3^{C,t}$ is motivated by the fact that, since the initial population is $P_2^C$, all the images in the population initially cause a DNN failure and, therefore, the fitness should drive the search algorithm to find a close image leading to a correct output. To do so, we should look for images that either increase the confidence of the DNN output (i.e., leading to a lower entropy) or are close to the border of the RCC. Concerning the latter, since a RCC characterizes unsafe images, we expect that images next to its border are similar to the ones in the RCC but are less likely to be failure-inducing. In addition, moving further away from the border is expected to lead to images that are increasingly different from the images in the RCC. 
\MAJOR{R2.20}{Such observations are captured by 
the absolute difference between 1 and $RCC_{\mathit{distance}}$ (i.e., how close the image is to the RCC border), which is  added to $\mathit{entropy}(i)$ to help generate safe images}.

The codomains of the second and the third sub-formula overlap as the former may return a fitness value above 3; this choice reflects the fact that 
an image being far away from the RCC border (when $|1 - RCC_\mathit{distance}(C,i)| > 1$) is unlikely to provide useful information
as it is not helpful to derive rules that characterize safe images that are not similar to the ones in the RCC. 
Therefore, such image would be as useless as an image that is not close to the target image (i.e., $P_2^C[t]$).
Also, if the DNN is highly uncertain about an image $i$ (i.e., $\mathit{entropy}(i)$ is close to $1$), it is unlikely for that image to help driving the algorithm towards a correct output. 
For the reasons above, the output of the second sub-formula falls into the codomain of the third sub-formula when the sum 
 $\hspace{1pt} \mathit{entropy}(i) \hspace{1pt} +  |1-\mathit{RCC}_{\mathit{distance}}(\mathit{RCC}_C,i)|$ $>$ 2.


The first sub-formula in  $F_3^{C,t}$, in the presence of images leading to correct DNN outputs, aims to minimize the distance from the t-th image; therefore, it returns the chromosome distance between the individual $i$ and the target image $P_2^C[t]$.


The algorithm \NSGAP terminates after $k$ iterations with a population $P_3^C$ including images that are likely safe and close to the images in $P_2^C$; \MAJOR{3.8}{though unlikely, images leading to DNN failures are  removed from $P_3^C$ before \APPR's Step 3.}

%

\subsection{Step 3. Characterize unsafe images}
\label{sec:approach:part}

In Step 3, 
we aim to generate an expression that characterizes unsafe images by using PART, to learn decision rules.
For example, we could learn that images likely lead to a DNN failure when
the parameters $HeadPose_X$ $>$ $10$ $and$ $HeadPose_Y$ $>$ $50$.

PART generates as output a list of mutually exclusive rules (see Section~\ref{sec:rule:learning}) that predict the class of a data point; in our context, these rules predict the DNN output generated for an image (i.e., wrong or correct output) based on the values of the simulator parameters used to generate the image.

Since in Step 2, for each RCC, \APPR generated a set of unsafe ($P_2^C$) and safe ($P_3^C$) images, associated with simulator parameters, 
to learn decision rules with PART, \APPR selects from $P_2^C$ all the images that belong to the RCC and are failing, and all the images from $P_3^C$ that are not failing.


To generate an expression from the PART output (hereafter, referred to as \emph{PART expression}), we implemented a procedure that, 
for all the rules leading to a DNN failure,  generates a subexpression that joins the negation of all the preceding rules with the current rule. The generated subexpressions are mutually exclusive. 
For the example in Fig.~\ref{example:part}, it leads to the following expression:

\begin{equation}
\label{equation:PART:expression}
\EquationsSize
  \begin{aligned}
& (HeadPose_Y > 50.34 ) \parallel \\
& (\ \neg  (HeadPose_Y > 50.34 ) \&\ \neg  ( HeadPose_Y < 13.34 ) \ \\ 
& \hspace{10pt} \& ( HeadPose_Z > 60 \ \& \  HeadPose_Y > 30 ) ) \parallel \\
& (\ \neg  (HeadPose_Y > 50.34 )\ \&\ \neg  ( HeadPose_Y < 13.34 ) \ \\ 
& \hspace{10pt} \&\ \neg  ( HeadPose_Z > 60 \ \& \  HeadPose_Y > 30 ) \\
& \hspace{10pt} \&\ \neg  ( HeadPose_Z <= 60 ) ) \\
  \end{aligned}
\end{equation}
%
%
%

%

Indeed, for Line~\ref{example:part:1} in Fig.~\ref{example:part}, \APPR simply reports the expression generated by PART (there are no preceding expressions), which leads to 
``$HeadPose_Y > 50.34$''.
For Line~\ref{example:part:2}, \APPR does not generate any subexpression because it concerns cases in which the DNN generates a correct output; the same happens for Line~\ref{example:part:4}.
For Line~\ref{example:part:3}, \APPR negates the preceding expressions (i.e., ``$HeadPose_Y > 50.34$'' and ``$HeadPose_Y < 13.34$'' ) and joins them with the PART expression of Line~\ref{example:part:3} (i.e., ``$HeadPose_Z > 60 \ \& \  HeadPose_Y > 30 $''). 
For Line~\ref{example:part:5}, \APPR simply generates a subexpression joining the negation of all the preceding expressions.

Unfortunately, PART does not generate expressions for parameters whose range is shared by the two classes under consideration (i.e., DNN failure and DNN correct).  This is an issue as \APPR generates safe images that are similar to the unsafe ones. For example, the PART output in Fig.~\ref{example:part} does not include an expression with ``$HeadPose_X > 10$'', which indicates that the head is looking top because this characteristic is observed in both safe and unsafe images. Such commonalities, however, might be important to characterize unsafe inputs; indeed, in our example, if the RCC includes images with a person looking top, safe images close to the generated unsafe images will likely be looking top.
 To address this issue, \APPR joins the PART expression with an expression that constrains the parameters not appearing in the PART expression. The additional expression is generated by identifying the min and max values assigned to the parameters for all the unsafe images.
For the example above, it joins the expression ``$HeadPose_X > 10$'' with the PART expression in Equation~\ref{equation:PART:expression}, thus leading to: 
\begin{equation}
\EquationsSize
  \begin{aligned}
& HeadPose_X > 10  \ \&   \\
& \hspace{10pt} ( (HeadPose_Y > 50.34 ) \parallel \\
& \hspace{20pt} ( \neg (HeadPose_Y > 50.34 ) \& \neg ( HeadPose_Y < 13.34 ) \ \\ 
& \hspace{30pt} \& ( HeadPose_Z > 60 \ \& \  HeadPose_Y > 30 ) ) \parallel \\
& \hspace{20pt} ( \neg (HeadPose_Y > 50.34 ) \& \neg ( HeadPose_Y < 13.34 ) \ \\ 
& \hspace{30pt} \& \neg ( HeadPose_Z > 60 \ \& \  HeadPose_Y > 30 ) \\
& \hspace{30pt} \&  \neg ( HeadPose_Z \leq 60 ) ) )\\
  \end{aligned}
\end{equation}

\subsection{Step 4. Generate unsafe images for retraining}
\label{sec:approach:retrain}

In Step 4, for each RCC, \APPR automatically generates a set $U$ of unsafe images by relying on the simulator. Thanks to the simulator, these images can indeed be automatically labeled and, consequently, can be used to retrain the DNN.  We refer to this set of images as the \emph{unsafe improvement set}. 
More precisely, we aim to train a new DNN model by relying on an extended training set that consists of part of the training set used to train the DNN under analysis and the unsafe improvement set.
The retraining process is configured as a fine-tuning process where the weights of the DNN to be trained are initially set to be equal to the weights of the DNN under analysis. 
Our rationale is that by retraining the DNN with an additional set of images belonging to the unsafe portion of the input space, we enable the DNN to better learn how to provide correct outputs for such inputs.

To avoid losing information derived from the original training process, the extended training set consists of $S\%$  of the simulator images and 
$R\%$ of the real-world images belonging to the original training set. 
We suggest configuring $S$ and $R$ such that the number of simulator images is not overwhelmingly larger than the number of real-world images (e.g., twice the number of real-world images, at most). More precisely, we should limit the number of simulator images because they may overly influence the DNN with characteristics that are not present in real-world images (e.g., faces with regular shapes); nevertheless, we suggest keeping a representative set of simulator-based images (e.g., $5\%$) because they may include a number of scenarios (in our case studies, head positions) not covered by real-world images. For real-world images, since they are generally low in number, we suggest keeping most or all of them to prevent the risk of losing any image that captures an under-represented scenario. For our experiments, we set $S$=$5\%$ and $R$=$100\%$ (see Section~\ref{sec:empirical:rq5}). We leave to future work the identification of a solution for the automated selection of the best configuration for $S$ and $R$ (e.g., by selecting the DNN with the best accuracy among the ones generated by repeating the retraining process with different values for $S$ and $R$). \MAJOR{new}{For example, recent results suggest that the proportion of simulator ($S\%$) and real-world ($R\%$) images may also depend on the degree of fidelity of the simulator: the more realistic simulator images are, the less real-world images are required~\cite{DiasDaCruz2022Uncertainty}}.

The accuracy of the DNN training process may depend on a wide range of factors, including the specific simulator images selected from the original training set, the images belonging to the unsafe improvement set, and other random factors (e.g., the order of images in the training batches). For this reason, \APPR repeats the retraining process (i.e., generate unsafe images in $U$ for each RCC and retrain the DNN) multiple times and keeps the DNN that provides the best output. 
Though repeating the training process obviously increases the training cost (e.g., buying additional GPU time in Cloud systems), \MAJOR{R3.50}{for safety-critical systems, such additional cost may be justified by improvements in accuracy.}

\section{Empirical Evaluation}
\label{sec:empirical} 

Our empirical evaluation addresses the following research questions:

\begin{itemize}

\MAJORSTARTS
\item[RQ1] \emph{How does \GAAlg fare, compared to \nsga and \dnsga, for the generation of diverse images belonging to RCCs?}
The generation of a sufficiently large and diverse set of images belonging to all the RCCs under analysis is essential to enable the characterization of DNN failures. To that end, \APPR integrates a dedicated genetic algorithm (\GAAlg) whose performance is evaluated by this RQ and compared to \nsga \MAJOR{R3.3}{and \dnsga} in terms of percentage of RCCs for which an individual has been generated, percentage of individuals belonging to each cluster, and image diversity within clusters.
\MAJORENDS

\item[RQ2] \emph{Does \APPR generate simulator images that are close to the center of each RCC?} 
The generation of images belonging to a RCC is a necessary step to generate expressions that characterize the RCC.
Since there is no guarantee of generating, using simulators, images that are similar to real-world images, this research question evaluates if the images generated by \APPR are closer to the center (medoid) of a cluster than random simulated, unsafe images. 

\item[RQ3] \emph{Does \APPR generate, for each RCC, a set of images sharing similar characteristics?}
To successfully characterize RCCs with PART, it is necessary, for each RCC, to rely on a set of images presenting similar characteristics while preserving diversity regarding other aspects. In other words, the generated images shall present similar values for a subset of the simulator parameters while being diverse regarding the other parameters.

\item[RQ4] \emph{Do the RCC expressions identified by \APPR delimit an unsafe space?} 
This research question aims to determine if the analyzed DNN underperforms when processing images matching  RCC expressions. In other words, is the DNN accuracy significantly lower for such images, as expected, than for random images from the whole input space?

\item[RQ5] \emph{How does \APPR compare to state-of-the-art DNN accuracy improvement practices?} 
This research question evaluates the effectiveness of \APPR (Step 4), in terms of accuracy improvements, compared to \HUDD and a baseline consisting of randomly generated images for each RCC (namely, random baseline).
\end{itemize}

\subsection{Subject Systems}
\label{sec:empirical:subjects} 
We consider DNNs that implement head pose detection and face landmarks detection. They are building blocks for in-car monitoring systems (e.g., driver's drowsiness detection) under study at IEE Sensing, our industry partner.

The head pose detection DNN (HPD) receives as input the cropped image of the head of a person and determines its pose according to nine classes (straight, turned bottom-left, turned left, turned top-left, turned bottom-right, turned right, turned top-right, reclined, looking up). 

The facial landmarks detection DNN (FLD) determines the location of the pixels corresponding to $27$ face landmarks delimiting seven face elements: nose ridge, left eye, right eye, left brow, right brow, nose, and mouth. Several face landmarks match each face element (e.g., there are four landmarks to outline a mouth). The accuracy of this regression DNN is computed as the percentage of images with landmarks being accurately predicted. For each landmark, we compute the prediction error as the Euclidean distance between the predicted and correct landmark pixel on the input image; an image is considered accurately predicted if the average error is below $4$ pixels, as suggested by IEE engineers.


Our DNNs have been trained with simulator images, fine-tuned with real-world images, and tested with real-world images, following the process in Fig.~\ref{fig:fineTuning}.
For our experiments we relied on two different simulators developed by \IEE and based on the 3D simulation of MakeHuman~\cite{makehuman} and MBLab~\cite{MBLab} using the rendering engine Blender~\cite{Blender}. 
The first simulator is called IEE-Faces, it relies on seven preset MakeHuman models of human faces developed by IEE; it provides $13$ parameters that enable controlling different characteristics of the image such as illumination angle and head orientation. Table~\ref{tab:faces_params} provides a description of all the parameters of IEE-Faces,  \MAJOR{R2.21}{where the first three rows capture nine parameters, three for each row}. IEE-Faces generates one image in 20 seconds, on the hardware used for our experiments.
The second simulator is called IEE-Humans and 
it generates images that are more realistic (i.e., have more details) than the ones generated by IEE-Faces. 
IEE-Humans provides $23$ different configuration parameters, which are described in Table~\ref{tab:humans_params}  \MAJOR{R2.21}{
where the first two rows capture six parameters, three for each row}.
It takes 100 seconds to generate an image.

We trained three DNNs in total, two HPD DNNs and one FLD DNN. One HPD DNN (hereafter, HPD-F) has been trained using IEE-Faces and another one (hereafter, HPD-H) using IEE-Humans. The FLD DNN has been trained using IEE-Faces\MAJOR{R3.12}{; we could not train FLD with IEE-Humans because it does not provide landmarks for the generated images}.

\begin{table}[tb]
\caption{IEE-Faces simulator parameters}
\footnotesize
\begin{tabular}{|@{\hspace{1pt}}p{3cm}|@{\hspace{1pt}}p{6cm}|}
\hline
\textbf{Parameter}&\textbf{Description}\\
\hline
Camera Direction&Direction of scenario camera (X, Y, Z)\\
Camera Location&Location of scenario camera (X, Y, Z)\\
Lamp Location&Location of scenario lamp source (X, Y, Z)\\
$HeadPose_X$&Vertical position of the head (degrees)\\
$HeadPose_Y$&Horizontal position of the head (degrees)\\
$HeadPose_Z$&Tilting position of the head (degrees)\\
Makehuman Model&Preset makehuman face model (9 face models)\\
\hline
\end{tabular}

\label{tab:faces_params}
\end{table}%

\begin{table}[tb]
\caption{IEE-Humans simulator parameters}
\footnotesize
\begin{tabular}{|@{\hspace{1pt}}p{3cm}|@{\hspace{1pt}}p{6cm}|}
\hline
\textbf{Parameter}&\textbf{Description}\\
\hline
Lamp Location&Location of scenario lamp source (X, Y, Z)\\
Lamp Direction&Direction of scenario lamp source (X, Y, Z)\\
Lamp Color&Color of scenario lamp source (R, G, B)\\
Camera Height&Height of scenario camera (pixels)\\
$HeadPose_X$&Vertical position of the head (degrees)\\
$HeadPose_Y$&Horizontal position of the head (degrees)\\
$HeadPose_Z$&Tilting position of the head (degrees)\\
Age&Age of the generated humanoid\\
Gender&Gender of the generated humanoid\\
Iris Size&Size of the humanoid iris\\
Pupil Size&Size of the humanoid pupil\\
Eye Saturation&Level of the humanoid eye saturation\\
Eye Color&Color of the humanoid eye\\
Eye Value&Controls the lightness level of iris\\
Skin Freckles&Amount of procedural freckles added to the skin\\
Skin Oil&Brightness of subtle oil effect on the skin\\
Skin Veins&Amount of procedural veins added to the skin\\
\hline
\end{tabular}

\label{tab:humans_params}
\end{table}%

To fine-tune and test the HPD DNNs we relied on the BIWI real-world dataset~\cite{BIWI}, which contains over $15,000$ pictures of $20$ people faces ($6$ females and $14$ males) annotated with head pose angle. People were recorded sitting in front of a Kinect~\cite{kinect}, which is a motion sensor add-on for the Xbox 360 gaming console, and were asked to turn their head around trying to span all possible yaw/pitch angles they could perform. For each image, the head pose angle is computed using FaceShift~\cite{faceshift}.
For our experiments, we automatically labeled each image with a head pose class derived from the provided angles; we considered 1,000 close-up pictures--- similar to what can be obtained with in-car, DNN-based sensors---belonging to two persons where one is used for training and the other one for testing.
\MAJOR{R3.13}{To generate close-up pictures we performed face detection and image cropping with the Dlib framework~\cite{DLIB}.}
\MAJOR{R3.13}{We could select only two persons from the BIWI dataset because the pictures belonging to other subjects were taken far from the camera and thus not useful to mimic a realistic situation with an in-car camera shooting a driver. Further, to simulate a real-world scenario where the DNN processes images of people never observed during training, we do not test the DNN using images belonging to the person used for training.}
To fine-tune and test the FLD DNN, since we require accurately annotated landmarks, we relied on a dataset provided by IEE.

Table~\ref{tab:dnns} provides the number of images used to fine-tune the DNNs along with the obtained accuracy (i.e., the percentage of images for which the DNN provides correct outputs).
\begin{table*}[t]
\centering
\smaller
\footnotesize
\caption{Case Study Systems}
\begin{tabular}{|p{9mm}|p{15mm}|p{7mm}|p{11mm}|p{11mm}|p{7mm}|p{11mm}|p{11mm}|p{11mm}|@{\hspace{1pt}}p{7mm}|}
\hline
DNN &  \multicolumn{2}{c|}{Data Source}& \multicolumn{3}{c|}{Simulator-based Training} & \multicolumn{4}{c|}{Fine-tuning} \\
& Simulator & Real-world & Training Set Size & Test Set Size& Epochs & Training Set Size& Simulator-based Test Set Size& Real-world Test Set Size& Epochs\\
&  &  & (Accuracy) & (Accuracy) &  &  (Accuracy) &  (Accuracy) & (Accuracy) & \\
\hline
FLD & IEE-Faces & IEE & 16,000 (99.92\%) & 2,750 (44.41\%) & 10 & 9,000 (95.44\%) 
& 2,825 (43.22\%) 
& 1,000 (80.06\%) 
&  50  \\
HPD-F & IEE-Faces & BIWI & 21,500 (91.20\%) & 2,750 (85.43\%) & 18 & 21,976 (95.35\%) 
& 2,200 (87.10\%) 
& 500  (51.65\%) 
& 9  \\
HPD-H & IEE-Humans & BIWI & 15,400 (85.38\%) & 3,000 (80.21\%) & 25 & 18,476 (90.23\%) 
& 2,750 (85.23\%) 
& 500 (51.03\%) 
& 28 \\
\hline
\end{tabular}

\label{tab:dnns}
\end{table*}%

Column \emph{Data Source} provides the names of both the simulator and the real-world dataset used to train and fine-tune the DNNs.
The columns under \emph{Simulator-based Training} and \emph{Fine-tuning} provide details about the size (i.e., number of images) of the training and test sets used in those phases along with the accuracy of the DNN.
The data set used for training the fine-tuned DNNs (FLD, HPD-F, HPD-H) consists of simulator images 
($3,000$ for FLD, $21,500$ for HPD-F, $18,000$ for HPD-H)
and real-world images 
($6,000$ for FLD, $476$ for HPD-F, $476$ for HPD-H); \MAJOR{R2.23}{in Table~\ref{tab:dnns}, we report the size of the whole fine-tuning training set, including both simulator and real-world images.}

\MAJOR{R2.23}{Columns \emph{Simulator-based Training -- Epochs} and \emph{Fine-tuning -- Epochs} report the number of epochs considered to train and fine-tune the DNNs, respectively. All the DNNs have been fine-tuned for a number of epochs sufficient to achieve an accuracy above $90\%$ with
a training set of simulator and real-world images.}


All the DNNs were implemented with PyTorch~\cite{PyTorch}. HPD follows the AlexNet architecture~\cite{AlexNet} which is commonly used for image classification tasks, while FLD follows the Hourglass architecture~\cite{Hourglass}, which is optimized for landmarks detection (regression tasks).

In \APPR Step 1, for each case study DNN, we process all the failure-inducing images of the case study with \HUDD to generate RCCs. \HUDD's execution led to $10$ RCCs for HPD-F DNN, $11$ RCCs for HPD-H DNN, and $10$ RCCs for FLD DNN.

Table~\ref{tab:sede} provides further information about the configuration of \APPR. \MAJOR{R3.G}{Since in Steps 2.2 and 2.3 each individual of the initial population matches the reference one (i.e., the individual that should be similar to the one identified in the final population), we use low crossover and mutation probability to mutate only a few chromosomes at a time (i.e., we depart from the initial individual slowly).}
\begin{table*}[t]
\centering
\smaller
\caption{\APPR Configuration}
\begin{tabular}{|c|c|c|c|c|ccc|}
\hline
Step & \begin{tabular}[c]{@{}c@{}}Population  Size\end{tabular} & \begin{tabular}[c]{@{}c@{}}Number  of iterations\end{tabular}
& Crossover&
Mutation
& \multicolumn{3}{c|}{\begin{tabular}[c]{@{}c@{}} Avg. Execution Time (hrs.) \end{tabular}}
\\ 
& & & probability & probability & FLD & HPD-F & HPD-H
\\ \hline 
2.1 & 25 & 100 & 0.7& 0.3 &  40 & 40 & 200
\\ \hline
2.2 & 25 & 100 &0.3& 0.3 &  20 & 20 & 100
\\ \hline
2.3 & 25 & 100 &0.3& 0.3 &  20 & 20 & 100
\\ \hline

\end{tabular}
\label{tab:sede}
\end{table*}

\subsection{RQ1. \emph{How does \GAAlg fare, compared to \nsga and \dnsga, for the generation of diverse images belonging to RCCs?}}
\label{sec:empirical:gaalg}

\subsubsection{Experiment Design.} 

We aim to demonstrate that \GAAlg, our dedicated algorithm to derive a diverse image population belonging to a RCC, performs better than \nsga \MAJOR{R3.3}{and \dnsga}.
\MAJOR{}{To this end, we measure the diversity in the populations generated by the three algorithms over time, for all the RCCs under analysis.}

We choose \nsga as a baseline for comparison since its crowding distance function is designed to preserve and optimize the diversity between individuals by prioritizing individuals being more distant from others, along with individuals at the boundaries of the objective space.

We configured \nsga with an objective function (Equation~\ref{eq:f:nsga2}) that drives the generation of individuals belonging to the RCC through their normalized distance from the cluster's medoid, as for \GAAlg (Equation~\ref{eq:D}).

\begin{equation}
\label{eq:f:nsga2}
F^C(i) = \mathit{RCC}_{\mathit{distance}}(C,i)
\end{equation}

\MAJOR{R3.3}{Additionally, we compare \GAAlg with \dnsga because the latter is a state-of-the-art solution to generate a population of individuals that are diverse. To this end, we extended the implementation provided for DeepJanus~\cite{Riccio2020}. For our experiments, we rely on the following fitness functions:}

\MAJORSTARTS

\begin{equation}
\label{eq:f:deepnsga2:2}
F_1(i) = \mathit{ChromosomeDistance}(i,\mathit{closest}_{i_{A}})
\end{equation}

\begin{equation}
\label{eq:f:deepnsga2:2}
F_2^C(i) = \mathit{RCC}_{\mathit{distance}}(C,i)
\end{equation}

$F_1$, consistent with DeepJanus and DeepMetis, is computed as the distance between individual $i$  and the closest individual in the archive ($\mathit{closest}_{i_{A}}$). $F_2^C$ measures the distance between an individual $i$ and the medoid of the RCC $C$ under analysis, thus enabling the algorithm to drive the search towards its objective: generating individuals that belong to $C$. We execute four independent runs of \dnsga for each RCC under analysis.
At every search iteration, individuals are added to the archive if they belong to the RCC under analysis (i.e., $F_2^C(i) \leq 1$); we consider a sparseness threshold of zero (i.e., we add to the archive any individual that differs from the ones already in the archive, which happens when $F_1(i) > 0$). Since, at the end of the search, the archive may contain a number of individuals larger than the one required for the next steps of the algorithm (i.e., $25$, in our experiments), consistent with \GAAlg, we select the $25$ individuals with the highest $F_1$.
\MAJORENDS

For all three compared algorithms (\nsga, \dnsga, and \GAAlg), to measure the diversity of the population, since every individual is represented by a vector with the simulator parameter values used to generate the image (chromosome), we compute the average of the chromosome distance across pairs of individuals in the population.
Individuals that do not belong to any cluster are ignored from the computation of diversity; indeed, such an individual might increase diversity but it would not be useful for \APPR.

We configured \GAAlg, \nsga, and \dnsga with the same time budget; precisely, we let \nsga and \dnsga execute for the duration required by our algorithm to perform $100$ iterations. It amounts to $~40$ hours for HPD-F and FLD, $~200$ hours for HPD-H; differences are due to the time taken by the simulators to generate a single image. We executed \GAAlg, \nsga and \dnsga for all the $31$ RCCs identified for our case study DNNs. To account for randomness, we ran the experiment four different times for each root-cause cluster. With a total of $31$ RCCs, the four experiment runs led to the collection of $124$ data points, thus enabling statistical analysis within practical execution time.

To compare the increase in diversity achieved over time by the three algorithms, for each RCC, we recorded the diversity achieved by the three algorithms every hour.
Also, we tracked the percentage of individuals belonging to any RCC---a larger number of such individuals is expected to lead to better results in later stages of \APPR--- and the percentage of covered clusters (i.e., RCCs with at least one individual belonging to them).




\GAAlg performs better than \nsga and \dnsga if, over time, the following conditions hold: 
(1) the diversity achieved by \GAAlg is significantly higher than that achieved by \nsga and \dnsga, which would indicate that \GAAlg fits better our purpose (i.e., generating a diverse population of individuals);
(2) \GAAlg generates a larger proportion of individuals belonging to any RCC, which is useful to characterize RCCs since PART rules can be expected to be more accurate if a larger set of data points is considered;
(3) \GAAlg covers a larger number of RCCs (i.e., \GAAlg generates at least one image belonging to each cluster), thus enabling the characterization of a larger number of RCCs in later \APPR steps.


\subsubsection{Results.} 
\begin{figure*}[htp]
    \centering
    \begin{subfigure}[b]{0.6\textwidth}
        \centering
        \includegraphics[width=\textwidth]{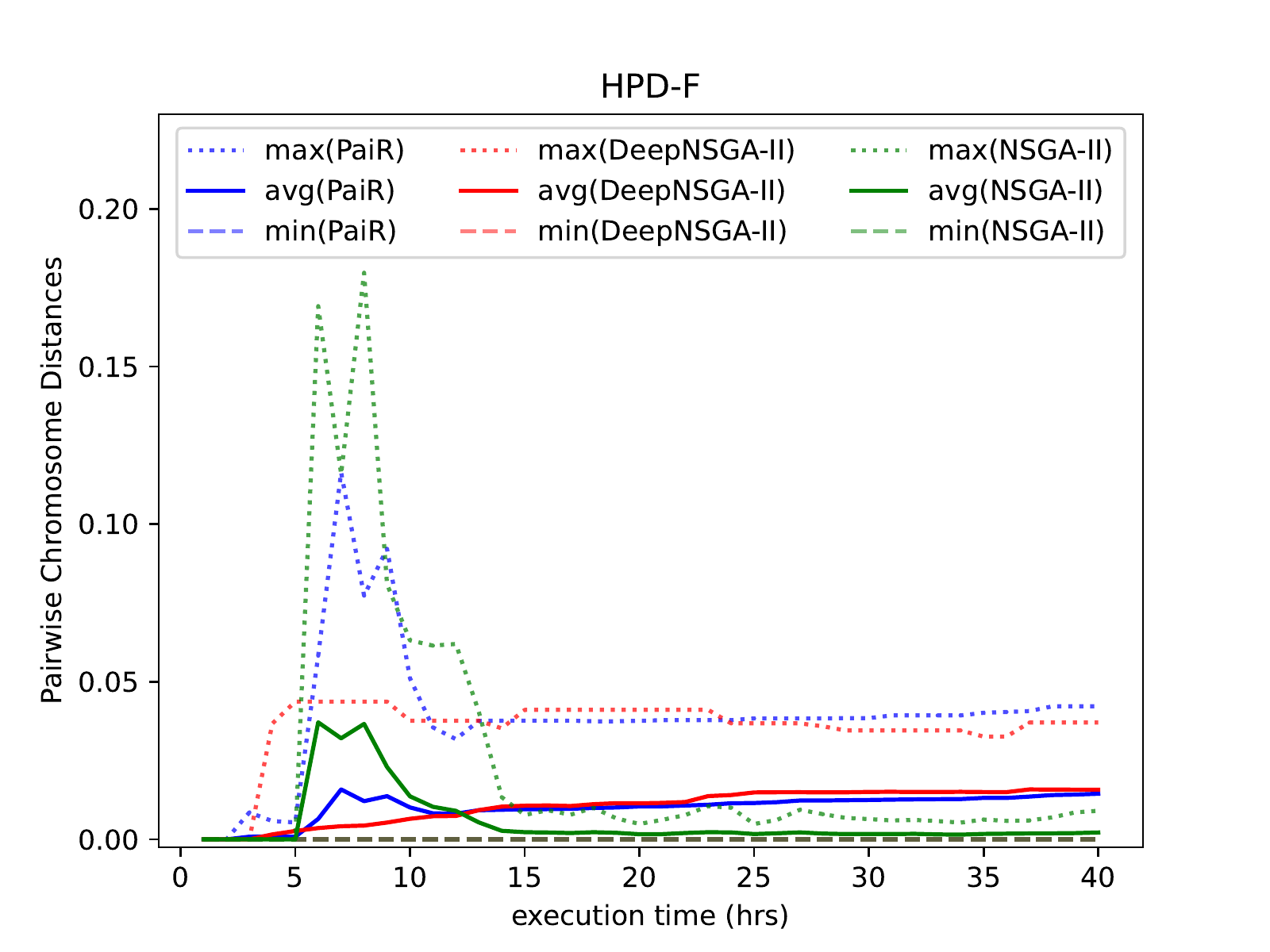}
    \end{subfigure}
    \begin{subfigure}[b]{0.6\textwidth}
        \centering
        \includegraphics[width=\textwidth]{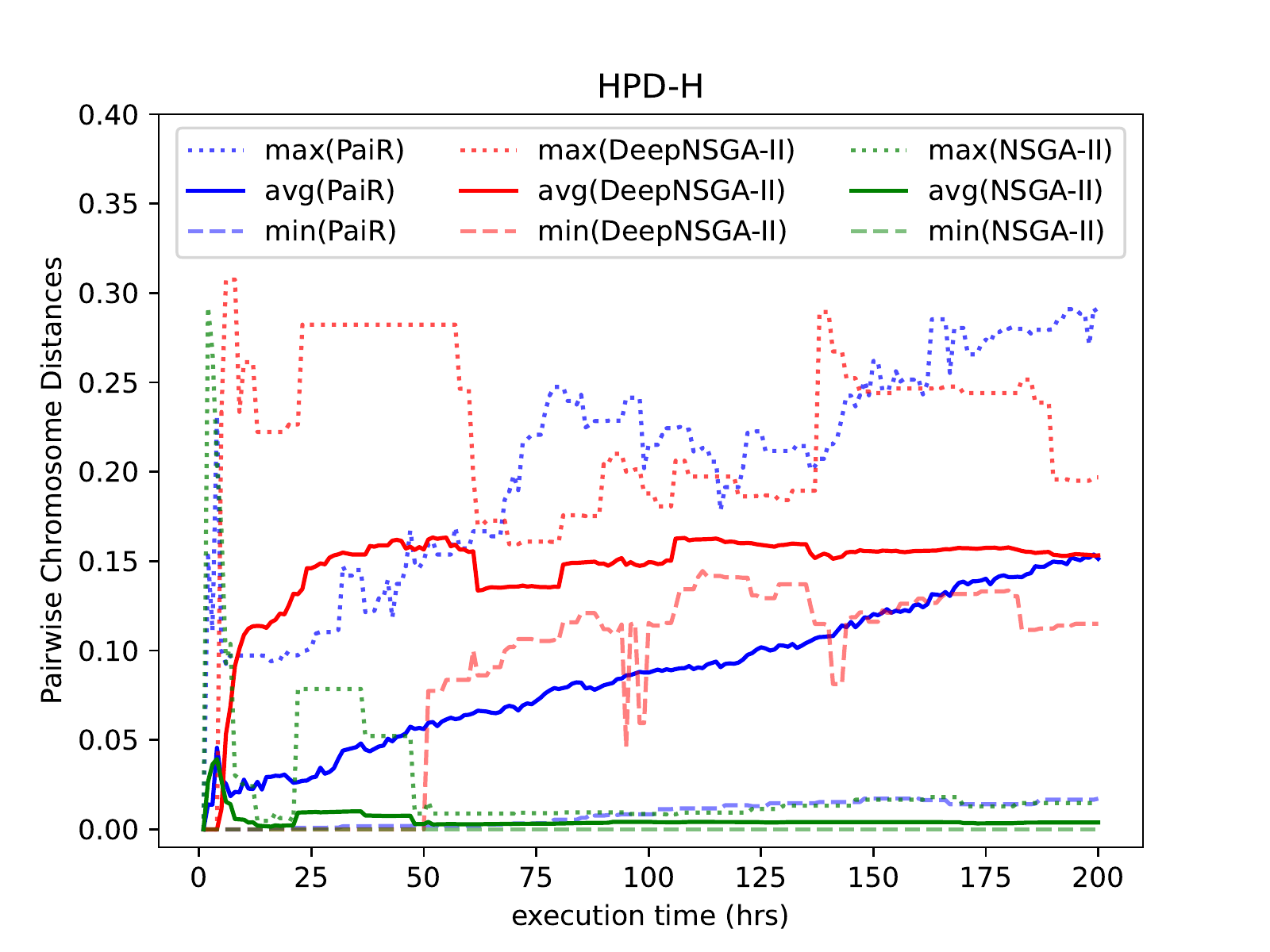}
    \end{subfigure}
    \begin{subfigure}[b]{0.6\textwidth}
        \centering
        \includegraphics[width=\textwidth]{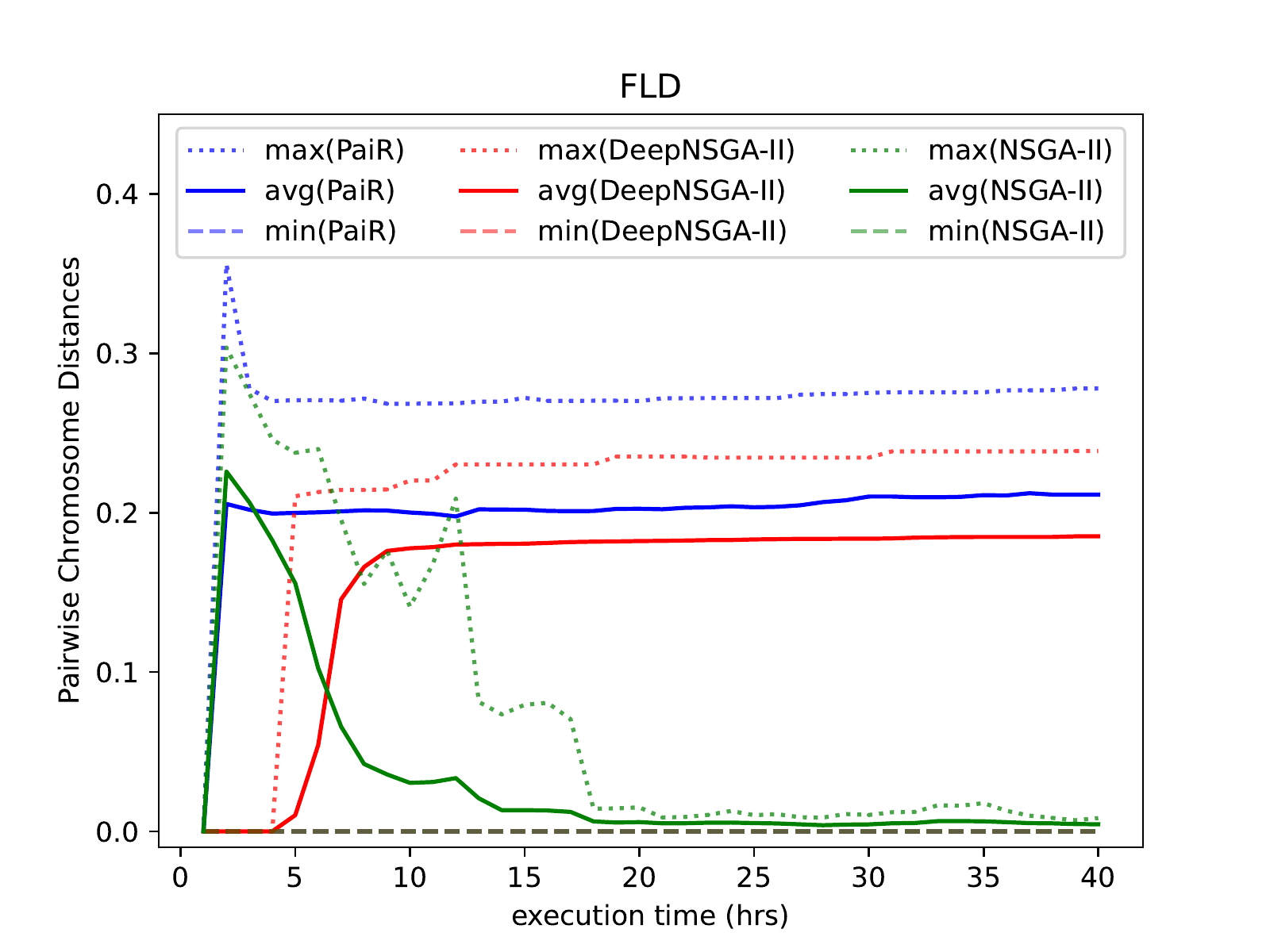}
    \end{subfigure}
\caption{Evolution of diversity achieved by \GAAlg compared to \nsga and \dnsga for HPD-F, HPD-H, and FLD DNNs}
\label{fig:RQ1-1}
\end{figure*}

Fig.~\ref{fig:RQ1-1} shows the evolution of diversity obtained with \GAAlg, \dnsga, and \nsga for our case study DNNs. To simplify visual comparisons, we plot the average, minimum, and maximum diversity observed at each timestamp (every hour); data points are  diversity values observed across the four executed runs for all RCCs. 

\begin{table}[t]
\centering
\footnotesize
\caption{RQ1: Average diversity across RCCs for HPD-F}
\begin{tabular}{|c|ccc|cccc|}
\hline
\begin{tabular}[c]{@{}c@{}}Time (hrs.)\end{tabular} & 
\multicolumn{3}{c|}{\begin{tabular}[c]{@{}c@{}}Avg. diversity (standard deviation)\end{tabular}} & 
\multicolumn{4}{c|}{\begin{tabular}[c]{@{}c@{}}Statistical test\end{tabular}} \\
& 
&
&
&
 \multicolumn{2}{c|}{NSGA-II} & \multicolumn{2}{c|}{DeepNSGA-II} \\ 
 &
\multicolumn{1}{C{1.2cm}|}{\GAAlg} & \multicolumn{1}{C{1.4cm}|}{NSGA-II} &  \multicolumn{1}{C{1.6cm}|}{DeepNSGA-II} & 
  \multicolumn{1}{c|}{\begin{tabular}[c]{@{}c@{}}\textit{p-value} \\(U-test)\end{tabular}} & \multicolumn{1}{c|}{\begin{tabular}[c]{@{}c@{}}Effect Size \\ ($\hat{A}_{12}$)\end{tabular}} &
  \multicolumn{1}{c|}{\begin{tabular}[c]{@{}c@{}}\textit{p-value} \\(U-test)\end{tabular}} & \multicolumn{1}{c|}{\begin{tabular}[c]{@{}c@{}}Effect Size \\ ($\hat{A}_{12}$)\end{tabular}}
 
\\ \hline
5 & \multicolumn{1}{c|}{0.001 (0.002)} & \multicolumn{1}{c|}{0.000 (0.000)} & \multicolumn{1}{c|}{0.003 (0.010)} & \multicolumn{1}{c|}{3.18e-03} & \multicolumn{1}{c|}{0.60} & \multicolumn{1}{c|}{0.159} & \multicolumn{1}{c|}{0.555} \\ \hline

10 & \multicolumn{1}{c|}{0.010 (0.017)} & \multicolumn{1}{c|}{0.014 (0.020)} & \multicolumn{1}{c|}{0.007 (0.011)} & \multicolumn{1}{c|}{0.870} & \multicolumn{1}{c|}{0.51} & \multicolumn{1}{c|}{0.190} & \multicolumn{1}{c|}{0.576} \\ \hline

15 & \multicolumn{1}{c|}{0.010 (0.013)} & \multicolumn{1}{c|}{0.002 (0.003)} & \multicolumn{1}{c|}{0.011 (0.012)} & \multicolumn{1}{c|}{8.53e-03} & \multicolumn{1}{c|}{0.66} & \multicolumn{1}{c|}{0.843} & \multicolumn{1}{c|}{0.512} \\ \hline

20 & \multicolumn{1}{c|}{0.010 (0.014)} & \multicolumn{1}{c|}{0.002 (0.002)} & \multicolumn{1}{c|}{0.011 (0.013)} & \multicolumn{1}{c|}{5.53e-04} & \multicolumn{1}{c|}{0.71} & \multicolumn{1}{c|}{0.843} & \multicolumn{1}{c|}{0.512} \\ \hline

25 & \multicolumn{1}{c|}{0.012 (0.013)} & \multicolumn{1}{c|}{0.002 (0.002)} & \multicolumn{1}{c|}{0.015 (0.014)} & \multicolumn{1}{c|}{6.18e-06} & \multicolumn{1}{c|}{0.78} & \multicolumn{1}{c|}{0.711} & \multicolumn{1}{c|}{0.476} \\ \hline

30 & \multicolumn{1}{c|}{0.013 (0.014)} & \multicolumn{1}{c|}{0.002 (0.002)} & \multicolumn{1}{c|}{0.015 (0.014)} & \multicolumn{1}{c|}{5.45e-05} & \multicolumn{1}{c|}{0.75} & \multicolumn{1}{c|}{0.771} & \multicolumn{1}{c|}{0.481} \\ \hline

35 & \multicolumn{1}{c|}{0.013 (0.014)} & \multicolumn{1}{c|}{0.002 (0.002)} & \multicolumn{1}{c|}{0.015 (0.014)} & \multicolumn{1}{c|}{6.18e-06} & \multicolumn{1}{c|}{0.78} & \multicolumn{1}{c|}{0.741} & \multicolumn{1}{c|}{0.479} \\ \hline

40 & \multicolumn{1}{c|}{0.014 (0.015)} & \multicolumn{1}{c|}{0.002 (0.003)} & \multicolumn{1}{c|}{0.016 (0.014)} & \multicolumn{1}{c|}{1.31e-05} & \multicolumn{1}{c|}{0.77} & \multicolumn{1}{c|}{0.640} & \multicolumn{1}{c|}{0.470} \\ \hline
\end{tabular}
\label{tab:rq1_1_hpdf_raw}
\end{table}
\begin{table}[h]
\centering
\footnotesize
\caption{RQ1: Average diversity across RCCs for HPD-H}
\begin{tabular}{|c|ccc|cccc|}
\hline
\begin{tabular}[c]{@{}c@{}}Time (hrs.)\end{tabular} & 
\multicolumn{3}{c|}{\begin{tabular}[c]{@{}c@{}}Avg. diversity (standard deviation)\end{tabular}} & 
\multicolumn{4}{c|}{\begin{tabular}[c]{@{}c@{}}Statistical test\end{tabular}} \\
& 
&
&
&
 \multicolumn{2}{c|}{NSGA-II} & \multicolumn{2}{c|}{DeepNSGA-II} \\
 &
\multicolumn{1}{C{1.2cm}|}{\GAAlg} & \multicolumn{1}{C{1.4cm}|}{NSGA-II} &  \multicolumn{1}{C{1.6cm}|}{DeepNSGA-II} & 
  \multicolumn{1}{c|}{\begin{tabular}[c]{@{}c@{}}\textit{p-value} \\(U-test)\end{tabular}} & \multicolumn{1}{c|}{\begin{tabular}[c]{@{}c@{}}Effect Size \\ ($\hat{A}_{12}$)\end{tabular}} &
  \multicolumn{1}{c|}{\begin{tabular}[c]{@{}c@{}}\textit{p-value} \\(U-test)\end{tabular}} & \multicolumn{1}{c|}{\begin{tabular}[c]{@{}c@{}}Effect Size \\ ($\hat{A}_{12}$)\end{tabular}}
 
\\ \hline

5 & \multicolumn{1}{c|}{0.028 (0.039)} & \multicolumn{1}{c|}{0.028 (0.056)} & \multicolumn{1}{c|}{0.011 (0.045)} & \multicolumn{1}{c|}{0.062} & \multicolumn{1}{c|}{0.612} & \multicolumn{1}{c|}{2.58e-08} & \multicolumn{1}{c|}{0.805} \\ \hline

10 & \multicolumn{1}{c|}{0.028 (0.039)} & \multicolumn{1}{c|}{0.005 (0.008)} & \multicolumn{1}{c|}{0.109 (0.083)} & \multicolumn{1}{c|}{5.43e-03} & \multicolumn{1}{c|}{0.669} & \multicolumn{1}{c|}{2.45e-03} & \multicolumn{1}{c|}{0.314} \\ \hline

15 & \multicolumn{1}{c|}{0.029 (0.037)} & \multicolumn{1}{c|}{0.002 (0.002)} & \multicolumn{1}{c|}{0.113 (0.077)} & \multicolumn{1}{c|}{5.45e-10} & \multicolumn{1}{c|}{0.880} & \multicolumn{1}{c|}{4.12e-04} & \multicolumn{1}{c|}{0.282} \\ \hline

20 & \multicolumn{1}{c|}{0.028 (0.035)} & \multicolumn{1}{c|}{0.002 (0.002)} & \multicolumn{1}{c|}{0.125 (0.067)} & \multicolumn{1}{c|}{5.66e-09} & \multicolumn{1}{c|}{0.860} & \multicolumn{1}{c|}{1.80e-06} & \multicolumn{1}{c|}{0.205} \\ \hline

25 & \multicolumn{1}{c|}{0.029 (0.036)} & \multicolumn{1}{c|}{0.010 (0.022)} & \multicolumn{1}{c|}{0.146 (0.061)} & \multicolumn{1}{c|}{4.06e-06} & \multicolumn{1}{c|}{0.785} & \multicolumn{1}{c|}{4.32e-10} & \multicolumn{1}{c|}{0.114} \\ \hline

50 & \multicolumn{1}{c|}{0.056 (0.049)} & \multicolumn{1}{c|}{0.003 (0.002)} & \multicolumn{1}{c|}{0.156 (0.047)} & \multicolumn{1}{c|}{2.30e-12} & \multicolumn{1}{c|}{0.934} & \multicolumn{1}{c|}{3.09e-11} & \multicolumn{1}{c|}{0.089} \\ \hline

75 & \multicolumn{1}{c|}{0.072 (0.061)} & \multicolumn{1}{c|}{0.003 (0.003)} & \multicolumn{1}{c|}{0.135 (0.014)} & \multicolumn{1}{c|}{1.20e-13} & \multicolumn{1}{c|}{0.959} & \multicolumn{1}{c|}{5.37e-10} & \multicolumn{1}{c|}{0.115} \\ \hline

100 & \multicolumn{1}{c|}{0.088 (0.063)} & \multicolumn{1}{c|}{0.004 (0.002)} & \multicolumn{1}{c|}{0.149 (0.015)} & \multicolumn{1}{c|}{1.89e-15} & \multicolumn{1}{c|}{0.992} & \multicolumn{1}{c|}{1.54e-06} & \multicolumn{1}{c|}{0.202} \\ \hline

125 & \multicolumn{1}{c|}{0.102 (0.071)} & \multicolumn{1}{c|}{0.004 (0.003)} & \multicolumn{1}{c|}{0.159 (0.012)} & \multicolumn{1}{c|}{6.37e-16} & \multicolumn{1}{c|}{1.0} & \multicolumn{1}{c|}{8.51e-04} & \multicolumn{1}{c|}{0.293} \\ \hline

150 & \multicolumn{1}{c|}{0.120 (0.083)} & \multicolumn{1}{c|}{0.004 (0.004)} & \multicolumn{1}{c|}{0.155 (0.020)} & \multicolumn{1}{c|}{6.37e-16} & \multicolumn{1}{c|}{1.0} & \multicolumn{1}{c|}{0.021} & \multicolumn{1}{c|}{0.357} \\ \hline

175 & \multicolumn{1}{c|}{0.140 (0.095)} & \multicolumn{1}{c|}{0.003 (0.003)} & \multicolumn{1}{c|}{0.157 (0.017)} & \multicolumn{1}{c|}{6.37e-16} & \multicolumn{1}{c|}{1.0} & \multicolumn{1}{c|}{0.049} & \multicolumn{1}{c|}{0.378} \\ \hline

200 & \multicolumn{1}{c|}{0.151 (0.093)} & \multicolumn{1}{c|}{0.004 (0.004)} & \multicolumn{1}{c|}{0.153 (0.016)} & \multicolumn{1}{c|}{6.37e-16} & \multicolumn{1}{c|}{1.0} & \multicolumn{1}{c|}{0.369} & \multicolumn{1}{c|}{0.444} \\ \hline
\end{tabular}
\label{tab:rq1_1_hpdh_raw}
\end{table}
\begin{table}[h]
\centering
\smaller
\caption{RQ1: Average diversity across RCCs for FLD}
\begin{tabular}{|c|ccc|cccc|}
\hline
\begin{tabular}[c]{@{}c@{}}Time (hrs.)\end{tabular} & 
\multicolumn{3}{c|}{\begin{tabular}[c]{@{}c@{}}Avg. diversity (standard deviation) \end{tabular}} & 
\multicolumn{4}{c|}{\begin{tabular}[c]{@{}c@{}}Statistical test\end{tabular}} \\
& 
&
&
&
 \multicolumn{2}{c|}{NSGA-II} & \multicolumn{2}{c|}{DeepNSGA-II} \\ 
 &
\multicolumn{1}{C{1.2cm}|}{\GAAlg} & \multicolumn{1}{C{1.4cm}|}{NSGA-II} &  \multicolumn{1}{C{1.6cm}|}{DeepNSGA-II} & 
  \multicolumn{1}{c|}{\begin{tabular}[c]{@{}c@{}}\textit{p-value} \\(U-test)\end{tabular}} & \multicolumn{1}{c|}{\begin{tabular}[c]{@{}c@{}}Effect Size \\ ($\hat{A}_{12}$)\end{tabular}} &
  \multicolumn{1}{c|}{\begin{tabular}[c]{@{}c@{}}\textit{p-value} \\(U-test)\end{tabular}} & \multicolumn{1}{c|}{\begin{tabular}[c]{@{}c@{}}Effect Size \\ ($\hat{A}_{12}$)\end{tabular}}
 
\\ \hline

5 & \multicolumn{1}{c|}{0.200 (0.077)} & \multicolumn{1}{c|}{0.156 (0.073)} & \multicolumn{1}{c|}{0.010 (0.045)} & \multicolumn{1}{c|}{4.01e-03} & \multicolumn{1}{c|}{0.685} & \multicolumn{1}{c|}{6.06e-13} & \multicolumn{1}{c|}{0.932} \\ \hline

10 & \multicolumn{1}{c|}{0.200 (0.077)} & \multicolumn{1}{c|}{0.030 (0.044)} & \multicolumn{1}{c|}{0.178 (0.061)} & \multicolumn{1}{c|}{4.36e-10} & \multicolumn{1}{c|}{0.905} & \multicolumn{1}{c|}{0.015} & \multicolumn{1}{c|}{0.657} \\ \hline

15 & \multicolumn{1}{c|}{0.202 (0.077)} & \multicolumn{1}{c|}{0.013 (0.023)} & \multicolumn{1}{c|}{0.181 (0.063)} & \multicolumn{1}{c|}{4.36e-10} & \multicolumn{1}{c|}{0.905} & \multicolumn{1}{c|}{0.028} & \multicolumn{1}{c|}{0.643} \\ \hline

20 & \multicolumn{1}{c|}{0.202 (0.077)} & \multicolumn{1}{c|}{0.006 (0.004)} & \multicolumn{1}{c|}{0.182 (0.063)} & \multicolumn{1}{c|}{4.36e-10} & \multicolumn{1}{c|}{0.905} & \multicolumn{1}{c|}{0.013} & \multicolumn{1}{c|}{0.660} \\ \hline

25 & \multicolumn{1}{c|}{0.203 (0.077)} & \multicolumn{1}{c|}{0.005 (0.003)} & \multicolumn{1}{c|}{0.183 (0.064)} & \multicolumn{1}{c|}{4.36e-10} & \multicolumn{1}{c|}{0.905} & \multicolumn{1}{c|}{0.028} & \multicolumn{1}{c|}{0.643} \\ \hline

30 & \multicolumn{1}{c|}{0.210 (0.080)} & \multicolumn{1}{c|}{0.004 (0.003)} & \multicolumn{1}{c|}{0.184 (0.064)} & \multicolumn{1}{c|}{4.36e-10} & \multicolumn{1}{c|}{0.905} & \multicolumn{1}{c|}{2.08e-03} & \multicolumn{1}{c|}{0.700} \\ \hline

35 & \multicolumn{1}{c|}{0.211 (0.080)} & \multicolumn{1}{c|}{0.006 (0.005)} & \multicolumn{1}{c|}{0.185 (0.065)} & \multicolumn{1}{c|}{4.36e-10} & \multicolumn{1}{c|}{0.905} & \multicolumn{1}{c|}{1.60e-03} & \multicolumn{1}{c|}{0.705} \\ \hline

40 & \multicolumn{1}{c|}{0.211 (0.080)} & \multicolumn{1}{c|}{0.004 (0.002)} & \multicolumn{1}{c|}{0.185 (0.065)} & \multicolumn{1}{c|}{4.36e-10} & \multicolumn{1}{c|}{0.905} & \multicolumn{1}{c|}{1.60e-03} & \multicolumn{1}{c|}{0.705} \\ \hline
\end{tabular}
\label{tab:rq1_1_fld_raw}
\end{table}

\MAJOR{R3.3}{In Fig.~\ref{fig:RQ1-1}, we can observe that, after 13 hours of execution, both \GAAlg and \dnsga outperform \nsga; also, with the largest test budget (i.e., 200 hours for HPD-H and 40 hours for HPD-F and FLD) \GAAlg performs either similar to \dnsga (for HPD-F and HPD-H) or outperforms it (FLD). } 
However, \GAAlg and \dnsga differ in performance across case studies: In the first 10 hours of execution, for the case study subjects relying on a low-fidelity simulator (HPD-F and FLD), PaiR shows better ($\textit{p-value} \leq 0.05$ for FLD) or slightly better ($\textit{p-value} > 0.05$ but $\hat{A}_{12} \geq 0.55$ for HPD-F) performance than \dnsga, while \dnsga yields better initial performance than \GAAlg for the case study with a high-fidelity simulator (HPD-H).
We believe that such result can be explained by the fact that, since the low-fidelity simulator provides less configuration parameters, it is more difficult to generate diverse images than with a high-fidelity simulator. Therefore, with a low-fidelity simulator it might be easier to achieve higher diversity by continuously evolving the same population (as done by \GAAlg) rather than by relying on repopulation (what is integrated into \dnsga). Indeed, repopulation forces the algorithm to use the test budget to re-generate images belonging to the RCC rather than diversifying images already belonging to a RCC. 

\begin{figure*}[htp]
    \centering
    \begin{subfigure}[b]{0.56\textwidth}
        \centering
        \includegraphics[width=\textwidth]{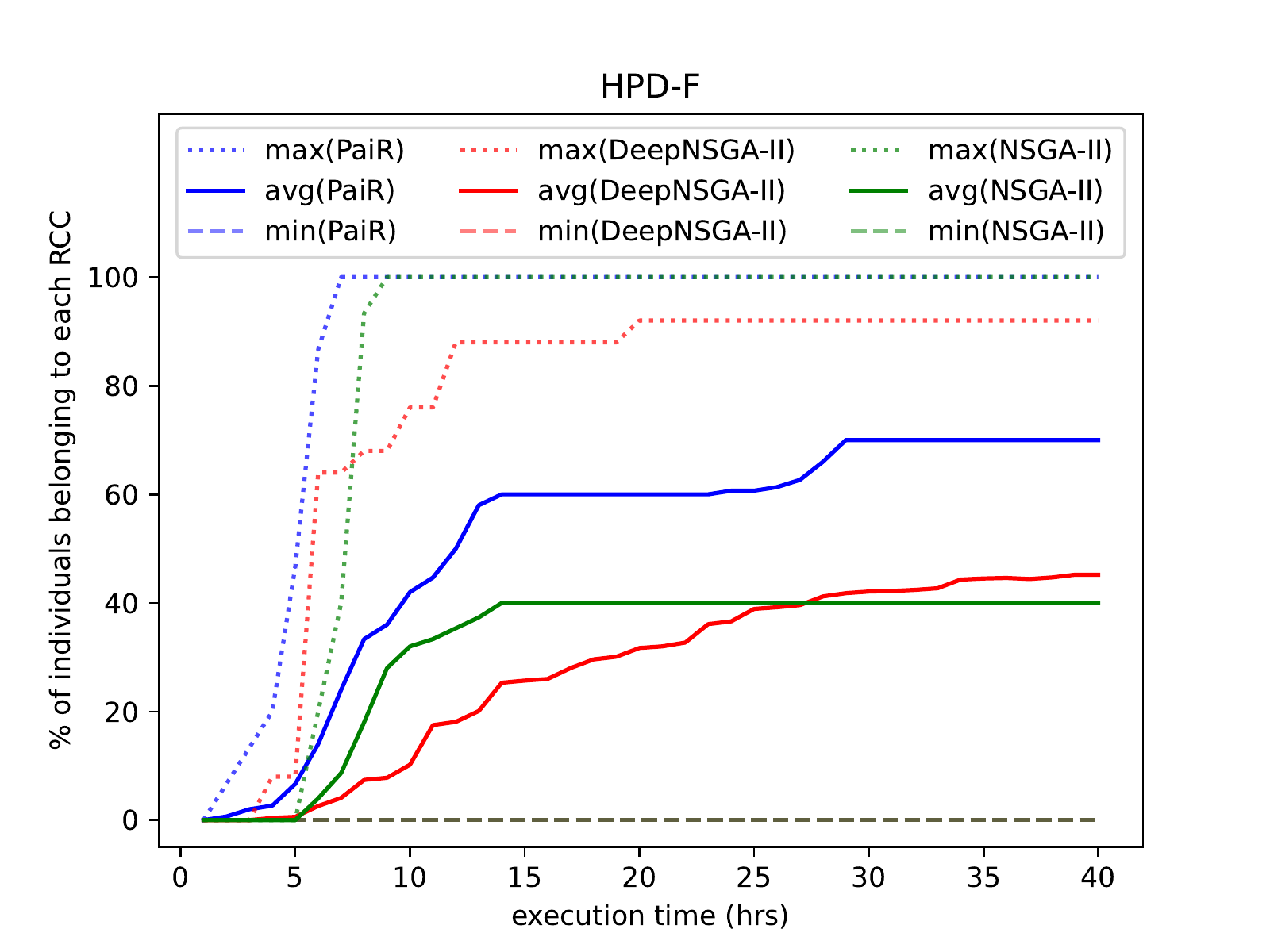}
    \end{subfigure}
    \begin{subfigure}[b]{0.56\textwidth}
        \centering
        \includegraphics[width=\textwidth]{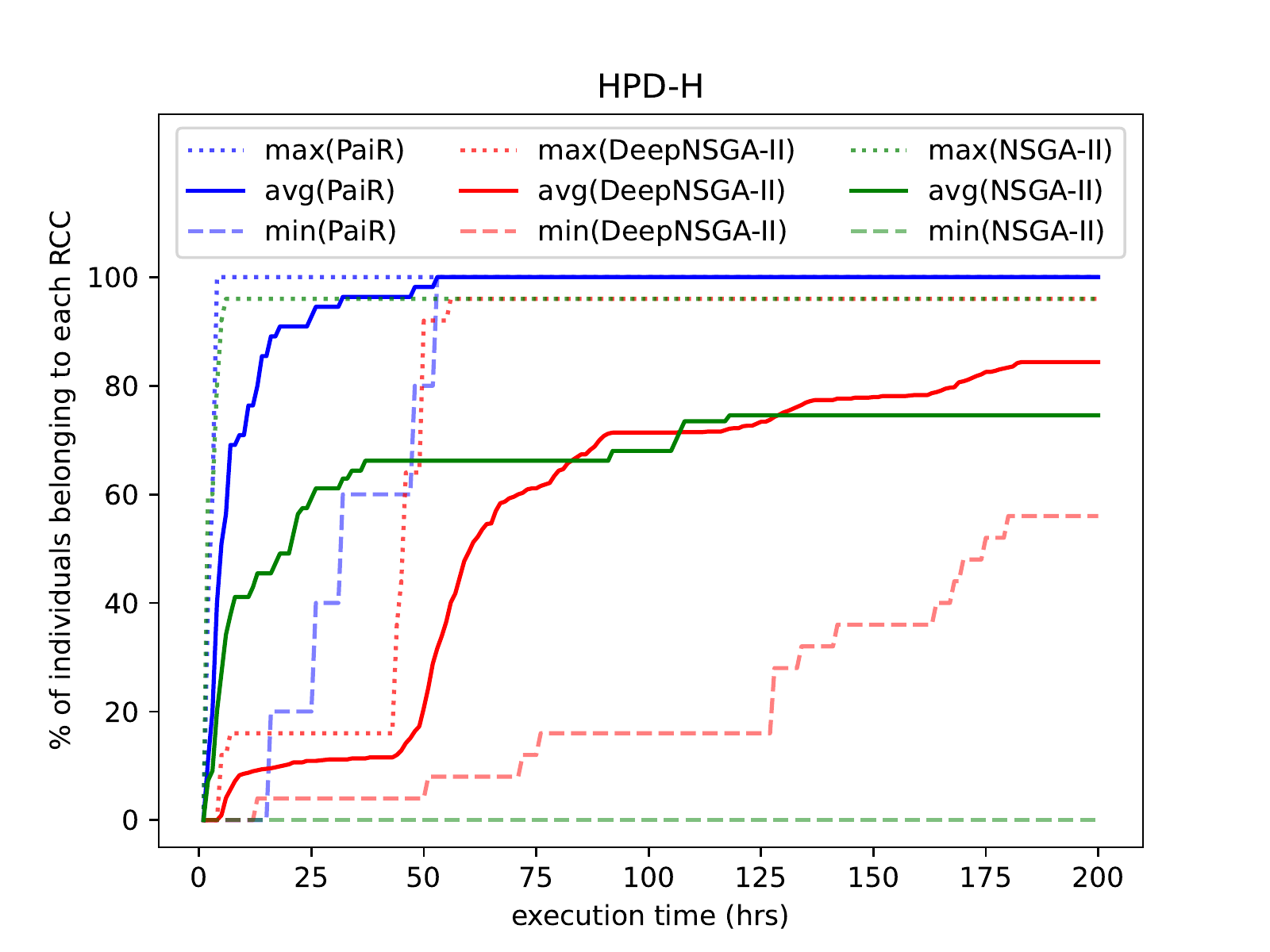}
    \end{subfigure}
    \begin{subfigure}[b]{0.56\textwidth}
        \centering
        \includegraphics[width=\textwidth]{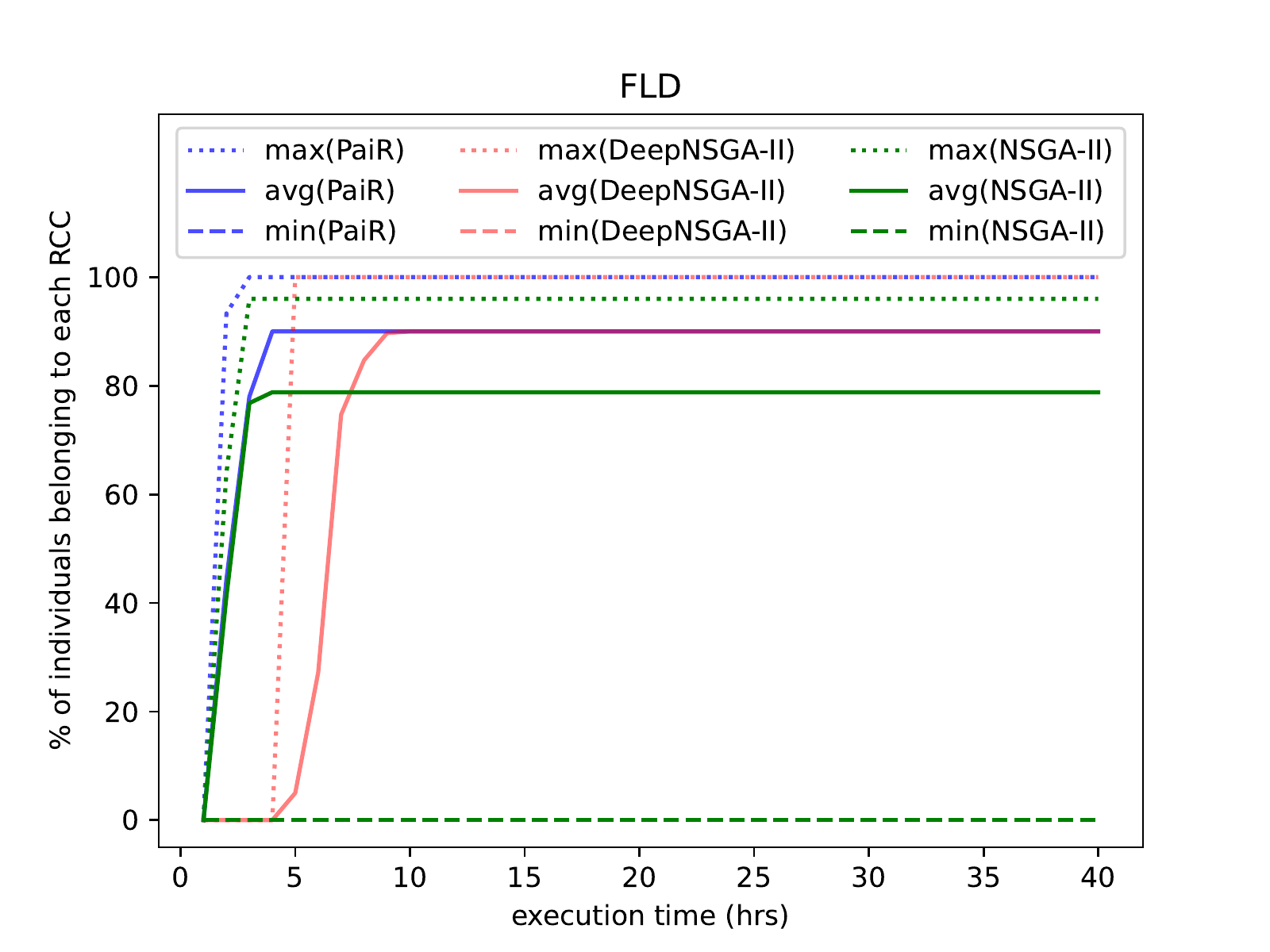}
    \end{subfigure}
\caption{Percentage of individuals belonging to a cluster generated by \GAAlg compared to \nsga and \dnsga for HPD-F, HPD-H, and FLD DNNs}
\label{fig:RQ1-2}
\end{figure*}

Further, we can observe that, during the first $10$ hours of execution, both \GAAlg and \nsga present a sharp peak for both  average and maximum diversity. 
This is mainly due to the fact that in the first hours of execution the two algorithms generate a limited number of images belonging to each RCC (see Fig.~\ref{fig:RQ1-2}), thus leading to high diversity regardless of the generation strategy.
After $10$ hours of execution and the initial peak, for \GAAlg, we can observe an up-trend in average diversity reaching $0.014$,  $0.151$, and $0.211$ for HPD-F, HPD-H, and FLD, respectively. 
In contrast, \nsga presents an average diversity that keeps decreasing till $15$ hours (HPD-F), $50$ hours (HPD-H), and $20$ hours (FLD), and then stabilizes around $0.002$, $0.004$, and $0.004$, at much lower levels than \GAAlg.
\MAJOR{R3.3}{Such initial peak is not observed with \dnsga,  likely because of repopulation. Indeed, repopulation slows down the identification of images belonging to a cluster:  \dnsga requires between 10 and 75 hours to generate the same number of RCC images generated by \GAAlg in five hours, as shown in Tables \ref{tab:rq1_1_hpdf_raw} to \ref{tab:rq1_1_fld_raw}, discussed below. However, repopulation leads to the identification of more diverse images since those generated by \dnsga in later iterations are not similar to the ones generated in previous iterations, as shown by the up-trending \dnsga curve.} 
For \dnsga, we observe an average diversity reaching $0.016$ (HPD-F), $0.153$ (HPD-H), and $0.185$ (FLD), respectively.



Tables \ref{tab:rq1_1_hpdf_raw}, \ref{tab:rq1_1_hpdh_raw}, and \ref{tab:rq1_1_fld_raw} report statistics about the diversity obtained for all RCCs, for \GAAlg, \nsga, \MAJOR{R3.3}{and \dnsga} \MAJOR{R3.18}{along with the standard deviation}. To compare the three techniques, we also provide the Vargha and Delaney's $\hat{A}_{12}$ effect size and p-values resulting from a non-parametric Mann-Whitney U-test, where each individual observation is the diversity for one RCC in one of the four runs. \GAAlg achieves a significantly higher diversity ($\textit{p-value} \leq 0.05$) compared to \nsga in all subjects and for all but two timestamps. 
For HPD-F, \GAAlg performs similarly to \nsga up to 10 hours of execution but then outperforms it with a large effect size \MAJOR{R2.25}{(i.e., $>$ $0.714$)\footnote{Please note that effect size is considered small when $\hat{A}_{12} > 0.556$, medium when $0.638 < \hat{A}_{12} < 0.714$, large when $\hat{A}_{12} \ge 0.714$~\cite{Kitchenham2017}.}}.
Given the safety-critical contexts in which the DNN under analysis should be adopted, we believe a budget of $20$ hours to be acceptable for DNN analysis.
\MAJOR{R3.3}{\GAAlg performs similarly to \dnsga (i.e., $\textit{p-value} > 0.05$), with a higher average diversity for \GAAlg up to 10 hours of execution.}

For HPD-H, \GAAlg significantly outperforms \nsga at all timestamps except one (i.e., $\textit{p-value} > 0.05$ for 5-hour execution). 
\MAJOR{R3.3}{However, \dnsga performs significantly better than \GAAlg for  all the timestamps except two. Indeed, at $5$ hours of execution \GAAlg performs significantly better than \dnsga with a large effect size and converges to a similar diversity after 200 hours.}

\MAJOR{R3.3}{For FLD, \GAAlg performs significantly better than both \nsga and \dnsga, with large and medium effect sizes, respectively. Compared to HPD-F, which relies on the same simulator used by FLD, the performance of \GAAlg could be attributed to the ease of generating images belonging to FLD clusters;
indeed, the radius of FLD clusters
range between [$27.33$, $74.65$] while this range for HPD-F is [$0.039$, $0.145$]. In such situation, \GAAlg can spend most of the test budget to maximize the diversity of the generated images.}


Fig.~\ref{fig:RQ1-2} shows the average, minimum, and maximum percentage of individuals belonging to one of the RCCs under analysis, observed at each timestamp over the four executed runs.
For \GAAlg, \nsga and \dnsga, the number of individuals belonging to any RCC increases over time, though this trend is much steeper for \GAAlg.
Indeed, on average, a larger proportion of the individuals generated by \GAAlg belongs to a RCC for a given execution time. This should result in better supporting the generation of PART rules at later stages of \APPR.

\begin{table}[t]
\centering
\footnotesize
\caption{RQ1: Percentage of individuals belonging to RCCs for HPD-F}
\begin{tabular}{|c|ccc|cccc|}
\hline
\begin{tabular}[c]{@{}c@{}}Time (hrs.)\end{tabular} & 
\multicolumn{3}{c|}{\begin{tabular}[c]{@{}c@{}}Avg. (standard deviation) \% of individuals \\ belonging to RCCs  \end{tabular}} & 
\multicolumn{4}{c|}{\begin{tabular}[c]{@{}c@{}}Statistical test\end{tabular}} \\
& 
&
&
&
 \multicolumn{2}{c|}{NSGA-II} & \multicolumn{2}{c|}{DeepNSGA-II} \\ 
 &
\multicolumn{1}{C{1.2cm}|}{\GAAlg} & \multicolumn{1}{C{1.4cm}|}{NSGA-II} &  \multicolumn{1}{C{1.6cm}|}{DeepNSGA-II} & 
  \multicolumn{1}{c|}{\begin{tabular}[c]{@{}c@{}}\textit{p-value} \\(U-test)\end{tabular}} & \multicolumn{1}{c|}{\begin{tabular}[c]{@{}c@{}}Effect Size \\ ($\hat{A}_{12}$)\end{tabular}} &
  \multicolumn{1}{c|}{\begin{tabular}[c]{@{}c@{}}\textit{p-value} \\(U-test)\end{tabular}} & \multicolumn{1}{c|}{\begin{tabular}[c]{@{}c@{}}Effect Size \\ ($\hat{A}_{12}$)\end{tabular}}
 
\\ \hline

5 & \multicolumn{1}{c|}{6.7\% (14.162)} & \multicolumn{1}{c|}{0.0\% (0.0)} & \multicolumn{1}{c|}{0.60\% (2.134)} & \multicolumn{1}{c|}{2.06e-04} & \multicolumn{1}{c|}{0.65} & \multicolumn{1}{c|}{8.72e-03} & \multicolumn{1}{c|}{0.616} \\ \hline

10 & \multicolumn{1}{c|}{42.0\% (47.27)} & \multicolumn{1}{c|}{32.0\% (44.36)} & \multicolumn{1}{c|}{10.2\% (22.72)} & \multicolumn{1}{c|}{0.309} & \multicolumn{1}{c|}{0.56} & \multicolumn{1}{c|}{7.01e-03} & \multicolumn{1}{c|}{0.656} \\ \hline

15 & \multicolumn{1}{c|}{60.0\% (49.61)} & \multicolumn{1}{c|}{40.0\% (49.61)} & \multicolumn{1}{c|}{25.7\% (33.77)} & \multicolumn{1}{c|}{0.076} & \multicolumn{1}{c|}{0.60} & \multicolumn{1}{c|}{7.48e-04} & \multicolumn{1}{c|}{0.705} \\ \hline

20 & \multicolumn{1}{c|}{60.0\% (49.61)} & \multicolumn{1}{c|}{40.0\% (49.61)} & \multicolumn{1}{c|}{31.7\% (37.10)} & \multicolumn{1}{c|}{0.076} & \multicolumn{1}{c|}{0.60} & \multicolumn{1}{c|}{7.48e-04} & \multicolumn{1}{c|}{0.705} \\ \hline

25 & \multicolumn{1}{c|}{60.7\% (48.82)} & \multicolumn{1}{c|}{40.0\% (49.61)} & \multicolumn{1}{c|}{38.9\% (38.84)} & \multicolumn{1}{c|}{0.024} & \multicolumn{1}{c|}{0.63} & \multicolumn{1}{c|}{9.77e-04} & \multicolumn{1}{c|}{0.706} \\ \hline

30 & \multicolumn{1}{c|}{70.0\% (46.41)} & \multicolumn{1}{c|}{40.0\% (49.61)} & \multicolumn{1}{c|}{42.1\% (40.36)} & \multicolumn{1}{c|}{7.48e-03} & \multicolumn{1}{c|}{0.65} & \multicolumn{1}{c|}{2.10e-05} & \multicolumn{1}{c|}{0.764} \\ \hline

35 & \multicolumn{1}{c|}{70.0\% (46.41)} & \multicolumn{1}{c|}{40.0\% (49.61)} & \multicolumn{1}{c|}{44.5\% (40.98)} & \multicolumn{1}{c|}{7.48e-03} & \multicolumn{1}{c|}{0.65} & \multicolumn{1}{c|}{2.09e-05} & \multicolumn{1}{c|}{0.764} \\ \hline

40 & \multicolumn{1}{c|}{70.0\% (46.41)} & \multicolumn{1}{c|}{40.0\% (49.61)} & \multicolumn{1}{c|}{45.2\% (41.17)} & \multicolumn{1}{c|}{7.48e-03} & \multicolumn{1}{c|}{0.65} & \multicolumn{1}{c|}{2.88e-05} & \multicolumn{1}{c|}{0.760} \\ \hline
\end{tabular}
\label{tab:rq1_2_hpdf}
\end{table}
\begin{table}[t]
\centering
\footnotesize
\caption{RQ1: Percentage of individuals belonging to RCCs for HPD-H}
\begin{tabular}{|c|ccc|cccc|}
\hline
\begin{tabular}[c]{@{}c@{}}Time (hrs.)\end{tabular} & 
\multicolumn{3}{c|}{\begin{tabular}[c]{@{}c@{}}Avg. (standard deviation) \% of individuals \\ belonging to RCCs  \end{tabular}} & 
\multicolumn{4}{c|}{\begin{tabular}[c]{@{}c@{}}Statistical test\end{tabular}} \\
& 
&
&
&
 \multicolumn{2}{c|}{NSGA-II} & \multicolumn{2}{c|}{DeepNSGA-II} \\ 
 &
\multicolumn{1}{C{1.2cm}|}{\GAAlg} & \multicolumn{1}{C{1.4cm}|}{NSGA-II} &  \multicolumn{1}{C{1.6cm}|}{DeepNSGA-II} & 
  \multicolumn{1}{c|}{\begin{tabular}[c]{@{}c@{}}\textit{p-value} \\(U-test)\end{tabular}} & \multicolumn{1}{c|}{\begin{tabular}[c]{@{}c@{}}Effect Size \\ ($\hat{A}_{12}$)\end{tabular}} &
  \multicolumn{1}{c|}{\begin{tabular}[c]{@{}c@{}}\textit{p-value} \\(U-test)\end{tabular}} & \multicolumn{1}{c|}{\begin{tabular}[c]{@{}c@{}}Effect Size \\ ($\hat{A}_{12}$)\end{tabular}}
 
\\ \hline
5 & \multicolumn{1}{c|}{50.91\% (39.87)} & \multicolumn{1}{c|}{27.27\% (35.60)} & \multicolumn{1}{c|}{0.91\% (2.84)} & \multicolumn{1}{c|}{1.24e-03} & \multicolumn{1}{c|}{0.694} & \multicolumn{1}{c|}{2.26e-12} & \multicolumn{1}{c|}{0.909} \\ \hline

10 & \multicolumn{1}{c|}{70.91\% (40.80)} & \multicolumn{1}{c|}{41.09\% (42.93)} & \multicolumn{1}{c|}{8.55\% (4.69)} & \multicolumn{1}{c|}{3.75e-06} & \multicolumn{1}{c|}{0.776} & \multicolumn{1}{c|}{8.78e-08} & \multicolumn{1}{c|}{0.824} \\ \hline

15 & \multicolumn{1}{c|}{85.45\% (32.38)} & \multicolumn{1}{c|}{45.45\% (43.00)} & \multicolumn{1}{c|}{9.45\% (4.32)} & \multicolumn{1}{c|}{1.24e-10} & \multicolumn{1}{c|}{0.880} & \multicolumn{1}{c|}{6.01e-12} & \multicolumn{1}{c|}{0.909} \\ \hline

20 & \multicolumn{1}{c|}{90.91\% (23.41)} & \multicolumn{1}{c|}{49.09\% (40.58)} & \multicolumn{1}{c|}{10.27\% (4.26)} & \multicolumn{1}{c|}{6.38e-12} & \multicolumn{1}{c|}{0.909} & \multicolumn{1}{c|}{4.36e-17} & \multicolumn{1}{c|}{1.0} \\ \hline

25 & \multicolumn{1}{c|}{92.73\% (23.26)} & \multicolumn{1}{c|}{59.27\% (36.63)} & \multicolumn{1}{c|}{10.91\% (3.89)} & \multicolumn{1}{c|}{2.71e-13} & \multicolumn{1}{c|}{0.929} & \multicolumn{1}{c|}{1.48e-17} & \multicolumn{1}{c|}{1.0} \\ \hline

50 & \multicolumn{1}{c|}{98.18\% (5.82)} & \multicolumn{1}{c|}{66.18\% (35.71)} & \multicolumn{1}{c|}{20.64\% (19.92)} & \multicolumn{1}{c|}{1.87e-14} & \multicolumn{1}{c|}{0.950} & \multicolumn{1}{c|}{3.41e-17} & \multicolumn{1}{c|}{1.0} \\ \hline

75 & \multicolumn{1}{c|}{100.0\% (0.0)} & \multicolumn{1}{c|}{66.18\% (35.71)} & \multicolumn{1}{c|}{61.09\% (34.37)} & \multicolumn{1}{c|}{5.02e-18} & \multicolumn{1}{c|}{1.0} & \multicolumn{1}{c|}{5.39e-18} & \multicolumn{1}{c|}{1.0} \\ \hline

100 & \multicolumn{1}{c|}{100.0\% (0.0)} & \multicolumn{1}{c|}{68.00\% (32.60)} & \multicolumn{1}{c|}{71.36\% (26.34)} & \multicolumn{1}{c|}{5.14e-18} & \multicolumn{1}{c|}{1.0} & \multicolumn{1}{c|}{5.48e-18} & \multicolumn{1}{c|}{1.0} \\ \hline

125 & \multicolumn{1}{c|}{100.0\% (0.0)} & \multicolumn{1}{c|}{74.55\% (29.29)} & \multicolumn{1}{c|}{73.36\% (24.22)} & \multicolumn{1}{c|}{4.44e-18} & \multicolumn{1}{c|}{1.0} & \multicolumn{1}{c|}{5.38e-18} & \multicolumn{1}{c|}{1.0} \\ \hline

150 & \multicolumn{1}{c|}{100.0\% (0.0)} & \multicolumn{1}{c|}{74.55\% (29.29)} & \multicolumn{1}{c|}{77.91\% (18.79)} & \multicolumn{1}{c|}{4.44e-18} & \multicolumn{1}{c|}{1.0} & \multicolumn{1}{c|}{5.24e-18} & \multicolumn{1}{c|}{1.0} \\ \hline

175 & \multicolumn{1}{c|}{100.0\% (0.0)} & \multicolumn{1}{c|}{74.55\% (29.29)} & \multicolumn{1}{c|}{82.55\% (13.89)} & \multicolumn{1}{c|}{4.44e-18} & \multicolumn{1}{c|}{1.0} & \multicolumn{1}{c|}{4.92e-18} & \multicolumn{1}{c|}{1.0} \\ \hline

200 & \multicolumn{1}{c|}{100.0\% (0.0)} & \multicolumn{1}{c|}{74.55\% (29.29)} & \multicolumn{1}{c|}{84.36\% (11.60)} & \multicolumn{1}{c|}{4.44e-18} & \multicolumn{1}{c|}{1.0} & \multicolumn{1}{c|}{4.76e-18} & \multicolumn{1}{c|}{1.0} \\ \hline
\end{tabular}
\label{tab:rq1_2_hpdh}
\end{table}
\begin{table}[t]
\centering
\footnotesize
\caption{RQ1: Percentage of individuals belonging to RCCs for FLD}
\begin{tabular}{|c|ccc|cccc|}
\hline
\begin{tabular}[c]{@{}c@{}}Time (hrs.)\end{tabular} & 
\multicolumn{3}{c|}{\begin{tabular}[c]{@{}c@{}}Avg. (standard deviation) \% of individuals \\ belonging to RCCs  \end{tabular}} & 
\multicolumn{4}{c|}{\begin{tabular}[c]{@{}c@{}}Statistical test\end{tabular}} \\
& 
&
&
&
 \multicolumn{2}{c|}{NSGA-II} & \multicolumn{2}{c|}{DeepNSGA-II} \\ 
 &
\multicolumn{1}{C{1.2cm}|}{\GAAlg} & \multicolumn{1}{C{1.4cm}|}{NSGA-II} &  \multicolumn{1}{C{1.6cm}|}{DeepNSGA-II} & 
  \multicolumn{1}{c|}{\begin{tabular}[c]{@{}c@{}}\textit{p-value} \\(U-test)\end{tabular}} & \multicolumn{1}{c|}{\begin{tabular}[c]{@{}c@{}}Effect Size \\ ($\hat{A}_{12}$)\end{tabular}} &
  \multicolumn{1}{c|}{\begin{tabular}[c]{@{}c@{}}\textit{p-value} \\(U-test)\end{tabular}} & \multicolumn{1}{c|}{\begin{tabular}[c]{@{}c@{}}Effect Size \\ ($\hat{A}_{12}$)\end{tabular}}
 
\\ \hline

5 & \multicolumn{1}{c|}{90.0\% (30.38)} & \multicolumn{1}{c|}{78.80\% (28.13)} & \multicolumn{1}{c|}{5.0\% (22.07)} & \multicolumn{1}{c|}{5.42e-11} & \multicolumn{1}{c|}{0.905} & \multicolumn{1}{c|}{4.06e-14} & \multicolumn{1}{c|}{0.925} \\ \hline

10 & \multicolumn{1}{c|}{90.0\% (30.38)} & \multicolumn{1}{c|}{78.80\% (28.13)} & \multicolumn{1}{c|}{90.0\% (30.38)} & \multicolumn{1}{c|}{5.42e-11} & \multicolumn{1}{c|}{0.905} & \multicolumn{1}{c|}{1.0} & \multicolumn{1}{c|}{0.5} \\ \hline

15 & \multicolumn{1}{c|}{90.0\% (30.38)} & \multicolumn{1}{c|}{78.80\% (28.13)} & \multicolumn{1}{c|}{90.0\% (30.38)} & \multicolumn{1}{c|}{5.42e-11} & \multicolumn{1}{c|}{0.905} & \multicolumn{1}{c|}{1.0} & \multicolumn{1}{c|}{0.5} \\ \hline

20 & \multicolumn{1}{c|}{90.0\% (30.38)} & \multicolumn{1}{c|}{78.80\% (28.13)} & \multicolumn{1}{c|}{90.0\% (30.38)} & \multicolumn{1}{c|}{5.42e-11} & \multicolumn{1}{c|}{0.905} & \multicolumn{1}{c|}{1.0} & \multicolumn{1}{c|}{0.5} \\ \hline

25 & \multicolumn{1}{c|}{90.0\% (30.38)} & \multicolumn{1}{c|}{78.80\% (28.13)} & \multicolumn{1}{c|}{90.0\% (30.38)} & \multicolumn{1}{c|}{5.42e-11} & \multicolumn{1}{c|}{0.905} & \multicolumn{1}{c|}{1.0} & \multicolumn{1}{c|}{0.5} \\ \hline

30 & \multicolumn{1}{c|}{90.0\% (30.38)} & \multicolumn{1}{c|}{78.80\% (28.13)} & \multicolumn{1}{c|}{90.0\% (30.38)} & \multicolumn{1}{c|}{5.42e-11} & \multicolumn{1}{c|}{0.905} & \multicolumn{1}{c|}{1.0} & \multicolumn{1}{c|}{0.5} \\ \hline

35 & \multicolumn{1}{c|}{90.0\% (30.38)} & \multicolumn{1}{c|}{78.80\% (28.13)} & \multicolumn{1}{c|}{90.0\% (30.38)} & \multicolumn{1}{c|}{5.42e-11} & \multicolumn{1}{c|}{0.905} & \multicolumn{1}{c|}{1.0} & \multicolumn{1}{c|}{0.5} \\ \hline

40 & \multicolumn{1}{c|}{90.0\% (30.38)} & \multicolumn{1}{c|}{78.80\% (28.13)} & \multicolumn{1}{c|}{90.0\% (30.38)} & \multicolumn{1}{c|}{5.42e-11} & \multicolumn{1}{c|}{0.905} & \multicolumn{1}{c|}{1.0} & \multicolumn{1}{c|}{0.5} \\ \hline
\end{tabular}
\label{tab:rq1_2_fld}
\end{table}

Tables \ref{tab:rq1_2_hpdf}, \ref{tab:rq1_2_hpdh}, and \ref{tab:rq1_2_fld} report statistics about the percentage of individuals belonging to any RCC, for \GAAlg, \nsga and \dnsga \MAJOR{R3.18}{along with the standard deviation}. We report the $\hat{A}_{12}$ statistics and  p-values resulting from a non-parametric Mann-Whitney U-test, where each observation is the percentage of individuals belonging to one cluster in one of the four runs. \GAAlg achieves a significantly higher number of individuals belonging to RCCs ($\textit{p-value} \leq 0.05$) compared to \nsga, for all subjects and timestamps except three. \MAJOR{R3.3}{Compared to \dnsga, \GAAlg identifies a significantly higher number of individuals belonging to RCCs for all the cases except FLD, where both techniques fare similarly.}
For HPD-F, around $10$ hours of execution, \GAAlg performs similarly to \nsga; both \GAAlg and \nsga perform significantly better than \dnsga. However, with a time budget of $25$ hours \GAAlg  performs significantly better than both \nsga and \dnsga, with \MAJOR{R2.26}{an effect size above  $0.60$}. 
\MAJOR{R3.3}{For HPD-H, \GAAlg significantly outperforms \nsga and \dnsga ($\textit{p-value} \leq 0.05$); effect size is always large (above $0.78$ in our cases) except for one case (medium, for 5-hour execution with \nsga). 
For HPD-F and HPD-H, we conjecture that \dnsga performs worse than \GAAlg because the images in a RCC are very similar to each other and several mutations of the best images in a population are needed to generate additional images belonging to a same RCC; such observation is supported by the plot for HPD-F, where \GAAlg lays in an intermediate plateau before reaching its best average result. The repopulation operator adopted by \dnsga may prevent \dnsga from generating individuals that belong to the RCC and are different from the already generated ones (e.g., after repopulation, \dnsga end-ups with images that match the ones already generated because they are simpler to generate).}
\MAJOR{R3.19}{For FLD, \GAAlg quickly reaches (in less than $5$ hours) a plateau for the average ($90\%$) and max ($100\%$) percentages of individuals per RCC. Though \nsga reaches its plateau in less than $5$ hours, its plateau is lower than \GAAlg's ($78.8\%$ vs. $90.0\%$).
Further, \dnsga reaches the same plateau values as \GAAlg but requires more time (i.e., $8$ hours for the peak of the average curve).
These results are mainly due to the ease of generating images belonging to most FLD clusters; indeed, all three algorithms quickly reach their peaks. However, both \GAAlg and \dnsga do not improve over $90\%$ because they do not cover one RCC. In other words, for FLD, when \GAAlg and \dnsga analyse a cluster that they can cover, both generate a population of images that fully belongs to the cluster. The same observation cannot be made for \nsga; indeed, although all three algorithms cover the same FLD clusters, \GAAlg and \dnsga achieve a higher percentage of individuals per cluster. \nsga probably gets stuck in local optima.}

\begin{figure*}[htp]
    \centering
    \begin{subfigure}[b]{0.495\textwidth}
        \centering
        \includegraphics[width=\textwidth]{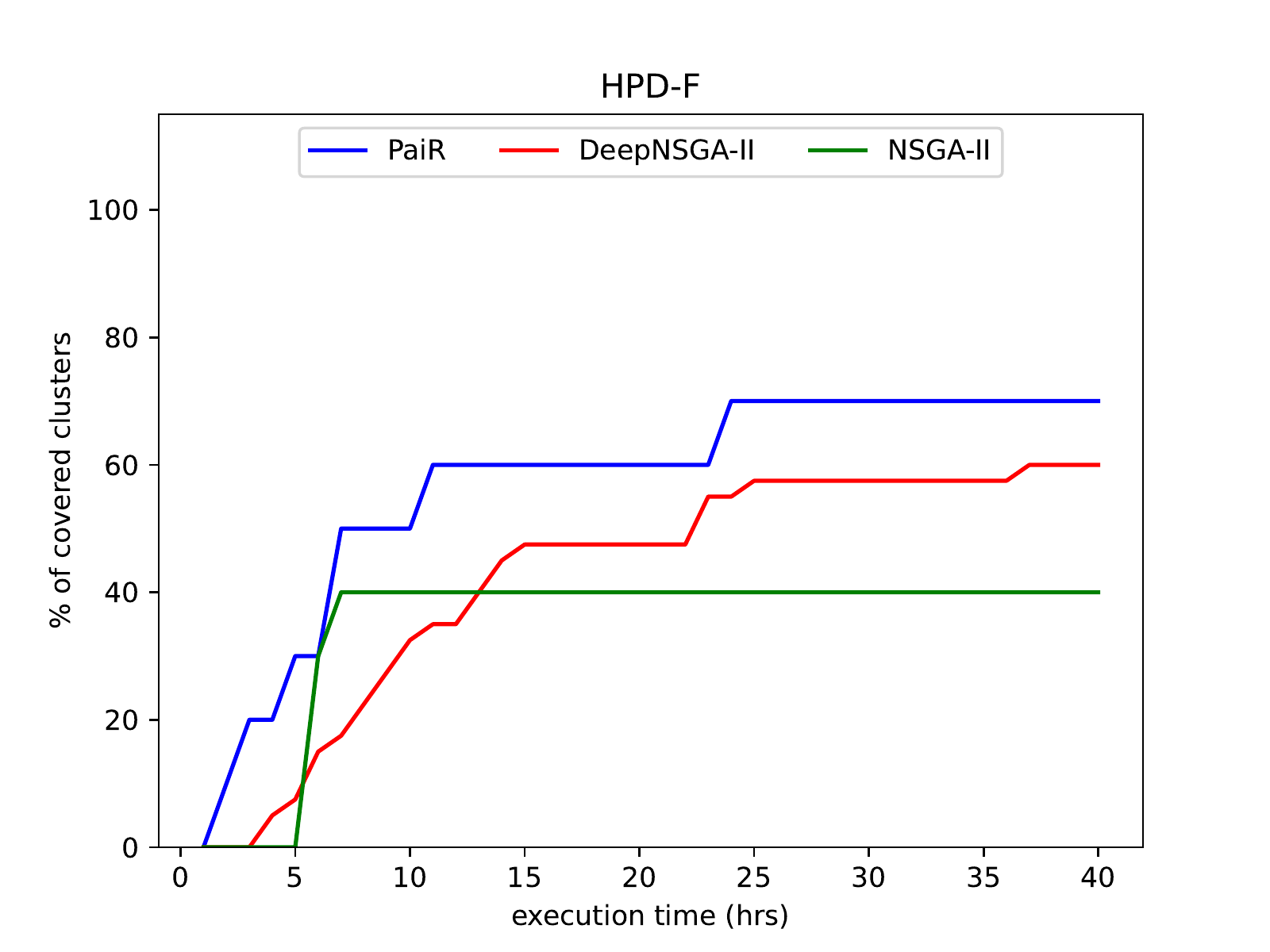}
        \label{fig:rq1-3-hpd-f}
    \end{subfigure}
    \begin{subfigure}[b]{0.495\textwidth}
        \centering
        \includegraphics[width=\textwidth]{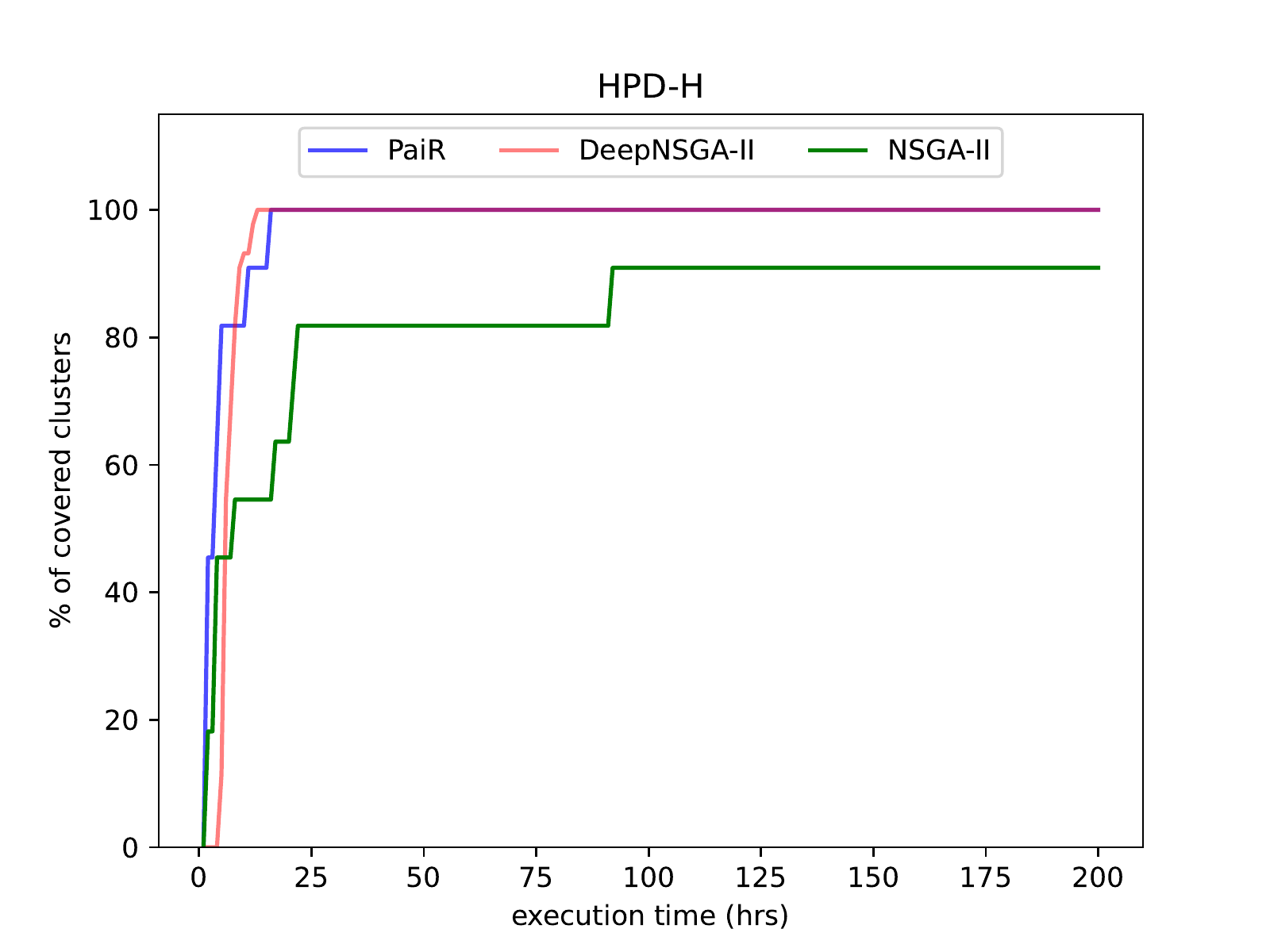}
        \label{fig:rq1-3-hpd-h}
    \end{subfigure}
    \begin{subfigure}[b]{0.495\textwidth}
        \centering
        \includegraphics[width=\textwidth]{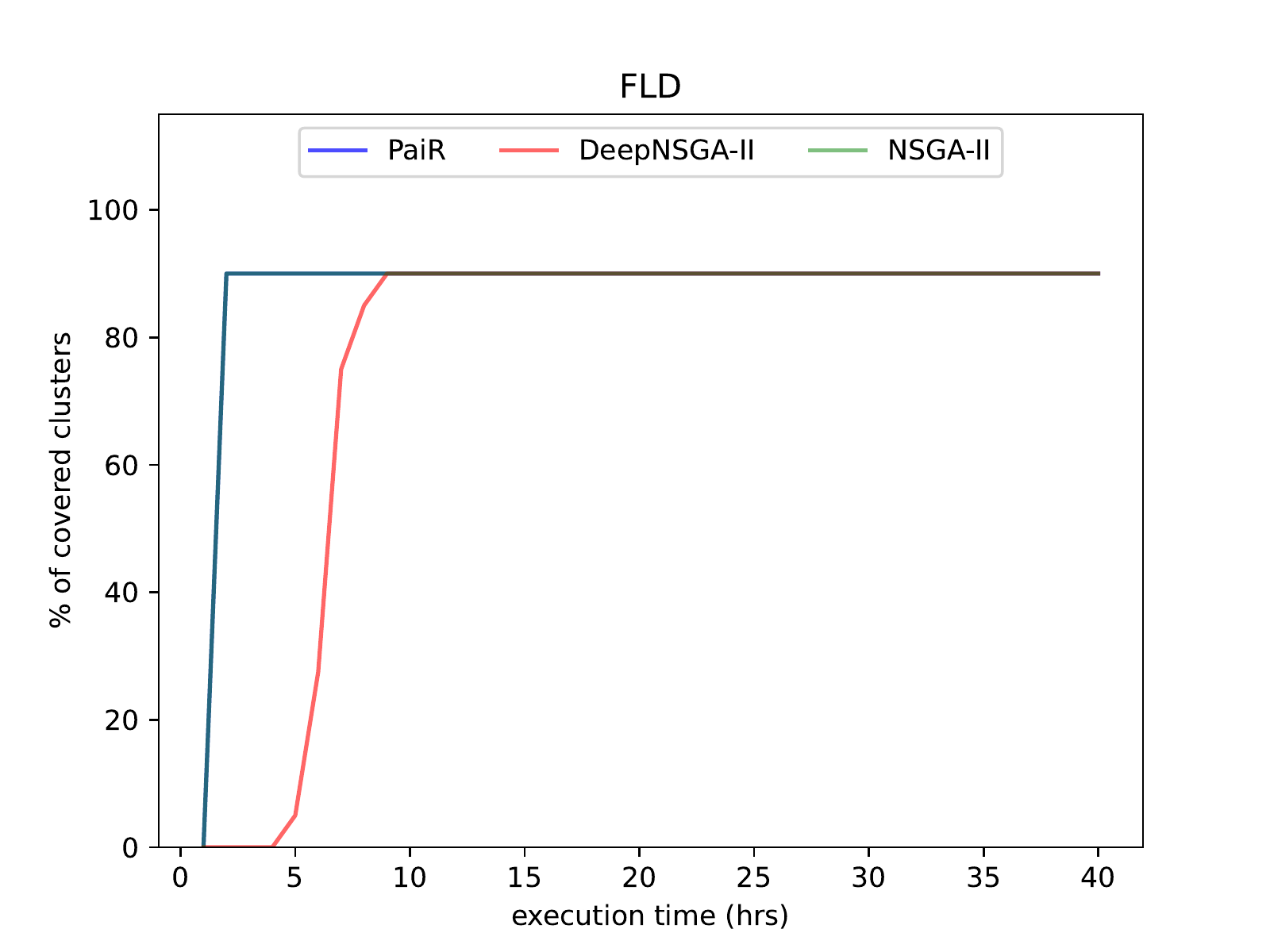}
        \label{fig:rq1-3-fld}
    \end{subfigure}
    \caption{Percentage of clusters covered by \GAAlg, \nsga and \dnsga for HPD-F, HPD-H, and FLD DNNs}
    \label{fig:RQ1-3}
\end{figure*}

Fig.~\ref{fig:RQ1-3} shows the percentage of clusters covered by \GAAlg, \nsga and \dnsga, for all subjects. \GAAlg is capable of covering (generating representative images of) a larger number of RCCs ($27$, in total): $7$ out of $10$ RCCs ($70.0\%$) for HPD-F, $11$ out of $11$ RCCs ($100.0\%$) for HPD-H, and $9$ out of $10$ RCCs ($90.0\%$) for FLD.
\MAJOR{R3.3}{\dnsga covers a lower number of clusters ($26$, in total): $6$ out of $10$ RCCs ($60.0\%$) for HPD-F, $11$ out of $11$ RCCs ($100.0\%$) for HPD-H, and $9$ out of $10$ RCCs ($90.0\%$) for FLD.}
\nsga, instead, covers only $23$ RCCs: $4$ out of $10$ RCCs ($40.0\%$) for HPD-F, $10$ out of $11$ RCCs ($90.9\%$) for HPD-H, and $9$ out of $10$ RCCs ($90.0\%$) for FLD.
Since the cases in which \GAAlg does not cover all the clusters are the ones involving a simulator with a lower number of configuration parameters (i.e., HPD-F and FLD), we believe that our imperfect results are due to our limited control of the simulator in use.
However, \GAAlg still performs better than \nsga and \dnsga, thus showing that it can better leverage the capabilities of the simulator.
We conclude that \GAAlg is a better choice than both \dnsga and \nsga since it helps explain a larger number of RCCs, $27$ ($87.1\%$) compared to $26$ ($83.8\%$) with \dnsga and $23$ ($74.1\%$) with \nsga.

%

\begin{table}[t]
\centering
\smaller
\caption{RQ1: Percentage of covered RCCs for HPD-F}
\begin{tabular}{|c|ccc|cc|}
\hline
\begin{tabular}[c]{@{}c@{}}Time (hrs.)\end{tabular} & 
\multicolumn{3}{c|}{\begin{tabular}[c]{@{}c@{}}Avg. \% of covered RCCs \end{tabular}} &  
\multicolumn{2}{c|}{\begin{tabular}[c]{@{}c@{}}\textit{p-value} (Fisher's Exact)\end{tabular}} \\
& 
\multicolumn{1}{C{1cm}|}{\GAAlg} & \multicolumn{1}{C{1cm}|}{NSGA-II} &  \multicolumn{1}{C{1.6cm}|}{DeepNSGA-II} 
& 
 \multicolumn{1}{C{1.6cm}|}{NSGA-II} & \multicolumn{1}{C{1.6cm}|}{DeepNSGA-II} 
 
\\ \hline

5 & \multicolumn{1}{c|}{30.0\%} & \multicolumn{1}{c|}{0.0\%} & \multicolumn{1}{c|}{7.5\%} & \multicolumn{1}{c|}{1.85e-04} & \multicolumn{1}{c|}{0.019} \\ \hline

10 & \multicolumn{1}{c|}{50.0\%} & \multicolumn{1}{c|}{40.0\%} & \multicolumn{1}{c|}{32.5\%} & \multicolumn{1}{c|}{0.5} & \multicolumn{1}{c|}{0.172} \\ \hline

15 & \multicolumn{1}{c|}{60.0\%} & \multicolumn{1}{c|}{40.0\%} & \multicolumn{1}{c|}{47.5\%} & \multicolumn{1}{c|}{0.117} & \multicolumn{1}{c|}{0.369} \\ \hline

20 & \multicolumn{1}{c|}{60.0\%} & \multicolumn{1}{c|}{40.0\%} & \multicolumn{1}{c|}{47.5\%} & \multicolumn{1}{c|}{0.117} & \multicolumn{1}{c|}{0.369} \\ \hline

25 & \multicolumn{1}{c|}{70.0\%} & \multicolumn{1}{c|}{40.0\%} & \multicolumn{1}{c|}{57.5\%} & \multicolumn{1}{c|}{0.012} & \multicolumn{1}{c|}{0.352} \\ \hline

30 & \multicolumn{1}{c|}{70.0\%} & \multicolumn{1}{c|}{40.0\%} & \multicolumn{1}{c|}{57.5\%} & \multicolumn{1}{c|}{0.012} & \multicolumn{1}{c|}{0.352} \\ \hline

35 & \multicolumn{1}{c|}{70.0\%} & \multicolumn{1}{c|}{40.0\%} & \multicolumn{1}{c|}{57.5\%} & \multicolumn{1}{c|}{0.012}  & \multicolumn{1}{c|}{0.352} \\ \hline

40 & \multicolumn{1}{c|}{70.0\%} & \multicolumn{1}{c|}{40.0\%} & \multicolumn{1}{c|}{60.0\%} & \multicolumn{1}{c|}{0.012} & \multicolumn{1}{c|}{0.482} \\ \hline
\end{tabular}
\label{tab:rq1_3_hpdf}
\end{table}
\begin{table}[t]
\centering
\smaller
\caption{RQ1: Percentage of covered RCCs for HPD-H}
\begin{tabular}{|c|ccc|cc|}
\hline
\begin{tabular}[c]{@{}c@{}}Time (hrs.)\end{tabular} & 
\multicolumn{3}{c|}{\begin{tabular}[c]{@{}c@{}}Avg. \% of covered RCCs \end{tabular}} & 
\multicolumn{2}{c|}{\begin{tabular}[c]{@{}c@{}}\textit{p-value} (Fisher's Exact)\end{tabular}} \\ &
\multicolumn{1}{C{1cm}|}{\GAAlg} & \multicolumn{1}{C{1cm}|}{NSGA-II} &  \multicolumn{1}{C{1.6cm}|}{DeepNSGA-II} 
& 
 \multicolumn{1}{C{1.6cm}|}{NSGA-II} & \multicolumn{1}{C{1.6cm}|}{DeepNSGA-II} 
 
\\ \hline
5 & \multicolumn{1}{c|}{81.82\%} & \multicolumn{1}{c|}{45.45\%} & \multicolumn{1}{c|}{11.36\%} & \multicolumn{1}{c|}{7.54e-04} & \multicolumn{1}{c|}{1.84e-11} \\ \hline

10 & \multicolumn{1}{c|}{81.82\%} & \multicolumn{1}{c|}{54.55\%} & \multicolumn{1}{c|}{93.18\%} & \multicolumn{1}{c|}{0.011} & \multicolumn{1}{c|}{0.195}  \\ \hline

15 & \multicolumn{1}{c|}{90.91\%} & \multicolumn{1}{c|}{54.55\%} & \multicolumn{1}{c|}{100.0\%} & \multicolumn{1}{c|}{2.27e-04} & \multicolumn{1}{c|}{0.116} \\ \hline

20 & \multicolumn{1}{c|}{100.0\%} & \multicolumn{1}{c|}{63.64\%} & \multicolumn{1}{c|}{100.0\%} & \multicolumn{1}{c|}{5.75e-06} & \multicolumn{1}{c|}{1.0} \\ \hline

25 & \multicolumn{1}{c|}{100.0\%} & \multicolumn{1}{c|}{81.82\%} & \multicolumn{1}{c|}{100.0\%} & \multicolumn{1}{c|}{5.51e-03} & \multicolumn{1}{c|}{1.0}  \\ \hline

50 & \multicolumn{1}{c|}{100.0\%} & \multicolumn{1}{c|}{81.82\%} & \multicolumn{1}{c|}{100.0\%} & \multicolumn{1}{c|}{5.51e-03} & \multicolumn{1}{c|}{1.0}\\ \hline

75 & \multicolumn{1}{c|}{100.0\%} & \multicolumn{1}{c|}{81.82\%} & \multicolumn{1}{c|}{100.0\%} & \multicolumn{1}{c|}{5.51e-03} & \multicolumn{1}{c|}{1.0} \\ \hline

100 & \multicolumn{1}{c|}{100.0\%} & \multicolumn{1}{c|}{90.91\%} & \multicolumn{1}{c|}{100.0\%} & \multicolumn{1}{c|}{0.116} & \multicolumn{1}{c|}{1.0} \\ \hline

125 & \multicolumn{1}{c|}{100.0\%} & \multicolumn{1}{c|}{90.91\%} & \multicolumn{1}{c|}{100.0\%} & \multicolumn{1}{c|}{0.116} & \multicolumn{1}{c|}{1.0} \\ \hline

150 & \multicolumn{1}{c|}{100.0\%} & \multicolumn{1}{c|}{90.91\%} & \multicolumn{1}{c|}{100.0\%} & \multicolumn{1}{c|}{0.116} & \multicolumn{1}{c|}{1.0}\\ \hline

175 & \multicolumn{1}{c|}{100.0\%} & \multicolumn{1}{c|}{90.91\%} & \multicolumn{1}{c|}{100.0\%} & \multicolumn{1}{c|}{0.116} & \multicolumn{1}{c|}{1.0}\\ \hline

200 & \multicolumn{1}{c|}{100.0\%} & \multicolumn{1}{c|}{90.91\%} & \multicolumn{1}{c|}{100.0\%} & \multicolumn{1}{c|}{0.116} & \multicolumn{1}{c|}{1.0} \\ \hline
\end{tabular}
\label{tab:rq1_3_hpdh}
\end{table}
\begin{table}[t]
\centering
\smaller
\caption{RQ1: Percentage of covered RCCs for FLD}
\begin{tabular}{|c|ccc|cc|}
\hline
\begin{tabular}[c]{@{}c@{}}Time (hrs.)\end{tabular} & 
\multicolumn{3}{c|}{\begin{tabular}[c]{@{}c@{}}Avg. \% of covered RCCs \end{tabular}} & 
\multicolumn{2}{c|}{\begin{tabular}[c]{@{}c@{}}\textit{p-value} (Fisher's Exact)\end{tabular}} \\ &
\multicolumn{1}{C{1cm}|}{\GAAlg} & \multicolumn{1}{C{1cm}|}{NSGA-II} &  \multicolumn{1}{C{1.6cm}|}{DeepNSGA-II} 
& 
 \multicolumn{1}{C{1.6cm}|}{NSGA-II} & \multicolumn{1}{C{1.6cm}|}{DeepNSGA-II} 
 
\\ \hline

5 & \multicolumn{1}{c|}{90.0\%} & \multicolumn{1}{c|}{90.0\%} & \multicolumn{1}{c|}{5.0\%} & \multicolumn{1}{c|}{1.0} & \multicolumn{1}{c|}{1.47e-15} \\ \hline

10 & \multicolumn{1}{c|}{90.0\%} & \multicolumn{1}{c|}{90.0\%} & \multicolumn{1}{c|}{90.0\%} & \multicolumn{1}{c|}{1.0} & \multicolumn{1}{c|}{1.0}  \\ \hline

15 & \multicolumn{1}{c|}{90.0\%} & \multicolumn{1}{c|}{90.0\%} & \multicolumn{1}{c|}{90.0\%} & \multicolumn{1}{c|}{1.0} & \multicolumn{1}{c|}{1.0} \\ \hline

20 & \multicolumn{1}{c|}{90.0\%} & \multicolumn{1}{c|}{90.0\%} & \multicolumn{1}{c|}{90.0\%} & \multicolumn{1}{c|}{1.0} & \multicolumn{1}{c|}{1.0}  \\ \hline

25 & \multicolumn{1}{c|}{90.0\%} & \multicolumn{1}{c|}{90.0\%} & \multicolumn{1}{c|}{90.0\%} & \multicolumn{1}{c|}{1.0}  & \multicolumn{1}{c|}{1.0}  \\ \hline

30 & \multicolumn{1}{c|}{90.0\%} & \multicolumn{1}{c|}{90.0\%} & \multicolumn{1}{c|}{90.0\%} & \multicolumn{1}{c|}{1.0}  & \multicolumn{1}{c|}{1.0}  \\ \hline

35 & \multicolumn{1}{c|}{90.0\%} & \multicolumn{1}{c|}{90.0\%} & \multicolumn{1}{c|}{90.0\%} & \multicolumn{1}{c|}{1.0} & \multicolumn{1}{c|}{1.0}  \\ \hline

40 & \multicolumn{1}{c|}{90.0\%} & \multicolumn{1}{c|}{90.0\%} & \multicolumn{1}{c|}{90.0\%} & \multicolumn{1}{c|}{1.0}  & \multicolumn{1}{c|}{1.0}  \\ \hline
\end{tabular}
\label{tab:rq1_3_fld}
\end{table}
\MAJOR{R3.20}{
Tables \ref{tab:rq1_3_hpdf}, \ref{tab:rq1_3_hpdh}, and \ref{tab:rq1_3_fld} report the percentage of covered RCCs across the four runs, for \GAAlg, \nsga and \dnsga.
Further, we report the p-value computed with a non-parametric Fisher's exact test where we compare the number of clusters covered by \GAAlg and the competing approaches. Unsurprisingly, there is no significant difference for FLD, where the three algorithms cover almost all the clusters because of the easiness of the task (see discussion above). Differences with \dnsga tend to be not significant, which indicates that covering a cluster is a simple task for both algorithms. However, differences are always significant for small time budget; indeed, up to 5 hours, \GAAlg covers more clusters. A higher number of significant differences is instead observed when comparing \GAAlg to \nsga.}


\MAJOR{R3.3}{To summarize, our results suggest that the adoption of \GAAlg is the best choice in our context; indeed, \GAAlg 
can (1) generate images for a larger number of RCCs than \nsga and \dnsga, 
(2) generate significantly more images belonging to each RCC, compared to both  \dnsga and \nsga, and, 
(3) achieve significantly higher image diversity than \nsga and similar diversity to that of \dnsga, except for one case study, where \GAAlg outperforms \dnsga.}


In the rest of our evaluation, we ignore the clusters for which \GAAlg was not able to identify representative images (Step 2.1), as described above. Also, we ignore one cluster for which \APPR did not identify unsafe images (Step 2.2), possibly because the hazard-triggering event is the presence of jewelry, which is not generated by our simulator. In total, we ignore three clusters in the case of HPD-F, and one cluster for HPD-H and FLD.

\subsection{RQ2. \emph{Does \APPR generate simulator images that are close to the center of each RCC?}}
\label{sec:empirical:rq2}

\subsubsection{Experiment Design.} This research question evaluates whether the images generated by \APPR for each RCC are closer to the medoid of the RCC than random failing images. Otherwise, we cannot claim that \APPR contributes to generating images that help characterize the RCC.

To measure the distance of an image from the RCC medoid we once again rely on the heatmap distance between the RCC medoid and the image (Equation~\ref{eq:HD}).

For each RCC, we compute the distance between the RCC medoid and every image generated by \APPR to characterize the RCC in Step 2.1. Also, we compute the distance of  the unsafe images in the simulator-based test set (i.e., random failing images) from the RCC medoid.
To positively answer our research question, for each RCC, the average distance from the medoid to the images generated by \APPR should be significantly smaller than the average distance obtained with random unsafe simulated images, based on a non-parametric Mann-Whitney U-test.


\subsubsection{Results.} 
\begin{figure}[htp]
\includegraphics[height=0.3\textheight]{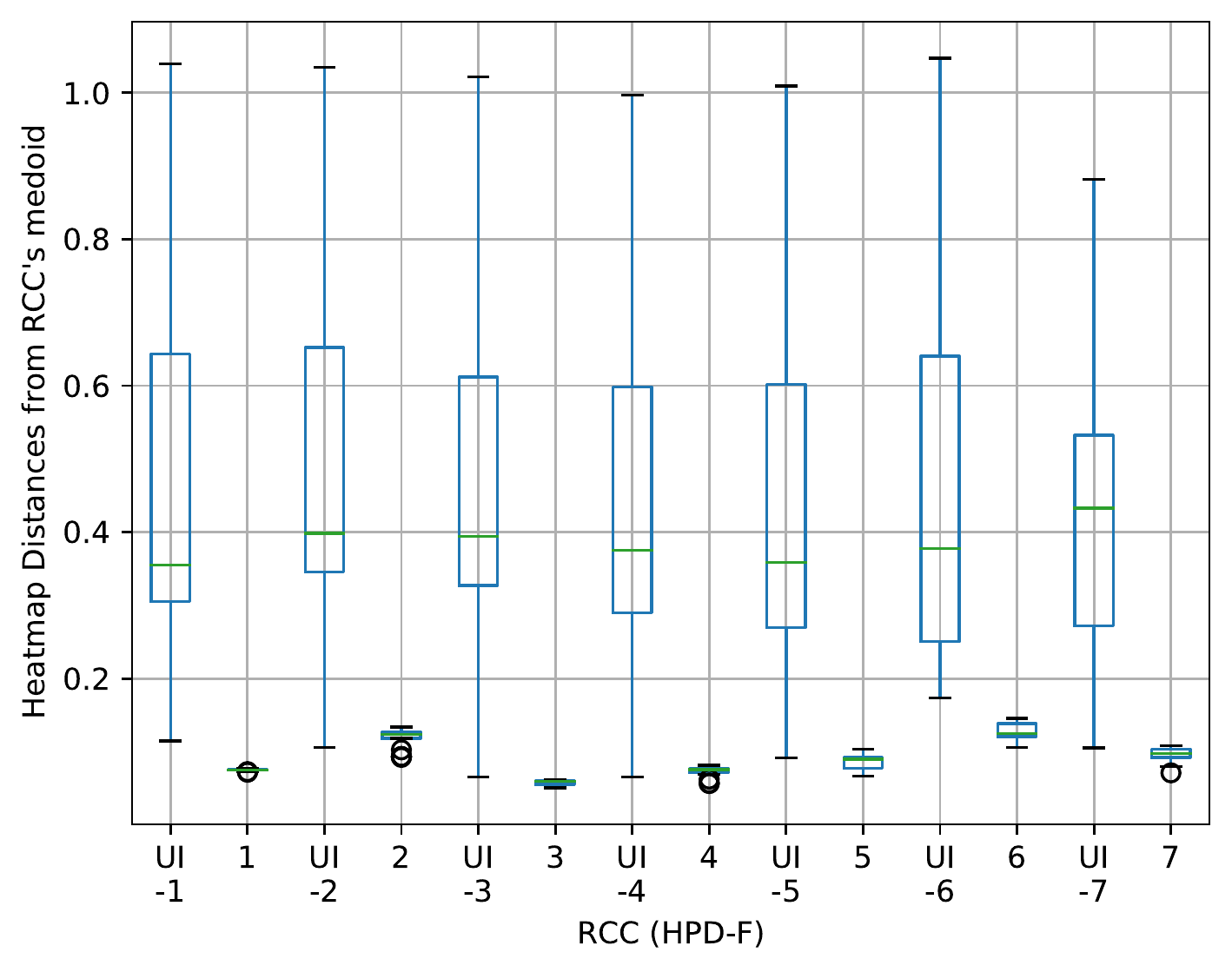}
\includegraphics[height=0.3\textheight]{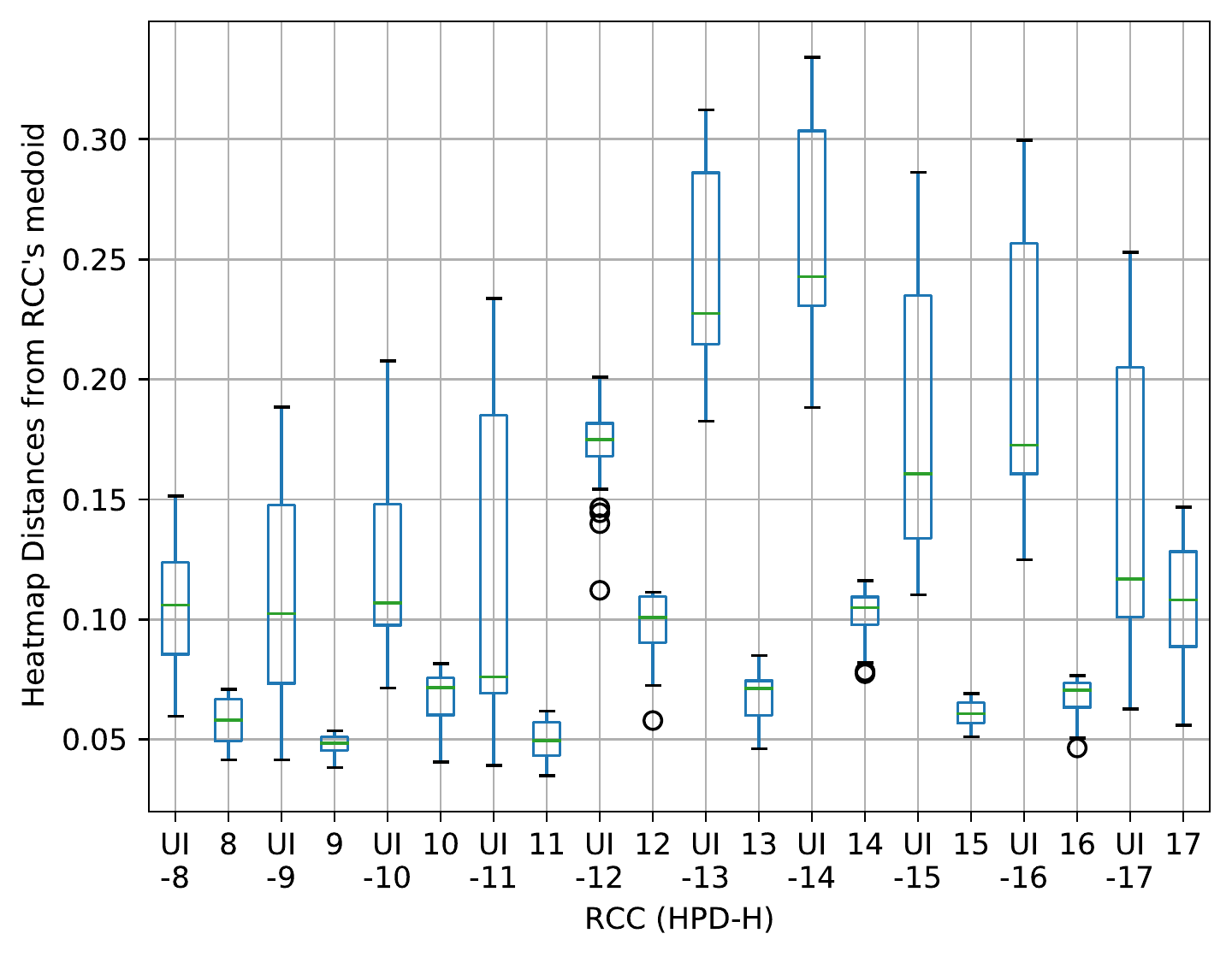}
\includegraphics[height=0.3\textheight]{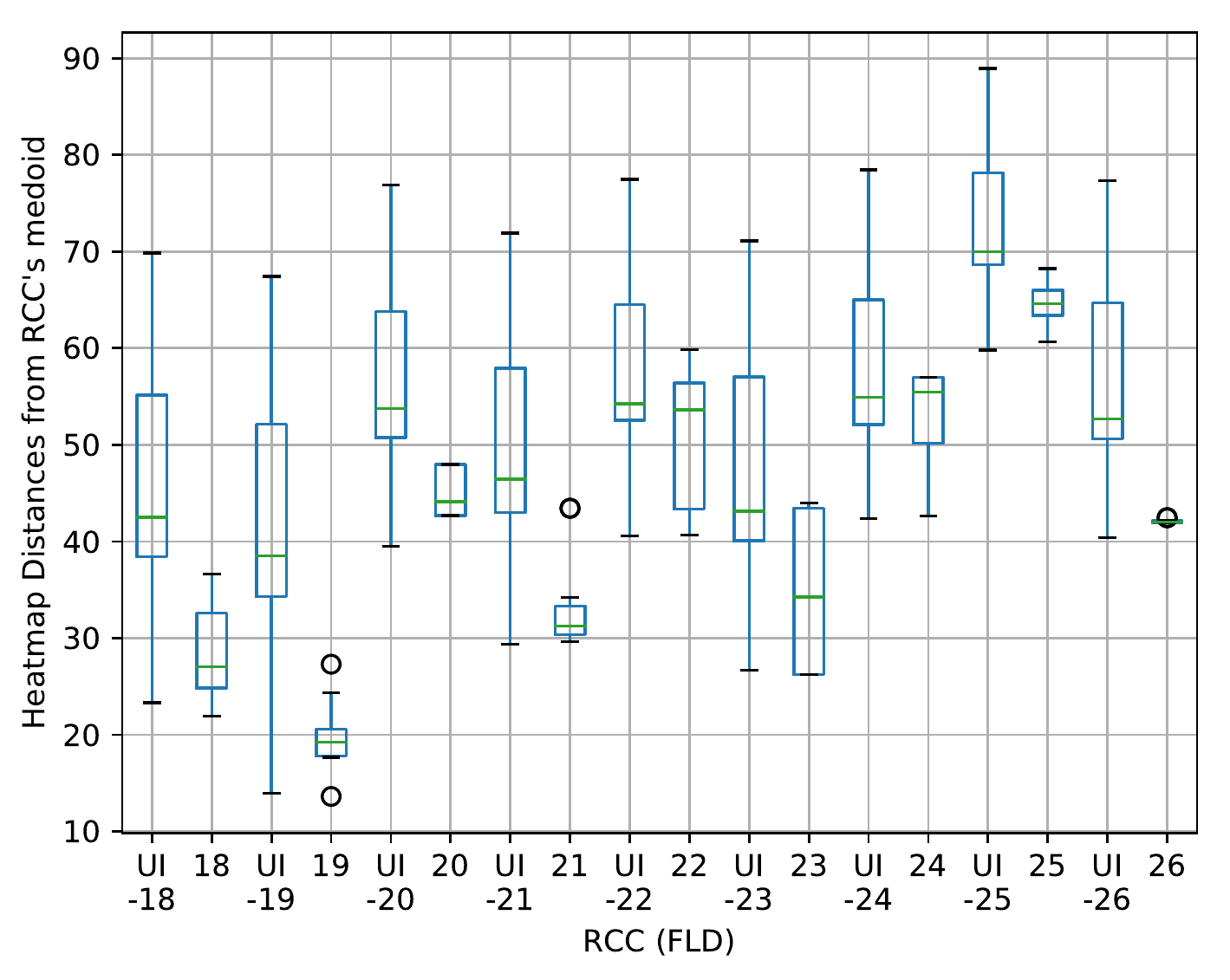}
\caption{Heatmap distances of \APPR images from the medoid of HPD-F (top), HPD-H (middle) and FLD (bottom) RCCs compared to unsafe test set images (UI)}
\label{fig:RQ2-HPD}
\end{figure}

Fig.~\ref{fig:RQ2-HPD} shows boxplots reporting, for all RCCs across DNNs, the heatmap distances from RCC medoids, for images in the unsafe test set and those generated by \APPR for every RCC in Step 2.1. 
Fig.~\ref{fig:RQ2-HPD} shows that the images generated by \APPR (i.e., boxplots with IDs $1$ to $26$) are much closer to the RCC medoid than random unsafe images (i.e., boxplots with IDs UI-1 to UI-26). Indeed, with \APPR, the median lays in the ranges [$0.056$, $0.125$] for HPD-F, [$0.048$, $0.113$] for HPD-H, and [$18.6$, $64.3$] for FLD. 
For random unsafe images, the median tends to be much higher and  
lays in the ranges [$0.38$, $0.48$] for HPD-F, [$0.08$, $0.24$] for HPD-H, and [$42.0$, $73.6$] for FLD.

Tables~\ref{tab:rq2_1} to ~\ref{tab:rq2_3} provide, for each RCC across DNNs, the average distance of \APPR images and random unsafe images, along with p-values and effect size.

On average, \APPR images are closer to the medoid of the clusters by $80.7\%$ for HPD-F, $54.6\%$ for HPD-H and $24.9\%$ for FLD. These differences with unsafe test set images are always statistically significant ($\textit{p-value} \leq 0.05$) with a large effect size.

\begin{table}[t]
\centering
\smaller
\caption{RQ2: Average distance from medoid for HPD-F}
\begin{tabular}{|c|cc|c|c|}
\hline
\begin{tabular}[c]{@{}c@{}}RCC\end{tabular} & 
\multicolumn{2}{c|}{\begin{tabular}[c]{@{}c@{}}Avg. distance from medoid\end{tabular}} & 
\multicolumn{1}{c|}{\begin{tabular}[c]{@{}c@{}}\textit{p-value} \\(U-test)\end{tabular}} & \begin{tabular}[c]{@{}c@{}}Effect Size \\ ($\hat{A}_{12}$)\end{tabular} \\
& 
\multicolumn{1}{c|}{\begin{tabular}[c]{@{}c@{}}\APPR  Images\end{tabular}} & \multicolumn{1}{c|}{\begin{tabular}[c]{@{}c@{}}Unsafe Images\end{tabular}} & 
 &
\\ \hline

1 & 
\multicolumn{1}{c|}{0.075} & \multicolumn{1}{c|}{0.469} &
\multicolumn{1}{c|}{5.11e-17} & \multicolumn{1}{c|}{1.0}
\\ \hline
2 & 
\multicolumn{1}{c|}{0.118} & \multicolumn{1}{c|}{0.480} &
\multicolumn{1}{c|}{4.52e-16} & \multicolumn{1}{c|}{1.0}
\\ \hline
3 & 
\multicolumn{1}{c|}{0.058} & \multicolumn{1}{c|}{0.476} &
\multicolumn{1}{c|}{7.17e-17} & \multicolumn{1}{c|}{1.0}
\\ \hline
4 & 
\multicolumn{1}{c|}{0.072} & \multicolumn{1}{c|}{0.459} &
\multicolumn{1}{c|}{2.12e-16} & \multicolumn{1}{c|}{1.0}
\\ \hline
5 & 
\multicolumn{1}{c|}{0.085} & \multicolumn{1}{c|}{0.457} &
\multicolumn{1}{c|}{6.45e-16} & \multicolumn{1}{c|}{1.0}
\\ \hline
6 & 
\multicolumn{1}{c|}{0.128} & \multicolumn{1}{c|}{0.470} &
\multicolumn{1}{c|}{1.48e-16} & \multicolumn{1}{c|}{1.0}
\\ \hline
7 & 
\multicolumn{1}{c|}{0.096} & \multicolumn{1}{c|}{0.439} &
\multicolumn{1}{c|}{5.43e-17} & \multicolumn{1}{c|}{1.0}
\\ \hline
\end{tabular}
\label{tab:rq2_1}
\end{table}
\begin{table}[t]
\centering
\smaller
\caption{RQ2: Average distance from medoid for HPD-H}
\begin{tabular}{|c|cc|c|c|}
\hline
\begin{tabular}[c]{@{}c@{}}RCC\end{tabular} & 
\multicolumn{2}{c|}{\begin{tabular}[c]{@{}c@{}}Avg. distance from medoid\end{tabular}} & 
\multicolumn{1}{c|}{\begin{tabular}[c]{@{}c@{}}\textit{p-value} \\(U-test)\end{tabular}} & \begin{tabular}[c]{@{}c@{}}Effect Size \\ ($\hat{A}_{12}$)\end{tabular} \\
& 
\multicolumn{1}{c|}{\begin{tabular}[c]{@{}c@{}}\APPR  Images\end{tabular}} & \multicolumn{1}{c|}{\begin{tabular}[c]{@{}c@{}}Unsafe Images\end{tabular}} & 
 &
\\ \hline

8 & 
\multicolumn{1}{c|}{0.057} & \multicolumn{1}{c|}{0.135} &
\multicolumn{1}{c|}{1.52e-14} & \multicolumn{1}{c|}{1.0}
\\ \hline
9 & 
\multicolumn{1}{c|}{0.047} & \multicolumn{1}{c|}{0.120} &
\multicolumn{1}{c|}{3.47e-14} & \multicolumn{1}{c|}{1.0}
\\ \hline
10 & 
\multicolumn{1}{c|}{0.067} & \multicolumn{1}{c|}{0.134} &
\multicolumn{1}{c|}{1.43e-08} & \multicolumn{1}{c|}{0.98}
\\ \hline
11 & 
\multicolumn{1}{c|}{0.050} & \multicolumn{1}{c|}{0.110} &
\multicolumn{1}{c|}{5.61e-12} & \multicolumn{1}{c|}{1.0}
\\ \hline
12 & 
\multicolumn{1}{c|}{0.098} & \multicolumn{1}{c|}{0.183} &
\multicolumn{1}{c|}{4.81e-17} & \multicolumn{1}{c|}{1.0}
\\ \hline
13 & 
\multicolumn{1}{c|}{0.068} & \multicolumn{1}{c|}{0.208} &
\multicolumn{1}{c|}{3.33e-17} & \multicolumn{1}{c|}{1.0}
\\ \hline
14 & 
\multicolumn{1}{c|}{0.101} & \multicolumn{1}{c|}{0.214} &
\multicolumn{1}{c|}{1.13e-13} & \multicolumn{1}{c|}{1.0}
\\ \hline
15 & 
\multicolumn{1}{c|}{0.060} & \multicolumn{1}{c|}{0.156} &
\multicolumn{1}{c|}{2.28e-17} & \multicolumn{1}{c|}{1.0}
\\ \hline
16 & 
\multicolumn{1}{c|}{0.067} & \multicolumn{1}{c|}{0.159} &
\multicolumn{1}{c|}{7.11e-17} & \multicolumn{1}{c|}{1.0}
\\ \hline
17 & 
\multicolumn{1}{c|}{0.107} & \multicolumn{1}{c|}{0.126} &
\multicolumn{1}{c|}{5.43e-03} & \multicolumn{1}{c|}{0.91}
\\ \hline
\end{tabular}
\label{tab:rq2_2}
\end{table}
\begin{table}[t]
\centering
\smaller
\caption{RQ2: Average distance from medoid for FLD}
\begin{tabular}{|c|cc|c|c|}
\hline
\begin{tabular}[c]{@{}c@{}}RCC\end{tabular} & 
\multicolumn{2}{c|}{\begin{tabular}[c]{@{}c@{}}Avg. distance from medoid\end{tabular}} & 
\multicolumn{1}{c|}{\begin{tabular}[c]{@{}c@{}}\textit{p-value} \\(U-test)\end{tabular}} & \begin{tabular}[c]{@{}c@{}}Effect Size \\ ($\hat{A}_{12}$)\end{tabular} \\
& 
\multicolumn{1}{c|}{\begin{tabular}[c]{@{}c@{}}\APPR  Images\end{tabular}} & \multicolumn{1}{c|}{\begin{tabular}[c]{@{}c@{}}Unsafe Images\end{tabular}} & 
 &
\\ \hline

18 & 
\multicolumn{1}{c|}{25.66} & \multicolumn{1}{c|}{46.28} &
\multicolumn{1}{c|}{1.79e-08} & \multicolumn{1}{c|}{0.985}
\\ \hline
19 & 
\multicolumn{1}{c|}{20.03} & \multicolumn{1}{c|}{41.90} &
\multicolumn{1}{c|}{2.87e-10} & \multicolumn{1}{c|}{1.0}
\\ \hline
20 & 
\multicolumn{1}{c|}{45.17} & \multicolumn{1}{c|}{56.84} &
\multicolumn{1}{c|}{4.12e-03} & \multicolumn{1}{c|}{0.910}
\\ \hline
21 & 
\multicolumn{1}{c|}{32.95} & \multicolumn{1}{c|}{49.99} &
\multicolumn{1}{c|}{1.22e-05} & \multicolumn{1}{c|}{0.965}
\\ \hline
22 & 
\multicolumn{1}{c|}{49.99} & \multicolumn{1}{c|}{57.64} &
\multicolumn{1}{c|}{0.037} & \multicolumn{1}{c|}{0.845}
\\ \hline
23 & 
\multicolumn{1}{c|}{35.63} & \multicolumn{1}{c|}{48.03} &
\multicolumn{1}{c|}{3.84e-03} & \multicolumn{1}{c|}{0.905}
\\ \hline
24 & 
\multicolumn{1}{c|}{51.10} & \multicolumn{1}{c|}{58.63} &
\multicolumn{1}{c|}{0.048} & \multicolumn{1}{c|}{0.645}
\\ \hline
25 & 
\multicolumn{1}{c|}{64.45} & \multicolumn{1}{c|}{72.75} &
\multicolumn{1}{c|}{5.12e-04} & \multicolumn{1}{c|}{0.940}
\\ \hline
26 & 
\multicolumn{1}{c|}{42.11} & \multicolumn{1}{c|}{53.30} &
\multicolumn{1}{c|}{4.32e-04} & \multicolumn{1}{c|}{0.945}
\\ \hline
\end{tabular}
\label{tab:rq2_3}
\end{table}


\subsection{RQ3. \emph{Does \APPR generate, for each RCC, a set of images sharing similar characteristics?}}
\label{sec:empirical:rq3}

\subsubsection{Experiment Design.} This research question assesses if the images generated by \APPR for each RCC, present similar characteristics (i.e., similar values for a subset of the simulator parameters). If this is true, for each RCC, we should observe a subset of parameters with a variance that is significantly lower than the variance observed in randomly generated images.
For each RCC, instead of comparing the variance of each parameter $p$, which depends on the parameter range, we can compare the variance reduction rate ($\mathit{VRR}$), which can be computed as the ratio of the variance for a parameter $p$ for a given RCC $C_{i}$ over that of a set of randomly generated images:

\begin{equation*}
\mathit{VRR}_{C_{i}}(p) = 1 - \frac { \mathit{variance}\ \mathit{of} p\ \mathit{for}\ \mathit{the}\ \mathit{images}\ \mathit{in}\ C_{i} } {\mathit{variance}\  \mathit{of}\  p\  \mathit{for}\ \mathit{a}\ \mathit{set}\ \mathit{of} \mathit{random}\ \mathit{images}}
\end{equation*}

For a parameter, a positive $\mathit{VRR}$ indicates that its values are likely to be constrained to a smaller range than the one of the input domain. A $0.5$ $\mathit{VRR}$ means that the variance is reduced by $50\%$, which is considered to be a high reduction rate~\cite{HUDD:TRel}.  

For each RCC, we compute $\mathit{VRR}$ for each parameter. We positively answer our research question if, for a large number of RCCs, a subset of parameters presents a large $\mathit{VRR}$ ($> 0.5$).


\subsubsection{Results.} 
\begin{figure*}[htp]
\includegraphics[height=0.33\textheight]{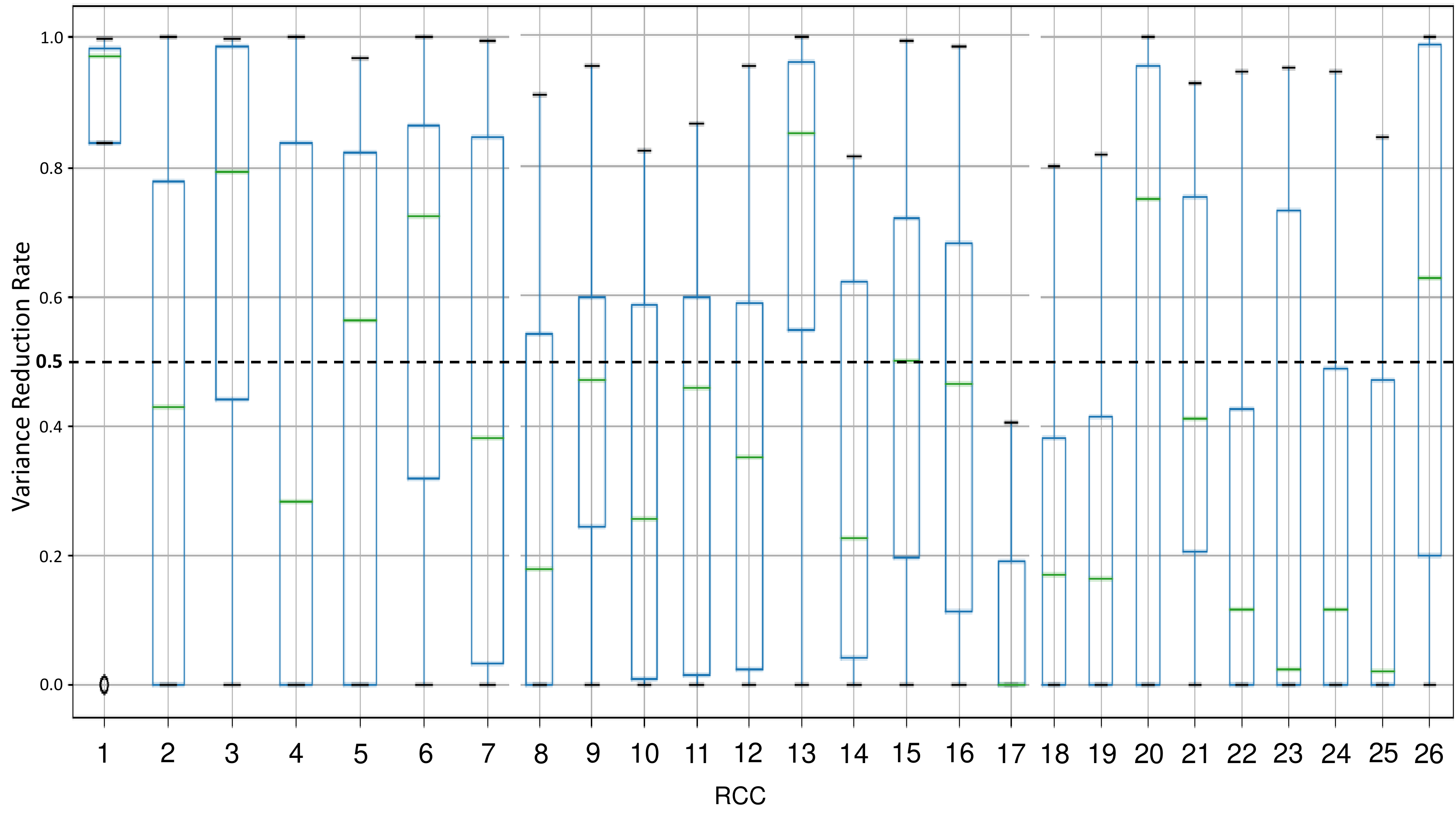}
\caption{Variance reduction for all simulator parameters associated with generated images in the root-cause clusters of HPD-F (boxplots $1$ to $7$), HPD-H (boxplots $8$ to $17$), and FLD (boxplots $18$ to $26$) DNNs}
\label{fig:RQ3}
\end{figure*}

Fig.~\ref{fig:RQ3} provides boxplots capturing the distribution of the variance reduction for each of the 26 RCCs. Each data point in a boxplot captures the variance reduction of one parameter. 

Fig.~\ref{fig:RQ3} shows that for all the RCCs except boxplot $17$, the top whisker is above $0.8$; since the top whisker reports the max value (excluding outliers\footnote{In our boxplots, $\mathit{upper whisker} = min(max(x), Q_3 + 1.5 * IQR)$, $\mathit{lower whisker} = max(min(x), Q_1 - 1.5 * IQR))$, with $IQR$ being the Inter Quartile Range and $Q_{x}$ the x-th quantile.}) observed for a parameter, such result indicates that for all RCCs except one, we can observe at least one parameter with a high variance reduction, thus enabling us to positively answer our research question. 
Also, for $20$ out of $26$ clusters, the third quartile is above $0.5$. Since it indicates the lowest value for the $25\%$ data points having the highest variance reduction, it means that in $20$ RCCs, $25\%$ of the parameters present a variance reduction rate above $0.5$. Therefore, for a large proportion of RCCs ($77\%$), more than one parameter present a large variance reduction, which indicates that the hazard-triggering event is captured by several parameters (i.e., a specific combination of parameter values is needed to make the DNN fail). 


\subsection{RQ4. \emph{Do the RCC expressions identified by \APPR delimit an unsafe space?}}
\label{sec:empirical:rq4}

\subsubsection{Experiment Design.} For each RCC, \APPR provides a set of expressions representing conjunctions of parameter ranges that characterize the unsafe images in the RCC.
This research question evaluates if these expressions actually characterize an unsafe space, which implies that images whose parameters match such an expression are more likely to trigger a DNN failure than those that do not. 

To address this research question, for each RCC, we generated $500$ images matching the RCC expression and computed the DNN accuracy obtained with these images.
To generate each image, for each simulator parameter, we selected a random value in the range provided in the expression.
Also, we considered a random input set consisting of $500$ images selected from the randomly generated simulator images used to test the fine-tuned DNN (see column Simulator-based Test Set in Table~\ref{tab:dnns}).

We can positively answer our research question if, for a large subset of RCCs, the images generated from RCC expressions lead to a significantly lower DNN accuracy\MAJOR{R3.21}{\footnote{Recall that accuracy is $100\%$ minus the percentage of inputs leading to DNN failures (i.e., if $90\%$ of the inputs generated by \APPR is failure-inducing, we will observe an accuracy of 10\%).}} than randomly generated images, for each RCC. Since, for each RCC, we compare two image groups (i.e., $500$ random images and $500$ images generated by \APPR) labeled with a categorical variable (i.e., indicating if the DNN produces a correct output), we rely on the Fisher's exact test to assess differences in proportions of images leading to DNN failures. 

\subsubsection{Results} 
\begin{figure}[htp]
\includegraphics[width=0.497\textwidth]{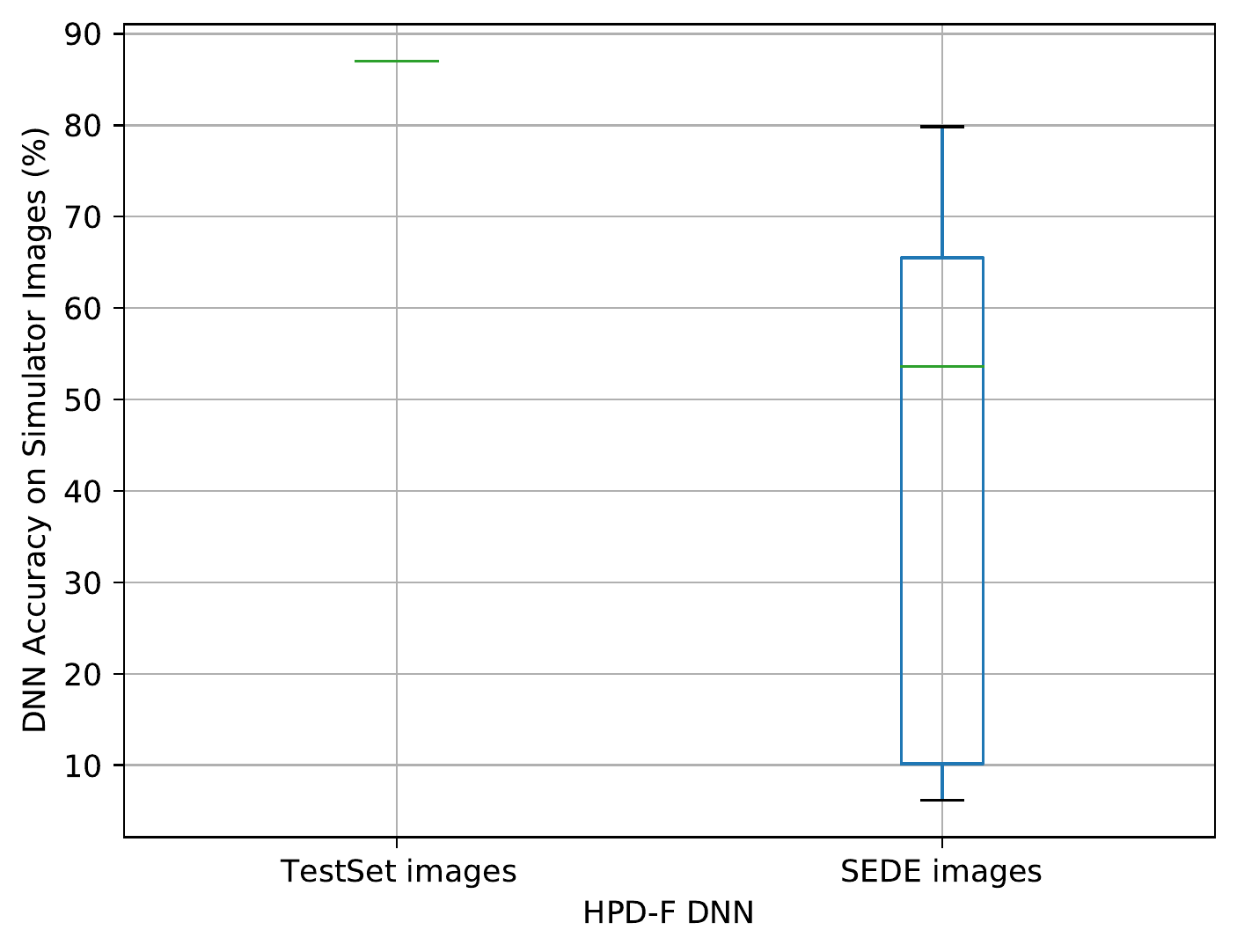}
\includegraphics[width=0.497\textwidth]{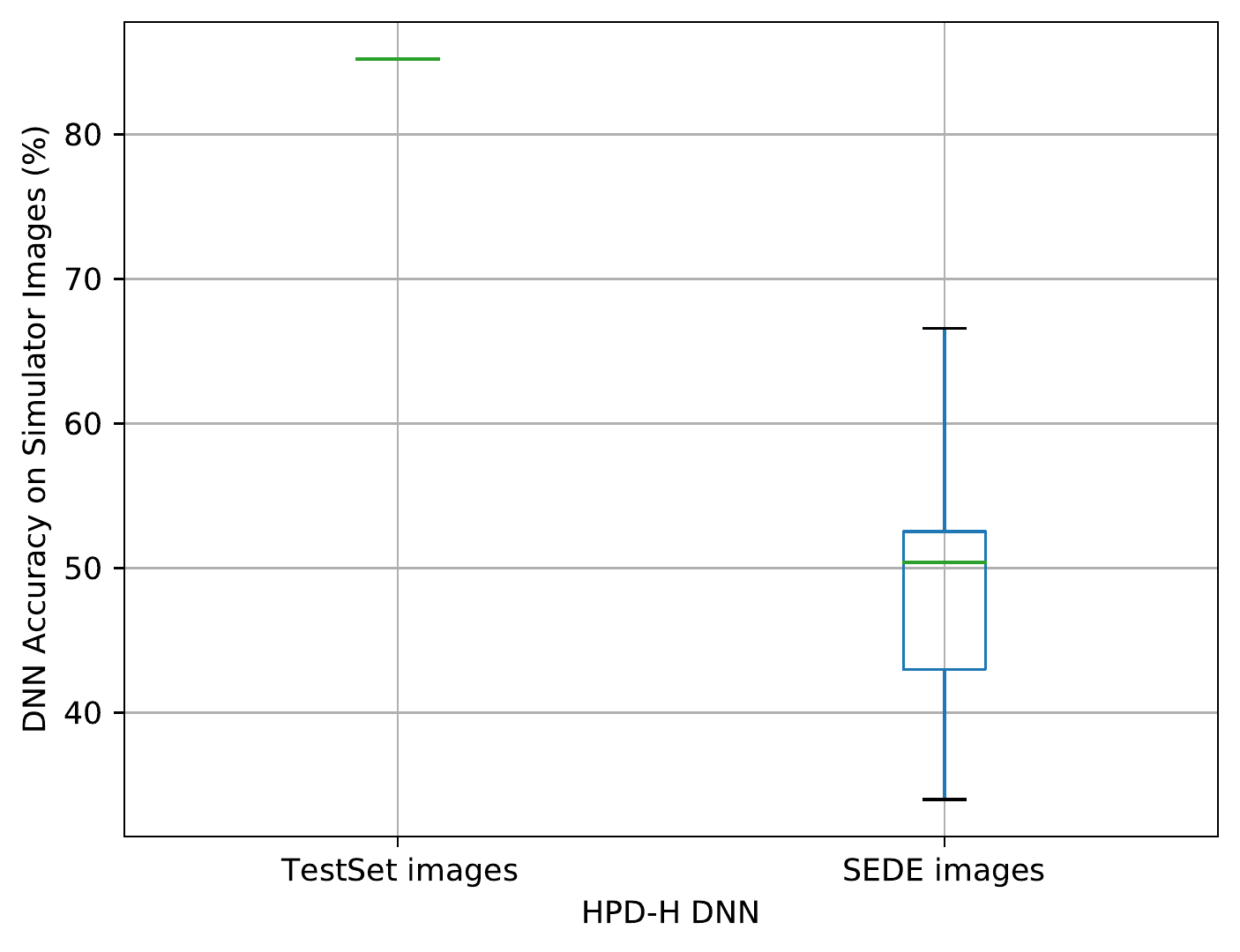}
\includegraphics[width=0.497\textwidth]{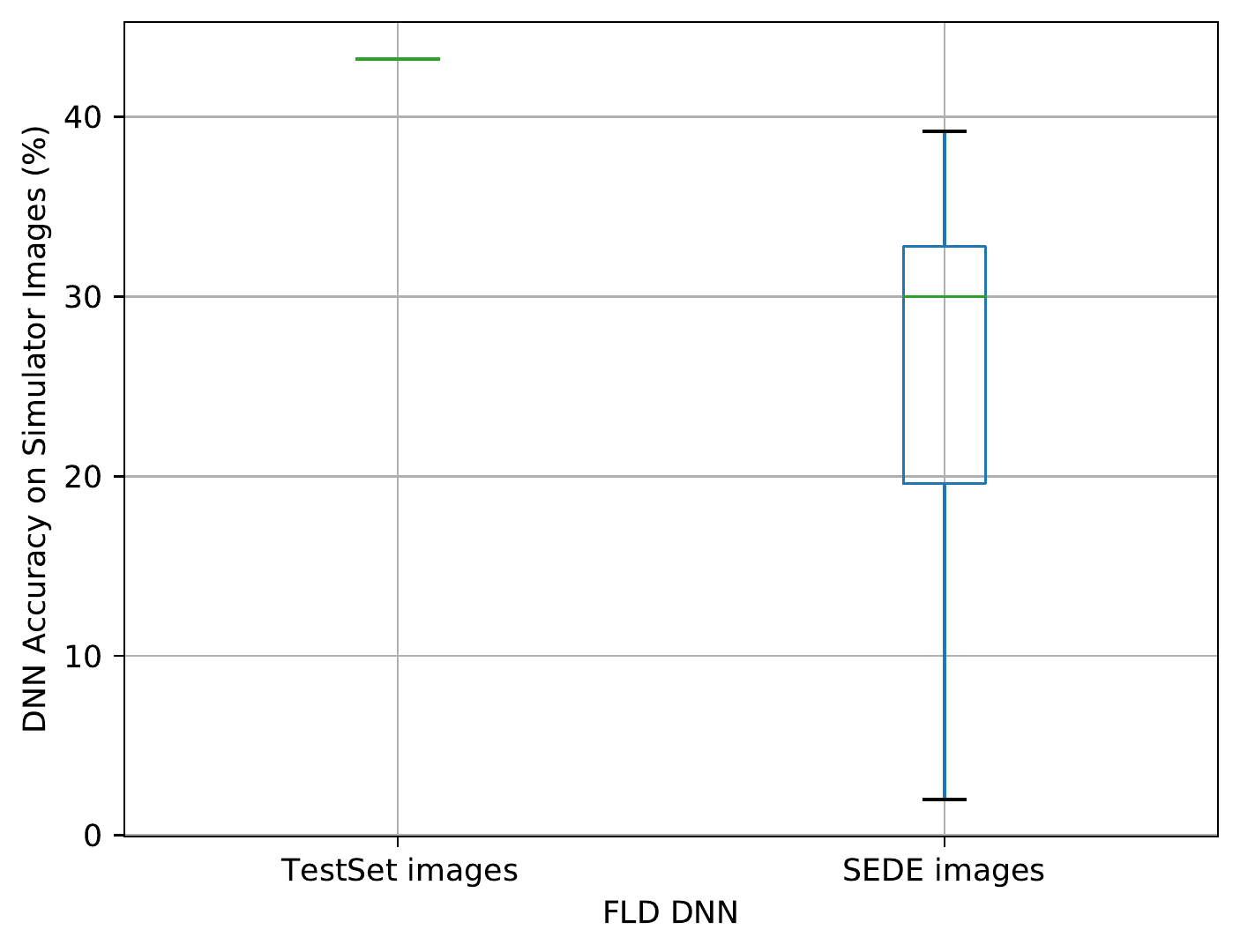}
\caption{Percentage of correctly classified images observed on \APPR images compared to the random test set images for HPD-F, HPD-H, and FLD DNNs}
\label{fig:RQ4}
\end{figure}

Fig.~\ref{fig:RQ4} provides boxplots capturing the accuracy obtained with the random test set images  and those generated according to RCC expressions. Each data point in the \emph{\APPR images} boxplots corresponds to the DNN accuracy of one of the 26 RCC expressions. 
The accuracy observed with the random test sets for HPD-F, HPD-H, and FLD DNNs is $87.0\%$, $85.2\%$, and $43.2\%$, respectively. 
In contrast, the images generated according to RCC expressions lead to much lower accuracy in the ranges [$6.2\%$, $79.8\%$], [$34.0\%$, $66.6\%$], and [$2.0\%$, $39.2\%$], for HPD-F, HPD-H and FLD, respectively. 
Moreover, for $25$ out of $26$ clusters, \APPR generates images leading to an accuracy that, based on a Fisher's exact test (see Tables~\ref{tab:rq4_1}, \ref{tab:rq4_2}, and \ref{tab:rq4_3}), significantly differs from the accuracy obtained with random images. Since in $25$ out of $26$ RCCs ($96.15\%$) the RCC expressions generated by \APPR clearly delimit an unsafe space, we can positively answer RQ4.



\begin{table}[t]
\centering
\smaller
\caption{RQ4: HPD-F accuracy within the unsafe space identified by \APPR compared to the whole input space}
\begin{tabular}{|c|cc|c|}
\hline
\begin{tabular}[c]{@{}c@{}}RCC\end{tabular} & 
\multicolumn{2}{c|}{DNN Accuracy} & 
\multicolumn{1}{c|}{\begin{tabular}[c]{@{}c@{}}\textit{p-value} \\(Fisher's Exact)\end{tabular}} \\ 
& 
\multicolumn{1}{c|}{\begin{tabular}[c]{@{}c@{}}SEDE Unsafe Set\end{tabular}} & \multicolumn{1}{c|}{\begin{tabular}[c]{@{}c@{}}Random Input Set\end{tabular}} & 
\\ \hline
1 & 
\multicolumn{1}{c|}{10.2\%} & \multirow{7}{*}{87.0\%} &
\multicolumn{1}{c|}{7.40e-73}
\\ \cline{0-0}
2 & 
\multicolumn{1}{c|}{67.4\%} & &
\multicolumn{1}{c|}{9.25e-04}
\\ \cline{0-0}
3 & 
\multicolumn{1}{c|}{6.2\%} & &
\multicolumn{1}{c|}{9.11e-88}
\\ \cline{0-0}
4 & 
\multicolumn{1}{c|}{79.8\%} & &
\multicolumn{1}{c|}{0.047}
\\ \cline{0-0}
5 & 
\multicolumn{1}{c|}{73.0\%} & &
\multicolumn{1}{c|}{0.019}
\\ \cline{0-0}
6 & 
\multicolumn{1}{c|}{59.8\%} & &
\multicolumn{1}{c|}{2.28e-06}
\\ \cline{0-0}
7 & 
\multicolumn{1}{c|}{53.6\%} & &
\multicolumn{1}{c|}{2.43e-09}
\\ \hline
\end{tabular}
\label{tab:rq4_1}
\end{table}
\begin{table}[t]
\smaller
\centering
\caption{RQ4: HPD-H accuracy within the unsafe space identified by \APPR compared to the whole input space}
\begin{tabular}{|c|cc|c|}
\hline
\begin{tabular}[c]{@{}c@{}}RCC\end{tabular} & 
\multicolumn{2}{c|}{DNN Accuracy} & 
\multicolumn{1}{c|}{\begin{tabular}[c]{@{}c@{}}\textit{p-value} \\(Fisher's Exact)\end{tabular}} \\ 
& 
\multicolumn{1}{c|}{\begin{tabular}[c]{@{}c@{}}SEDE Unsafe Set\end{tabular}} & \multicolumn{1}{c|}{\begin{tabular}[c]{@{}c@{}}Random Input Set\end{tabular}} & 
\\ \hline
8 & 
\multicolumn{1}{c|}{39.4\%} & \multirow{10}{*}{85.2\%} &
\multicolumn{1}{c|}{1.26e-19}
\\ \cline{0-0}
9 & 
\multicolumn{1}{c|}{50.6\%} & &
\multicolumn{1}{c|}{1.19e-10}
\\ \cline{0-0}
10 & 
\multicolumn{1}{c|}{34.0\%} & &
\multicolumn{1}{c|}{4.84e-25}
\\ \cline{0-0}
11 & 
 \multicolumn{1}{c|}{49.0\%} & &
\multicolumn{1}{c|}{1.31e-11}
\\ \cline{0-0}
12 & 
 \multicolumn{1}{c|}{41.0\%} & &
\multicolumn{1}{c|}{7.47e-18}
\\ \cline{0-0}
13 & 
\multicolumn{1}{c|}{51.8\%} & &
\multicolumn{1}{c|}{5.66e-10}
\\ \cline{0-0}
14 & 
 \multicolumn{1}{c|}{66.6\%} & &
\multicolumn{1}{c|}{1.73e-03}
\\ \cline{0-0}
15 & 
 \multicolumn{1}{c|}{52.8\%} & &
\multicolumn{1}{c|}{2.19e-09}
\\ \cline{0-0}
16 & 
\multicolumn{1}{c|}{56.6\%} & &
\multicolumn{1}{c|}{2.32e-07}
\\ \cline{0-0}
17 & 
\multicolumn{1}{c|}{50.2\%} & &
\multicolumn{1}{c|}{7.14e-10}
\\ \hline
\end{tabular}
\label{tab:rq4_2}
\end{table}
\begin{table}[t]
\centering
\smaller
\caption{RQ4: FLD accuracy within the unsafe space identified by \APPR compared to the whole input space}
\begin{tabular}{|c|cc|c|}
\hline
\begin{tabular}[c]{@{}c@{}}RCC\end{tabular} & 
\multicolumn{2}{c|}{DNN Accuracy} & 
\multicolumn{1}{c|}{\begin{tabular}[c]{@{}c@{}}\textit{p-value} \\(Fisher's Exact)\end{tabular}} \\ 
& 
\multicolumn{1}{c|}{\begin{tabular}[c]{@{}c@{}}SEDE Unsafe Set\end{tabular}} & \multicolumn{1}{c|}{\begin{tabular}[c]{@{}c@{}}Random Input Set\end{tabular}} & 
\\ \hline
18 & 
\multicolumn{1}{c|}{39.2\%} & \multirow{9}{*}{43.2\%} &

\multicolumn{1}{c|}{0.30}
\\ \cline{0-0}
19 & 
\multicolumn{1}{c|}{19.6\%} &  &
\multicolumn{1}{c|}{4.77e-13}
\\ \cline{0-0}
20 & 
\multicolumn{1}{c|}{30.0\%} &  &
\multicolumn{1}{c|}{9.19e-88}
\\ \cline{0-0}
21 & 
\multicolumn{1}{c|}{32.8\%} &  &
\multicolumn{1}{c|}{4.88e-03}
\\ \cline{0-0}
22 & 
\multicolumn{1}{c|}{19.6\%} &  &
\multicolumn{1}{c|}{4.39e-13}
\\ \cline{0-0}
23 & 
\multicolumn{1}{c|}{32.8\%} & &
\multicolumn{1}{c|}{4.12e-03}
\\ \cline{0-0}
24 & 
\multicolumn{1}{c|}{27.0\%} & &
\multicolumn{1}{c|}{2.82e-06}
\\ \cline{0-0}
25 & 
\multicolumn{1}{c|}{2.0\%} &   &
\multicolumn{1}{c|}{2.49e-58}
\\ \cline{0-0}
26 & 
\multicolumn{1}{c|}{31.0\%} & &
\multicolumn{1}{c|}{6.31e-04}
\\ \hline
\end{tabular}
\label{tab:rq4_3}
\end{table}

\subsection{RQ5. \emph{How does \APPR compare to state-of-the-art DNN accuracy improvement practices?}} 
\label{sec:empirical:rq5}

\subsubsection{Experiment Design.} This research question aims to determine if \APPR performs better than state-of-the-art solutions. We compare \APPR with \HUDD since it is the only approach that works with real-world images and addresses both root cause explanations and retraining (Section~\ref{sec:related}). \MAJOR{R3.22}{Further, other retraining approaches~\cite{Feldt19,Gao:Sensei:ICSE:20,Engstrom:2019,DeepFault,RobOT,Xie2019} generate retraining images through pixel value transformation (e.g., image contrast, image brightness, image blur, and image noise), affine transformation (e.g., image translation, image scaling, image shearing, and image rotation), or adversarial modification (e.g., FGSM~\cite{Goodfellow2015}, PGD~\cite{Madry2018}, C\&W~\cite{Carlini2017}). None of these approaches, however, aims to generate images similar to real-world, failure-inducing ones targeted by \APPR.} In addition, DNNs might be inaccurate because either trained for a limited number of epochs or with a limited number of inputs; in such cases, retraining the DNN for additional epochs with a larger set of randomly selected inputs would suffice. For this reason, we consider a baseline approach which consists of retraining the DNN with an additional set of randomly generated images.


Regarding \APPR, we followed the procedure described in Section~\ref{sec:approach:retrain}. The third column of Table~\ref{tab:rq5} provides the size of the training set considered to retrain each DNN; precisely, we report the total number of simulator images generated according to \APPR expressions (\emph{Unsafe Set}), the number of simulator images retained from the simulator training set (\emph{Training Set Sim.}) and the number of real-world images retained from the set used for fine-tuning (\emph{Training Set Real}). For each RCC, we have generated, with \APPR, $50$ images to be used for retraining.
As anticipated in Section~\ref{sec:approach:retrain}, we retained $5\%$ of the images in the original simulator-based training set, and all the real-world images used to fine-tune the DNN.

In the case of \HUDD, we applied its selection algorithm to sample, from an improvement set, $50$ images for each RCC (\HUDD selects the images that are closer to the RCC centroid). 
We generated an improvement set with random face images (i.e., generated by randomly selecting simulator parameter values). To avoid bias, we selected for each case study DNN the same number of images as that generated by \APPR (i.e., $50$ images for each RCC), which led to a total of $450$ images for FLD, $350$ for HPD-F, and $500$ for HPD-H.


Concerning the random baseline, for each case study DNN, we generated a retraining set consisting of a number of randomly generated simulator images and real-world images. To avoid unfair comparisons, the randomly generated images match in number the images selected by \APPR.


Further, for all the three approaches described above (i.e., \APPR, \HUDD, and random baseline), we followed the \APPR retraining process, which consists of retraining the DNN for multiple times and selecting the DNN yielding the best test set accuracy. 
In our experiments, since each retraining task takes approximately three hours, we performed three retraining tasks since they can be performed overnight, which is acceptable in practice.
For all the approaches, we constructed the retraining data set in the same way as  \APPR. It consists of the images selected by the approach under analysis, a random selection of images from the original simulator-based training set ($5\%$ of the whole set), and all the real-world images used to fine-tune the DNN. To account for randomness, for each approach, we repeated the experiment ten times (i.e., for ten times, we selected the best DNN generated out of three runs). For each approach, we generate ten DNNs. 



To positively answer our research questions, \APPR should lead to retrained DNNs with an accuracy that is significantly higher than the one observed when retraining the DNN using either \HUDD or the random baseline. 

\subsubsection{Results.} 

Columns five to seven (i.e., RBL, HUDD, and \APPR) in Table ~\ref{tab:rq5} show the average accuracy of the three approaches across ten runs along with the delta with respect to the original DNN model. 
Also, for \APPR, we report the delta with respect to the best competing approach (i.e., HUDD or random baseline). Finally, we report p-values using a non-parametric Mann–Whitney U-test, for statistical significance, and the $\hat{A}_{12}$ statistics, for effect size. 

\APPR, on average, improves the DNN for FLD, HPD-F and HPD-H by $+6.07\%$, $+4.50\%$ and $+18.65\%$; the other two approaches, instead, do not improve the HPD-F and FLD DNNs but decrease their performance. For HPD-H, HUDD and the random baseline led to an improvement that is lower than \APPR's, $+9.62\%$ and $+4.54\%$, respectively.
We believe that in the case of HPD-F and FLD,
the retraining based on HUDD and random decrease the DNN performance because the simulator being used generates images that are less realistic; such characteristic leads to the generation of images that are unlikely to include hazard-triggering events and, in turn, the retraining set selected by HUDD and random is not likely to include unsafe images. Since the retraining is performed using (1) the images selected by the approach under analysis, (2) a subset of the original training set, and (3) the real-world training set, safe cases will be over-represented in the training set. In general, unsafe images that are present in the original training set may not be retained for retraining.


Similar to the above, we believe that, since it is complicated for IEE-Faces to generate unsafe images, in the case of HPD-F and FLD, \APPR led to a more limited improvement in accuracy than in the case of HPD-H:  $+4.50\%$ and $+6.07\%$ vs $+18.65\%$.

Finally, \APPR, on average, improves the DNN results by $9$ percentage points over the best competing approach (i.e., $+6.19\%$ for FLD, $+10.35\%$ for HPD-F, and $+9.03\%$ for HPD-H). The difference is always statistically significant ($\textit{p-value} \leq 0.05$) with a large effect size.

\begin{table*}[t]
\centering
\smaller
\footnotesize
\caption{RQ5: Unsafe set size and the accuracy improvement of \APPR compared to HUDD and random baseline (RBL)}
\makebox[0.9\textwidth]{\begin{tabular}{|C{0.9cm}|C{1.2cm}|cc|C{1cm}|C{1cm}|cc|cc|}
\hline
DNN & Original Model & 
\multicolumn{2}{c|}{Retraining Set Size} & 
RBL  & HUDD  & 
\multicolumn{2}{c|}{\APPR} &
\multicolumn{2}{c|}{Stat. Sig.} 
\\ 
& Accuracy  & 
\begin{tabular}[c]{@{}c@{}}Training Set\\ (Sim. + Real)\end{tabular} & \begin{tabular}[c]{@{}c@{}}Unsafe\\ Set\end{tabular} & 
\begin{tabular}[c]{@{}c@{}}Accuracy \\(Gain)\end{tabular} & \begin{tabular}[c]{@{}c@{}}Accuracy \\(Gain)\end{tabular} & 
\begin{tabular}[c]{@{}c@{}}Accuracy \\(Gain)\end{tabular} & \begin{tabular}[c]{@{}c@{}}Gain over\\ best baseline\end{tabular} &
\textit{p-value} & $\hat{A}_{12}$
\\ \hline

FLD & 80.06\% & 
\multicolumn{1}{c|}{150 + 6,000} & \multicolumn{1}{c|}{450} &
77.41\% (\textbf{-2.65\%}) & 79.94\% (\textbf{-0.11\%}) & 
\multicolumn{1}{c|}{\begin{tabular}[c]{@{}c@{}}86.14\% \\(\textbf{+6.07\%})\end{tabular}} & \multicolumn{1}{c|}{\textbf{+6.19\%}}
& 1.84e-04 & 1.0
\\ \hline

HPD-F & 51.65\% & 
\multicolumn{1}{c|}{1,075 + 476} & \multicolumn{1}{c|}{350} &
44.33\% (\textbf{-7.32\%}) & 45.80\% (\textbf{-5.85\%}) & 
\multicolumn{1}{c|}{\begin{tabular}[c]{@{}c@{}}56.15\% \\(\textbf{+4.50\%})\end{tabular}} &  \multicolumn{1}{c|}{\textbf{+10.35\%}} 
& 4.43e-04 & 0.94
\\ \hline

HPD-H & 51.03\% & 
\multicolumn{1}{c|}{900 + 476} & \multicolumn{1}{c|}{500} &
55.57\% (\textbf{+4.54\%})& 60.65\% (\textbf{+9.62\%}) &
\multicolumn{1}{c|}{\begin{tabular}[c]{@{}c@{}}69.68\% \\(\textbf{+18.65\%})\end{tabular}} &  \multicolumn{1}{c|}{\textbf{+9.03\%}} 
& 1.12e-04 & 1.0
\\ \hline

\end{tabular}}
\label{tab:rq5}
\end{table*}

\begin{figure*}[htp]
\includegraphics[width=0.497\textwidth]{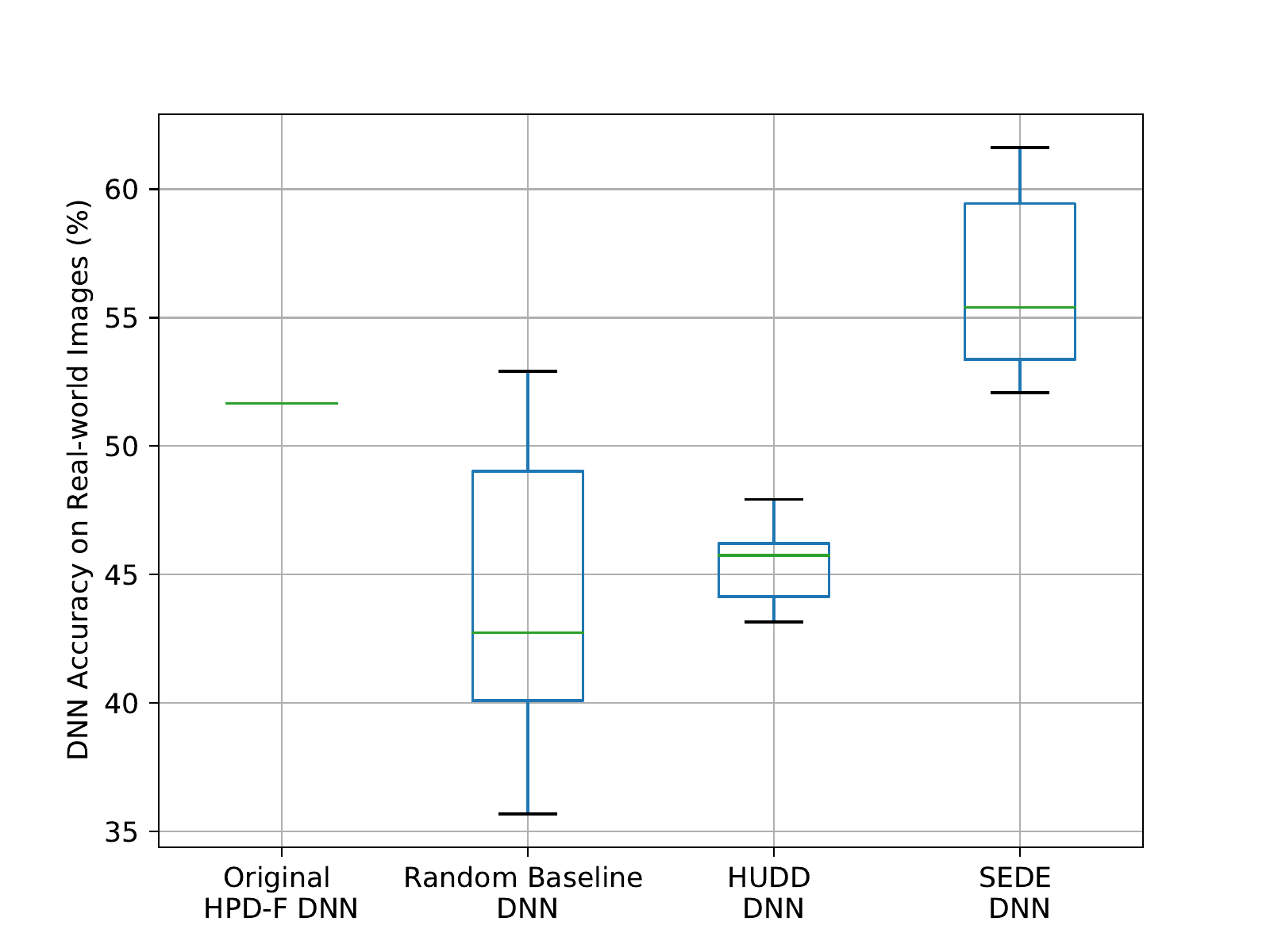}
\includegraphics[width=0.497\textwidth]{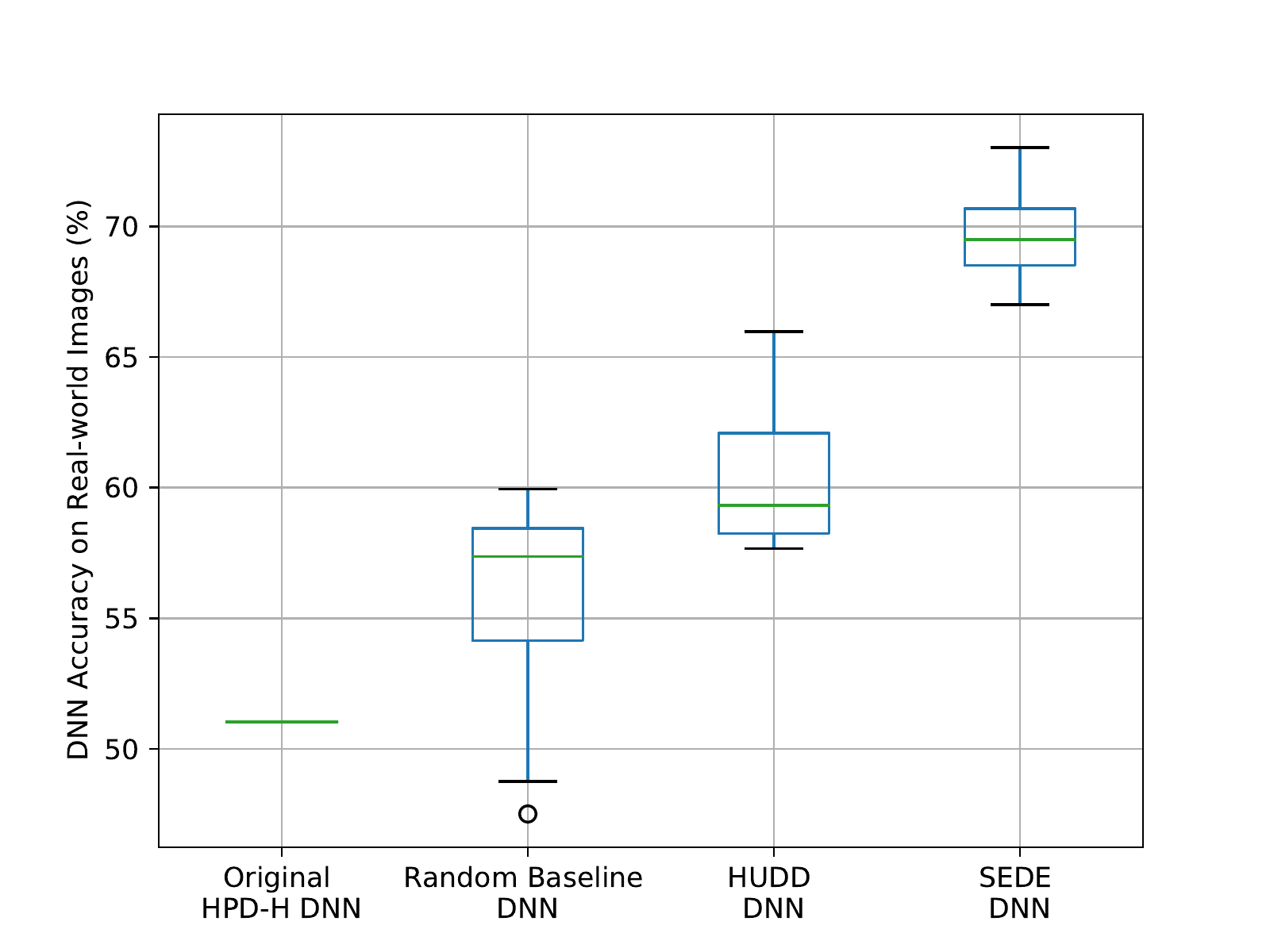}
\includegraphics[width=0.497\textwidth]{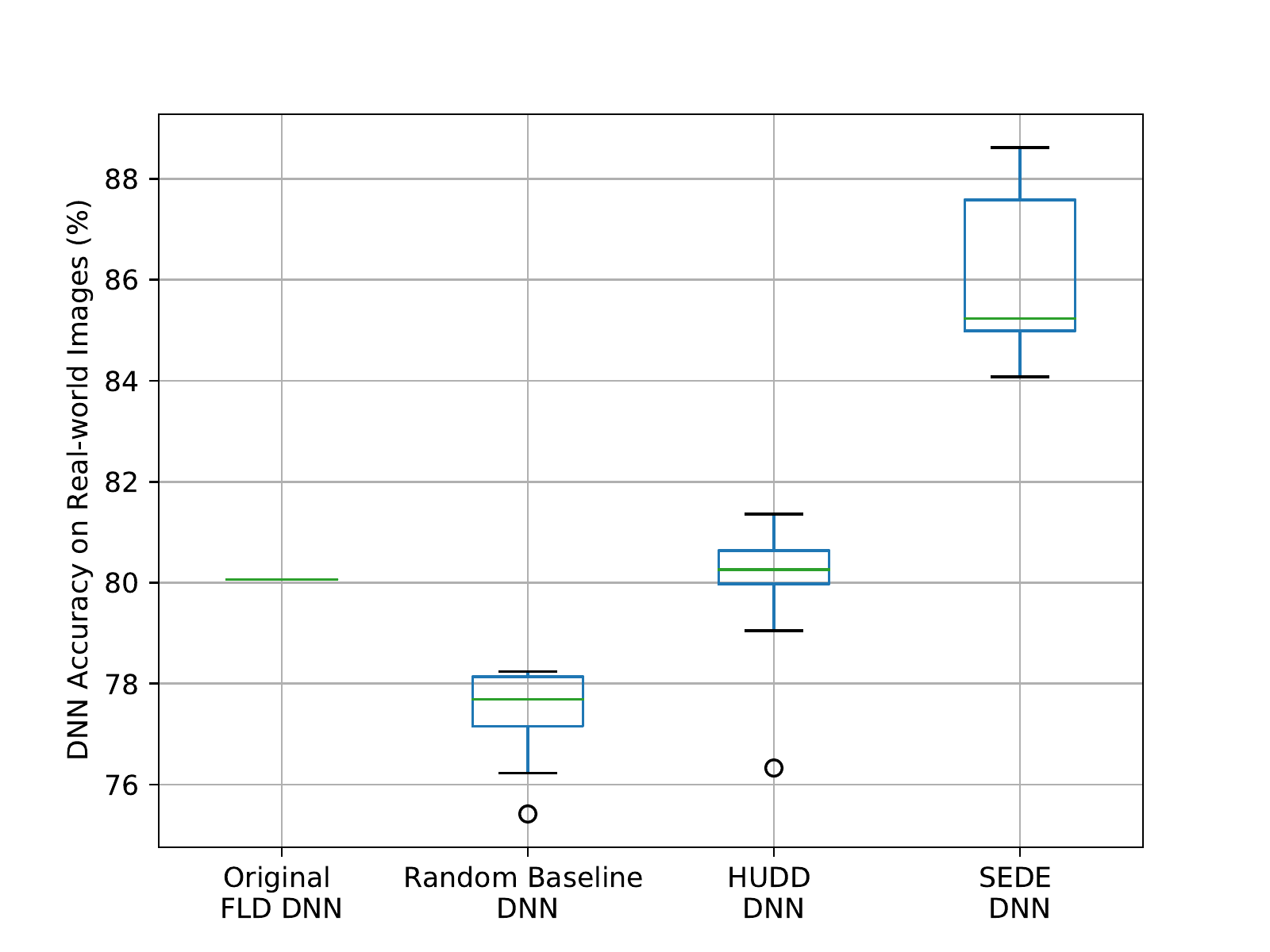}
\caption{DNN accuracy on real-world test set images after retraining by \APPR compared to \HUDD and random baseline for HPD-F, HPD-H and FLD case study DNNs}
\label{fig:RQ5-HPD1}
\end{figure*}

Figure ~\ref{fig:RQ5-HPD1} shows boxplots with the accuracy of the DNNs retrained using \APPR, \HUDD, and the random baseline. Each data point captures the accuracy obtained in one of the ten retraining runs.
The boxplots of \APPR overlap with the boxplot of a competing approach only in one case (i.e., the max value for the random baseline), thus suggesting that \APPR performs significantly better with a strong effect size; indeed,  competing approaches never lead to a better DNN than \APPR.

\subsection{Threats to validity} 

\subsubsection{Internal validity}

In our work, we rely on clusters to capture different root causes of a DNN failure where the quality of clusters largely affects the characterization of an unsafe space. 
This threat is mitigated by empirical results demonstrating that HUDD clusters include images with similar characteristics~\cite{HUDD:TRel}.


\subsubsection{External validity}

The selection of the case study DNNs and simulators may affect the generalizability of results. We alleviate this issue 
by selecting subject DNNs that implement classification and regressions tasks motivated by IEE business needs and addressing problems that are quite common in the automotive industry.

Moreover, we rely on two simulators that differ in their fidelity and number of configuration parameters. As we have seen, these characteristics affect the effectiveness of retraining and \APPR's capacity to identify hazard-triggering events. 
We focus on DNNs processing human faces because they are the focus of our project partners at IEE; we did not consider DNNs performing other tasks (e.g., processing road images) because an industrial case study on that topic was not available for our project. 

\subsubsection{Conclusion validity}

To avoid violating parametric assumptions in our statistical analysis, we rely on a non-parametric test and effect size measure (i.e., Mann Whitney U-test and the Vargha and Delaney’s $\hat{A}_{12}$ statistics, respectively) to evaluate the statistical and practical significance of differences in results. When comparing classification results (i.e., RQ4) we applied the Fisher's exact test, which is commonly used in similar contexts.

Further, for RQ1, we executed the competing approaches for four runs, for each of the $31$ RCCs under analysis, to account for randomness and eliminate bias.
For RQ2 to RQ4, we compared the results obtained with a large number of images. 
For RQ2, we considered $25$ \APPR images for each RCC and $5,470$ unsafe test set images. 
For RQ3, we considered $650$ \APPR images and $1,500$ randomly generated images. 
For RQ4, we considered $500$ \APPR images for each RCC and $1,500$ random test set images. 
For RQ5, because of the stochastic nature of \APPR (e.g., DNN retraining), the experiments were executed over thirty runs. 




\subsubsection{Construct validity}

In RQ1 we aim to evaluate if our approach achieves a higher diversity than a state-of-the-art approach. We rely on the chromosome distance computed for each pair of images generated for a RCC, which is based on the parameter values used to generate images with a simulator. By relying on simulator parameter values we can 
objectively measure diversity.

In RQ2 we evaluate if the images generated by \APPR are close to the medoid of the RCC under analysis. 
We rely on the heatmap distance since it is the distance metric adopted to generate RCCs.

For RQ3, since simulator parameters drive the selection of specific image characteristics (e.g., head direction), we rely on the variance reduction rate for parameter values to objectively assess the similarity across images belonging to a RCC; variance reduction has also been used to evaluate HUDD~\cite{HUDD:TRel}.

RQ4 evaluates if \APPR expressions are useful for delimiting an unsafe space. 
RQ5 evaluates the ability of \APPR to improve a DNN. For both,
we relied on the accuracy metric (i.e., the percentage of correctly classified images), which is also suggested by safety standards as a mean to evaluate if DNNs can be used for safety-related tasks~\cite{SOTIF}.


\subsection{Data Availability}

\MAJOR{R3.E}{Our implementation of \APPR, the IEE simulators, and the data generated to address our research questions are available online~\cite{REPLICABILITY,SEDE:Github}. We cannot share the real-world images used for FLD experiments because they were collected by IEE and protected by privacy agreements.}

\section{Related Work}
\label{sec:related}

\MAJOR{R3.24}{INNvestigate~\cite{innvestigate} and TorchRay~\cite{TorchRay} are well known tools supporting DNN explanation.
However, for explanations concerning DNNs that process images, they generate
one heatmap for every failure-inducing input image; each heatmap
must be visually inspected by engineers, which makes the investigation of many DNN failures highly expensive.}
The cost of the manual inspection of heatmaps is one of the problems addressed by HUDD~\cite{HUDD:TRel}, which groups together similar images thus simplifying the identification of the root causes of DNN failures; however, HUDD still requires domain experts capable of visually spotting the commonalities across images, which is no longer needed with \APPR.

Similarly to HUDD, MODE automatically identifies the images to be used to retrain a DNN~\cite{Ma2018}. 
However, it cannot identify the root causes of DNN failures;
further, in contrast to \APPR, it cannot automatically generate the images to be used for retraining. Like HUDD, MODE's effectiveness remains limited by the availability of an improvement set including unsafe images.
Finally, a reusable tool implementing MODE is not available.




DeepJanus characterizes the frontier of DNN misbehaviours by identifying pairs of inputs that are close to each other, with one input leading to a correct DNN output and the other to a DNN failure \MAJOR{R2.27}{~\cite{Riccio2020}.} It relies on the popular \nsga algorithm extended with an archive (to keep the best individuals found in the search)
and with repopulation (to escape from stagnation by replacing the most dominated individuals with random ones). Like \APPR, DeepJanus relies on a fitness function that includes a measure of sparseness of the solutions; sparseness is measured as the distance from the closest input in the archive, which is populated with inputs having a distance above a given threshold. Different from DeepJanus, \APPR does not require the configuration of a threshold value, which might be particularly expensive in our context (see Section~\ref{sec:diversity}).
Also, \APPR provides explicit explanations for DNN failures represented using expressions constraining parameter values; DeepJanus, instead, presents example images to end-users thus requiring the visual inspection of images like HUDD.
Finally, DeepJanus cannot relate failures observed with images generated by a simulator to failures observed with real-world data, which is a key contribution of \APPR.

\emph{Anchors} are if-then rules that constrain a subset of the input features so that changes to the unconstrained features do not influence the output of the model to be explained~\cite{Ribeiro:18}. The Anchors algorithm constructs an explanation rule iteratively, by interacting with the model. At each iteration it alters the values associated to one input feature, till it identifies a range within which the accuracy is above a given threshold.
The Anchors algorithm works with the test dataset, not simulators; also, when applied to DNN processing images, it does not generate expressions but identifies the image chunks influencing the DNN output, similarly to heatmap-based approaches.


Kim et al.~\cite{kimRules} process images generated with simulators and rely on rule extraction algorithms to characterize correctly and incorrectly classified images in terms of simulator parameters. 
Unlike \APPR, the approach of Kim et al. cannot be applied to real-world images, which are instead necessary to ensure the applicability of the DNN in the field.

Other works~\cite{Feldt19,Gao:Sensei:ICSE:20,Engstrom:2019,DeepFault,RobOT,Xie2019}
concern DNN retraining approaches that aim to improve the DNN robustness by relying on either image transformations (e.g., rotations) or adversarial inputs. \APPR focuses on DNN accuracy not robustness; also, instead of relying on image transformations, it is the first approach relying on simulators to improve the DNN accuracy observed when the DNN is applied to real-world images.

Some DNN testing approaches can provide explanations for portions of the input space in which DNN failures are observed~\cite{abdessalem2018testing,Fitash:ISSTA:2021,Haq:EMSE:2021,zohdinasab2021deephyperion}.
For instance, Abdessalem et al. \cite{abdessalem2018testing} rely on evolutionary algorithms to search for test inputs using simulators and, to maximize test effectiveness, decision trees are used during the search process to learn the regions of the input space that are likely unsafe and, hence, should be targeted by testing. Finally, engineers are presented with decision tree leaves that characterize such portions.
Further, recent work studies the effectiveness of decision trees in characterizing the input space of the simulator-based testing process~\cite{Fitash:ISSTA:2021,Haq:EMSE:2021}.
Finally, DeepHyperion~\cite{zohdinasab2021deephyperion} configures a generative model using a metaheuristic search algorithm directed towards generating test inputs in a specific dimension of the inputs space and provides a set of feature maps which visualize the degree of accuracy obtained for different values of dimensions pairs.
Different from \APPR, these DNN testing approaches can only be used to characterize simulated scenarios; \MAJOR{R3.25}{indeed, they do not integrate any solution (e.g., fitness function based on heatmaps) allowing them to generate inputs that are similar to real-world scenarios.}

To summarize, \APPR is the first approach to automatically derive expressions that characterize the unsafe portion of the input space, based on the outputs obtained with real-world images leading to DNN failures. Also, it is the first approach that leverages the generated explanations in order to retrain and improve the DNN.


\section{Conclusion}
\label{sec:conclusion}

The identification of hazard-triggering events is a safety engineering practice that enables engineers to evaluate the risk associated to potentially hazardous behaviors of a system.
In this paper, we address the problem of characterizing hazard-triggering events affecting the correct execution of DNN-based systems.
We introduced \APPR, a novel approach based on evolutionary algorithms, which generates expressions that 
characterize hazard-triggering events observed in real-world images processed by DNNs.
Such expressions constrain the configuration parameters of a simulator capable of generating images similar to the real-world images under analysis. In turn, they characterize the unsafe portions of the input space.

To identify the unsafe portions of the input space from real-world images, 
\APPR relies on a state-of-the-art approach, HUDD, that generates clusters (i.e., root cause clusters, RCCs) containing images sharing a common set of characteristics. Such commonalities capture the hazard-triggering events to be investigated by engineers.
For each RCC, \APPR performs three distinct executions of evolutionary algorithms; as a result, it generates two sets of images belonging to the RCC, one leading to DNN failures while the other leading to correct DNN outputs. The two sets are then processed by PART, a rule extraction algorithm, to automatically derive \APPR expressions. 
The evolutionary algorithms employed by \APPR are a modified version of \nsga and \GAAlg, an algorithm introduced in this paper to efficiently generate, using a simulator, images that belong to a RCC and are diverse.


Additionally, \APPR improves the DNN under analysis by retraining it with images that match the generated expressions. 

Empirical results conducted with representative case studies in the automotive domain show that
(a) our genetic algorithm, \GAAlg, can generate images for a larger number ($87.1\%$) of RCCs, it generates a larger number of images for each RCC, 
and the generated images have a significantly higher diversity than those generated by \nsga (for all case studies) and \dnsga (for one case study),
(b) the evolutionary searches employed by \APPR lead to images belonging to the RCCs and having similar characteristics, 
(c) the expressions generated by \APPR successfully characterize the unsafe input space (i.e., they lead to images showing a DNN accuracy decreased by at least $30\%$ points, on average), and, 
(d) the retraining process employed by \APPR increases the DNN accuracy up to $18$ percentage points with a gain over the best baseline of at least $6$ percentage points. 



\begin{acks}
This project has received funding from IEE Luxembourg, Luxembourg’s National Research Fund (FNR) under grant BRIDGES2020/IS/14711346/FUNTASY, the European Research Council (ERC) under the European Union’s Horizon 2020 research and innovation programme (grant agreement No 694277), and NSERC of Canada under the Discovery and CRC programs. Authors would like to thank Jun Wang for his contribution to IEE simulators. The experiments presented in this paper were carried out using the HPC facilities of the University of Luxembourg (see http://hpc.uni.lu).
\end{acks}


\bibliographystyle{ACM-Reference-Format}
\bibliography{DNNexplanation}

\end{document}